\documentclass[12pt]{article}
\usepackage{amsmath,amsfonts}
\usepackage{amssymb}
\usepackage{amsthm}
\usepackage{times,txfonts}

\usepackage{accents}

\newtheorem{thm}{Theorem}[section]
\newtheorem{prop}[thm]{Proposition}

\newtheorem{lemma}[thm]{Lemma}
\newtheorem{dfn}[thm]{Definition}

\newtheorem{remark}[thm]{\it Remark}
\newtheorem{example}[thm]{\it Example}

\numberwithin{equation}{section}

\def\pf{\noindent{\it Proof.} \ }

\def\qed{\hfill $\square$}

\newcount\N  \newcount\M  \newdimen\SIZE  \newdimen\INC
\def\YGBOX#1#2#3{
      \N=#1  \M=1  \INC=#2pt  \advance\INC by .#3pt  
      \vbox{
         \loop\ifnum\M>0
            \M=\N
            \divide\N by 10         
            \multiply\N by 10
            \advance\M by -\N
            \divide\N by 10
            \SIZE=\INC
            \multiply\SIZE by \M    
            \advance\SIZE by .#3pt
             \hrule  width \SIZE  height .#3pt
              \hbox{\loop\ifnum\M>0                      
                        \vrule  height #2pt  width .#3pt 
                        \hskip #2pt
                        \advance\M by -1  \repeat
                        \vrule  width .#3pt }
             \hrule  width \SIZE  height .#3pt
            \vskip -.#3pt
         \repeat } }

\def\young#1{{             
       \mathchoice{\YGBOX{#1}61}{\YGBOX{#1}61}{\YGBOX{#1}41}{\YGBOX{#1}31}}}
\newcount\N  \newcount\M  \newdimen\SIZE  \newdimen\INC
\def\YGBOXC#1#2#3{
      \N=#1  \M=1  \INC=#2pt  \advance\INC by .#3pt  
      \vcenter{\vbox{
         \loop\ifnum\M>0
            \M=\N
            \divide\N by 10         
            \multiply\N by 10
            \advance\M by -\N
            \divide\N by 10
            \SIZE=\INC
            \multiply\SIZE by \M    
            \advance\SIZE by .#3pt
             \hrule  width \SIZE  height .#3pt
              \hbox{\loop\ifnum\M>0                      
                        \vrule  height #2pt  width .#3pt 
                        \hskip #2pt
                        \advance\M by -1  \repeat
                        \vrule  width .#3pt }
             \hrule  width \SIZE  height .#3pt
            \vskip -.#3pt
         \repeat } } }

\title{From KP/UC hierarchies to  Painlev\'e equations}

\author{Teruhisa Tsuda   \\
Faculty of Mathematics, Kyushu University, \\  
Fukuoka 819-0395, Japan.}
\date{August 15, 2009 (June 1, 2011 revised)}

\begin{document}
\maketitle

\begin{abstract}

We study
the underlying relationship between
Painlev\'e equations 
and infinite-dimensional integrable systems,
such as the KP and UC hierarchies.
We show that a certain  
reduction of these hierarchies 
by requiring homogeneity and periodicity 
yields Painlev\'e equations,
including their higher order generalization.
This result allows us to clearly understand
various aspects of the equations, e.g.,
Lax formalism, 
Hirota bilinear relations for
$\tau$-functions, 
Weyl group symmetry,  
and algebraic solutions in terms of  
the character polynomials, i.e.,
the Schur function and the universal character.
\end{abstract}

\renewcommand{\thefootnote}{\fnsymbol{footnote}}

\tableofcontents

\section{Introduction}
The present article is aimed 
to develop the study of Painlev\'e equations
by means of a viewpoint of
infinite-dimensional integrable systems.
First of all, to explain our motivation, we recall the special polynomials associated with Painlev\'e equations.
For example, 
the second one
($P_{\rm II}$):
\[
 \frac{{\rm d}^2 q}{{\rm d}x^2}=2 q^3 + x q +a  
 \]
has a particular solution $q \equiv 0$ when $a=0$.
Furthermore
$P_{\rm II}$ has
 the B\"acklund  
transformations  
generated by 
(see, e.g., \cite{gls02})
\begin{align*}
\pi & : q \mapsto -q, \quad a \mapsto -a,\\
r_1 & : q \mapsto q+\frac{a-\frac{1}{2}}{ q^2-  \frac{{\rm d} q}{{\rm d}x}+\frac{x}{2}},
\quad a \mapsto 1-a.
\end{align*}
It follows that
$P_{\rm II}$ 
has a rational solution if 
$a$ is an integer. 
Interestingly enough, the factors appearing in 
the denominator and numerator of the rational solution
form monic polynomials 
with integer coefficients;
we call them the {\it Yablonskii--Vorob'ev polynomials}
\cite{vor65}.
The first few are 
$T_1(x)=x$, 
$T_2(x)=x^3+4$,
$T_3(x)=x^6+20x^3-80$, 
$T_4(x)=x(x^{9}+60 x^3 +11200)$,
etc.
An effective way to understand the nature of these polynomials
is provided by the connection with soliton theory.
Let us consider the (modified) KdV equation
\begin{equation}  \label{eq:mkdv}
4 \frac{\partial v}{\partial t}= -6 v^2 \frac{\partial v}{\partial x} + \frac{\partial^3 v}{\partial x^3},
\end{equation}
which is a typical soliton equation.
Since  
(\ref{eq:mkdv})
is a homogeneous equation (of degree $-4$)
by counting the degree of variables as $\deg x=1$, $\deg t=3$, and $\deg v=-1$,
it admits a similarity solution of the form
$v(x,t)=(-3t/4)^{-1/3}  q((-3t/4)^{-1/3}x)$.
The function $q=q(x)$ thus 
satisfies $P_{\rm II}$; 
see \cite{as77}.
On the other hand, we have previously known 
that
(\ref{eq:mkdv}) has
a rational and similarity solution 
written by the use of the Schur functions attached to staircase partitions.
Finally,
the Yablonskii--Vorob'ev polynomial turns out to be a specialization of the Schur function;
see \cite{ko96}.
The emergence of the Schur function is
significant from our standpoint
because
as shown by Sato \cite{sat}
the KP hierarchy, 
which is the most basic class of soliton equations,
is exactly
an infinite-dimensional integrable system characterized by 
the Schur function.

Besides the second one,
such special polynomials associated with algebraic or rational solutions
have been defined also for other Painlev\'e equations;
they are referred to as the {\it Okamoto polynomial} for $P_{\rm IV}$
and the  {\it Umemura polynomials} collectively for 
 $P_{\rm III}$, $P_{\rm V}$, and $P_{\rm VI}$.
These polynomials are known to
possess interesting features
 from both combinatorial and representation-theoretical point 
 of view (see  \cite{noou}),
 and thus may be regarded 
as `nonlinear analogues' of the classical orthogonal polynomials 
typified by the Jacobi polynomial. 
Along the same lines,
the Okamoto polynomial is expressible in terms of
 the Schur function
attached to a three-core partition,
based on the fact that 
$P_{\rm IV}$ coincides with a certain similarity constraint of the Boussinesq equation
(which belongs to the category of KP hierarchy as well as KdV);
see \cite{ko98, ny99}.
But, however, 
it takes on a different posture concerning the Umemura polynomials for 
$P_{\rm V}$ and $P_{\rm VI}$.
As was discovered by Masuda et al. 
\cite{mas03, mok02},
they are in fact expressed by 
the {\it universal character} (attached to a pair of staircase partitions).

The universal character $S_{[\lambda,\mu]}$, 
defined by Koike \cite{koi89},
is a polynomial attached to a pair of partitions $[\lambda,\mu]$
and is a generalization of the Schur function $S_\lambda$.
While the latter, as is well known, 
describes the character of 
an irreducible polynomial representation 
of the general linear group,
the former does that of a rational one.  
Inspired by the connection 
between the KP hierarchy and the Schur function,
the author proposed in \cite{tsu04} 
an extension of the KP hierarchy,
called the {\it UC hierarchy},
 as an infinite-dimensional integrable system characterized by the universal character.
 \begin{center}
 \begin{tabular}{ccc}
 \hline
 Character polynomials & 
 versus 
 & Infinite integrable systems
 \\
 \hline
 Schur function $S_{\lambda}$
 &&
 KP hierarchy 
 \\
 $\cap$ &&$\cap$
 \\
 Universal character $S_{[\lambda,\mu]}$
 &&
 UC hierarchy  \\
 \hline 
 \end{tabular}
 \end{center}

In this paper,
expanding  
our subject to the UC hierarchy beyond the KP hierarchy,
we present a unified derivation of Painlev\'e equations
(including their higher order analogues)
from 
infinite-dimensional integrable systems
via a certain similarity reduction.
As a corollary 
we clarify the origin not only 
of the special polynomials 
but  also
of various aspects of Painlev\'e equations, e.g.,
bilinear relations for $\tau$-functions,  
Weyl group symmetry, and Lax formalism.

The KP hierarchy was originally introduced as 
a series of nonlinear partial differential equations
associated with an auxiliary linear problem.
In this paper, however, we adopt 
an equivalent definition of the KP hierarchy
due to Date--Jimbo--Kashiwara--Miwa (see \cite{mjd}) 
by a single functional equation
(called the {\it bilinear identity})
for an unknown function 
$\tau=\tau({\boldsymbol x})$
in infinitely many time variables 
${\boldsymbol x}=(x_1,x_2,\ldots)$.
Likewise, the UC hierarchy can be 
defined by
 a system of two equations 
for an unknown function 
$\tau=\tau({\boldsymbol x})$
in  
$({\boldsymbol x}, {\boldsymbol y})=(x_1,x_2,\ldots, y_1,y_2,\ldots)$.
 It should be noted that 
 if we count the degree of each variable as $\deg x_n=n$ and $\deg y_n=-n$, then both hierarchies 
 come out to be homogeneous, 
and thereby admit similarity solutions. 
 In other words, 
 the UC hierarchy 
 well
 generalizes  the KP hierarchy
 by taking the negative time evolutions into account
  besides the positive ones
while keeping its homogeneity.
Remarkably, the homogeneous polynomial solutions of the KP  (resp. UC) hierarchy 
are filled with the Schur functions (resp. universal characters).
We summarize below some fundamental data to illustrate 
difference between the KP and UC hierarchies.
For details to Sects.~\ref{sect:kp} and  \ref{sect:uc}.
\begin{center}
\small {\bf Table 1}. Comparison of  KP and UC hierarchies
\\ 
\normalsize
\begin{tabular}{ccc}
\hline
  & KP hierarchy & UC hierarchy\\
\hline
Time variables & \begin{tabular}{c}
${\boldsymbol x}=(x_1,x_2,\ldots)$
\\
$\deg x_n=n$
\end{tabular}
&  \begin{tabular}{c}
 $({\boldsymbol x},{\boldsymbol y})=(x_1,x_2,\ldots, y_1,y_2,\ldots)$
 \\
$\deg x_n=n$, $\deg y_n=-n$
\end{tabular}
\\
\hline
Dependent variables &  $\tau=\tau({\boldsymbol x})$ &
$\tau=\tau({\boldsymbol x},{\boldsymbol y})$
\\
\hline
Bilinear identities & 
$ \sum_{i+j=-1} X^{-}_i \tau \otimes X^{+}_j \tau =0$ 
& 
$\begin{array}{l}
\sum_{i+j=-1} X^{-}_i \tau \otimes X^{+}_j \tau \\
=
\sum_{i+j=-1} Y^{-}_i \tau \otimes Y^{+}_j \tau =0
\end{array}
$
\\
\hline
Vertex operators &
$\sum_{n \in {\mathbb Z}} X_n^\pm z^n
=e^{ \pm  \xi({\boldsymbol x},z)}e^{ \mp \xi(\tilde{\partial}_{\boldsymbol x},z^{-1})}$
&
$ \begin{array}{c}
\sum_{n \in {\mathbb Z}} X_n^\pm z^n
=e^{ \pm  \xi({\boldsymbol x}-\tilde{\partial}_{\boldsymbol y},z)}e^{ \mp \xi(\tilde{\partial}_{\boldsymbol x},z^{-1})}, \\
\sum_{n \in {\mathbb Z}} Y_n^\pm z^{-n}
=e^{ \pm  \xi({\boldsymbol y}-\tilde{\partial}_{\boldsymbol x},z^{-1})}e^{ \mp \xi(\tilde{\partial}_{\boldsymbol y},z)}
\end{array}$
\\
\hline
\begin{tabular}{c}
Homogeneous \\ polynomial  solutions 
\end{tabular}
& $S_\lambda=S_{\lambda}({\boldsymbol x})$
& $S_{[\lambda,\mu]}=S_{[\lambda,\mu]}({\boldsymbol x},{\boldsymbol y})$ 
\\
\hline
Phase space &
$SGM$: Sato Grassmannian 
&
$SGM \times SGM$
\\
\hline
\end{tabular}
\end{center}

Moreover, 
we can derive  
from the original ones similar bilinear identities
among solutions generated by successive application of vertex operators;
let
$\tau_n=\tau_n({\boldsymbol x})$ 
and $\tau_{m,n}=\tau_{m,n}({\boldsymbol x},{\boldsymbol y})$
denote
such sequences of solutions for the
KP and UC hierarchies, respectively. 
In particular, the set of functional equations 
satisfied by the contiguous solutions 
is called  the {\it modified} hierarchy. 
For instance,
the modified KP hierarchy consists of 
the equation
\[  
\sum_{i+j=-2} X^{-}_i \tau_n \otimes X^{+}_j \tau_{n+1} =0.
\]
Also, a typical equation of the modified UC hierarchy is 
(cf. Example~\ref{example:muc})
\[
\tau_{m,n} \otimes \tau_{m+1,n+1}
= \sum_{i+j=0} X_i^{-} \tau_{m+1,n} \otimes X_j^{+} \tau_{m,n+1}.
\]
The modified hierarchies play an essential role in 
the
relationship to Painlev\'e equations.

Principle ingredients of the similarity reduction 
are {\it homogeneity}, {\it periodicity}, and {\it specialization}.
Firstly, 
since
the (modified) KP and UC hierarchies are homogeneous, 
it is possible to restrict them to the
self-similar solutions,
i.e.,
ones that satisfy
\begin{equation}
E \tau_{n}=d_n \tau_n \quad \text{and} \quad E' \tau_{m,n}=d_{m,n} \tau_{m,n}
\label{eq:intro_euler}
\end{equation}
with some constants 
$d_n, d_{m,n} \in {\mathbb C}$,
where 
\[
E= \sum_{n=1}^\infty
 n x_n \frac{\partial }{\partial x_n} 
\quad   \text{and}  \quad
E'=\sum_{n=1}^\infty 
\left(
n x_n \frac{\partial }{\partial x_n} -n y_n \frac{\partial }{\partial y_n} \right)
\]
for the KP and UC cases, respectively.
Secondly
we reduce the dimension of phase space
by imposing the periodic condition on the dependent variables
as
$\tau_{n+\ell}=\tau_n$ 
and 
$\tau_{m+\ell_1,n+\ell_2}=\tau_{m,n}$. 
Finally,
in order to obtain 
ordinary differential equations,
we need to choose 
a suitable
direction of
 time flow,
i.e., 
specialization of independent variables.
The result is stated as follows:
\begin{thm}  \label{thm:main}
The {\rm(}higher order{\rm)} Painlev\'e equations
$P_{\rm II}$, $P(A_{\ell-1}^{(1)})$, 
and $P_{\rm III}$-chain of order $2 \ell-2$
can be obtained as a certain similarity reduction of the modified KP
hierarchy 
with period of order
$2$, $\ell$ $(\geq 3)$, and $\ell$ $(\geq 2)$, 
respectively.
Likewise, both
$P(A_{2 \ell - 1}^{(1)})$ and  $P_{\rm VI}$-chain of order 
$2 \ell-2$ $(\ell \geq 2)$
can be obtained as that of the modified UC hierarchy 
with $(\ell,\ell)$-periodicity.
\end{thm}
\noindent
(See Tables 2 and 3 below.)

Here
the symbol $P(A_{\ell -1}^{(1)})$ represents 
the higher order Painlev\'e equation of type $A_{\ell-1}^{(1)}$
(\cite{ny98}) or, equivalently,  
the Darboux chain 
with period $\ell$
(\cite{adl94, vs93});
this is a further generalization of $P_{\rm IV}$ and $P_{\rm V}$,
and indeed recovers the original ones if $\ell=3$ and $\ell=4$, respectively.
The $P_{\rm III}$-chain is a 
higher order analogue of $P_{\rm III}$
(\cite{asy00, sha, w-h}).
According to its Lax pair (see Sect.~\ref{subsect:p6_lax}),
the $P_{\rm VI}$-chain is identified as 
a certain subfamily of the Schlesinger systems
(\cite{sch12}).
Both $P_{\rm III}$- and $P_{\rm VI}$-chains literally 
include the originals as their lowest order members.
Note that part of Theorem~\ref{thm:main}
about the reduction of the KP hierarchy to $P_{\rm II}$
and $P(A_{\ell -1}^{(1)})$ has been known from \cite{as77}
and \cite{n, schiff94}; see also 
Remark~\ref{remark:tables}.
The present result generalizes these previous ones to cover all the classical six Painlev\'e equations 
(including their higher order analogues)
by involving the UC hierarchy besides the KP hierarchy.
We emphasize that our study is based {\it only} on
the bilinear identities of the KP and UC hierarchies,
which play the central role as 
master equations
in soliton theory.

Now, let us describe some advantages of the similarity reduction
in Painlev\'e equations.
\begin{itemize}

\item[(i)] (Bilinear form of Painlev\'e equations).
In contrast with its original nonlinear one,
an alternative quadratic expression of a Painlev\'e equation
is often called the 
({\it Hirota}) {\it bilinear form}.
Via the similarity reduction, 
the bilinear forms can be reduced directly 
from 
the bilinear identities of the KP and UC hierarchies,
which are {\it a priori} 
quadratic relations.

\item[(ii)] (Algebraic solutions in terms of the character polynomials).
Since the Schur function and universal character are the
homogeneous polynomial solutions of the 
KP and UC hierarchies, respectively,
they are consistent with  the similarity reduction
and thus give rise to algebraic solutions
of Painlev\'e equations.

\item[(iii)] (Weyl group symmetry).
A sequence of homogeneous solutions of the KP hierarchy
admits an action of 
a Weyl group of type $A$
generated by a permutation 
of two serial vertex operators at each site.
Likewise, 
for the UC case we have
a commutative pair of 
Weyl group actions of type $A$;
this distinction, by the way,
 reflects 
the presence of two kinds of vertex operators 
$\{X_i^\pm\}$ and $\{Y_i^\pm\}$, 
which commute with each other. 
The above explains 
one origin of Weyl group symmetry of Painlev\'e
equations.

\item[(iv)] (Lax formalism).
The KP hierarchy amounts to the complete integrability condition
of a system of linear equations,
whose dependent variables
are 
\[
\psi_n({\boldsymbol x},k)= \frac{ \tau_n ({\boldsymbol x}-[k^{-1}])}{\tau_n({\boldsymbol x})} e^{\xi({\boldsymbol x}, k)},
\]
where $[k]=(k,k^2/2,k^3/3,\ldots)$ and 
$\xi({\boldsymbol x},k)=\sum_{n=1}^\infty x_n k^n$.
We call $\psi_n=\psi_n({\boldsymbol x},k)$ a {\it wave function}
and an extra parameter $k$ a {\it spectral variable};
see, e.g., \cite{mjd}.
The wave functions are extendedly 
defined 
for the UC hierarchy also as
\[
\psi_{m,n}({\boldsymbol x},{\boldsymbol y},k)
= \frac{ \tau_{m,n-1}({\boldsymbol x}-[k^{-1}],{\boldsymbol y}-[k])  }{ \tau_{m,n}({\boldsymbol x},{\boldsymbol y}) }e^{\xi({\boldsymbol x},k)}.
\]
The bilinear identities generate the linear 
equations for the wave functions.
Under the similarity reduction
they naturally induce an associated linear system 
with a Painlev\'e equation,
i.e., a {\it Lax pair};
one of which is the linear ordinary differential equation
with respect to the spectral variable $k$ and the other governs
its monodromy preserving deformation.
Note that, by this means, compatibility of the Lax pair 
is {\it a priori} established.
\end{itemize}
Among the most importance is the Lax formalism  
because: it enables us 
not only to classify the resulting Painlev\'e equation 
by singularity type of its associated linear equation
but also to detect 
an appropriate dependent variable
to translate its bilinear form into nonlinear one.

The following tables 2 and 3 show
 the corresponding choice to each individual 
 Painlev\'e equation, i.e., the periodicity 
and specialization of time variables 
imposed on the KP and UC hierarchies.
We also indicate
(in the fourth column of each table)
the number of singularities,
 and {\it Poincar\'e rank} for an irregular one,
of the associated linear 
ordinary differential equation with respect to the spectral variable.
Concerning the specialization 
(see the second column),
we note that 
$s$ or $t$
is converted to the independent variable of a Painlev\'e equation
through the similarity reduction;
in parallel,
$a$ and $b$, 
together with $d_n$ and $d_{m,n}$ 
appearing in (\ref{eq:intro_euler}), 
go over to constant parameters of it.

\begin{center}
\small {\bf Table 2}. From KP hierarchy to Painlev\'e equations 
\\ 
\normalsize
\begin{tabular}{ccccc}
\hline
Period  & Specialization on $\boldsymbol x$ & Painlev\'e eq. & Linear ODE in $z=k^\ell$ & Ref.
\\
\hline
$\ell = 2$ & $x_n=0$ $(n \neq 1,3)$ & $P_{\rm II}$ & 
\begin{tabular}{l} 
$2 \times 2$-system with\\
1 reg. sing.
\\
1 irreg. sing. ($\text{rk} = 3/2$)
\end{tabular}
&
\begin{tabular}{l}
Sect.~\ref{sect:p2}; \\
cf.  \cite{as77}
 \end{tabular}
 \\
\hline
$\ell$ $(\geq 3)$ & $x_n=0$ $(n \neq 1,2)$ &
\begin{tabular}{l}
$P(A_{\ell -1}^{(1)})$: \\
$\ell=3$ $\Rightarrow$ $P_{\rm IV}$, \\
$\ell=4$ $\Rightarrow$ $P_{\rm V}$
\end{tabular}
& 
\begin{tabular}{l} 
$\ell  \times \ell$-system with \\
1 reg. sing.
\\
1 irreg. sing. ($\text{rk} = 2/\ell$)
\end{tabular}
& 
\begin{tabular}{l}
Sect.~\ref{sect:p45};
\\
cf. \cite{n} 
\end{tabular} 
 \\
\hline
$\ell$ $(\geq 2)$ & 
\begin{tabular}{l} 
$x_1= s+a$, \\ 
$x_n=a/n$ $(n \neq 1)$
\end{tabular}
& 
\begin{tabular}{l}
$P_{\rm III}$-chain: 
 \\
$\ell=2$ $\Rightarrow$ $P_{\rm III}$
\end{tabular}
& 
\begin{tabular}{l} 
$\ell  \times \ell$-system with \\
2 reg. sing.
\\
1 irreg. sing. ($\text{rk} = 1/\ell$)
\end{tabular}
& Sect.~\ref{sect:p3}
 \\
\hline
\end{tabular}
\end{center}

\begin{center}
\small {\bf Table 3}. From UC hierarchy to Painlev\'e equations
\\ 
\normalsize
\begin{tabular}{ccccc}
\hline
Period   & Specialization on $({\boldsymbol x}, {\boldsymbol y})$ & Painlev\'e eq. & Linear ODE in $z=k^\ell$ & Ref.\\
\hline
$(\ell, \ell)$ & 
\begin{tabular}{l} 
$x_n=s+a/n$, 
\\ 
$y_n=-s+a/n$
\end{tabular}
&
\begin{tabular}{l}
 $P(A_{2\ell -1}^{(1)})$: \\
 $\ell=2$ $\Rightarrow$ $P_{\rm V}$
 \end{tabular}
   & 
\begin{tabular}{l} 
$\ell  \times \ell$-system with \\
2 reg. sing.
\\
1 irreg. sing. ($\text{rk} = 1$)
\end{tabular}
& Sect.~\ref{sect:p5}
   \\
\hline
$(\ell, \ell)$ &
\begin{tabular}{l} 
$x_n= (a+b t^n)/n$, 
\\ 
$y_n= (a+b t^{-n})/n$
\end{tabular}  
& \begin{tabular}{l}
$P_{\rm VI}$-chain:
 \\
$\ell=2$ $\Rightarrow$ $P_{\rm VI}$
\end{tabular}
 & 
 \begin{tabular}{l} 
$\ell  \times \ell$-system with \\
4 reg. sing. (Fuchsian)
\end{tabular}
& Sect.~\ref{sect:p6}
 \\
\hline
\end{tabular}
\end{center}

\begin{remark} \label{remark:tables}
\rm
Since the pioneering work of
Ablowitz and Segur \cite{as77},
the similarity reduction from  
the KdV equation to $P_{\rm II}$ 
has been well known.
The connection between
$P(A_{\ell-1}^{(1)})$ and  the KP hierarchy
 was first pointed out by 
 Schiff \cite{schiff94},
 and it was studied independently by
 Noumi and Yamada 
 (see, e.g., \cite{n})
 from a group-theoretical point of view.
Their theory still has been developed 
with involving the Drinfel'd--Sokolov hierarchy
(\cite{ds85})
and achieved various higher order 
Painlev\'e equations;
see \cite{fs06, fs08, fs09, kk07,  ny02, sas06}.
It would be an  interesting and important
problem
to examine 
their relevance to our present results.
\end{remark}

\begin{remark} \rm
The present work
mainly deals with nonlinear
ordinary differential equations of 
isomonodromic type.
However, 
our approach remains valid for the partial differential case.
In \cite{tsu09b},
we explore 
the similarity reduction from the UC hierarchy to
a broad class of the Schlesinger systems 
including the $P_{\rm VI}$-chain and 
the Garnier system.
See also \cite{tsu10} for their hypergeometric solutions.
\end{remark}

In the rest of this paper, 
we investigate each of Painlev\'e equations
on the basis of the
KP and UC hierarchies.
In the next section, we begin by a brief review about 
the KP hierarchy
and then 
construct a sequence of its homogeneous solutions 
applying the vertex operator technique.
We present a Weyl group symmetry of type $A$ acting on this sequence
and arrange some useful formulae for the following three sections.  
In Sect.~\ref{sect:p3}, we derive
 the $P_{\rm III}$-chain
 from the (modified) KP hierarchy
 through a similarity reduction.
We show how its bilinear form, Weyl group symmetry, rational solutions expressed in terms of the Schur function, and Lax formalism are systematically created.
Similarity reductions of the KP hierarchy to 
$P(A_{\ell -1}^{(1)})$ and $P_{\rm II}$ are the subjects of
Sects.~\ref{sect:p45} and \ref{sect:p2}, respectively;
though the result itself is essentially known,  
we demonstrate concisely the reduction 
to clarify our simple idea that various aspects of Painlev\'e equations
originate from the bilinear identities.
Section~\ref{sect:uc} provides
an introduction of the UC hierarchy,
which is an extension of the KP hierarchy,
with preliminaries to the 
last two sections.
Similarity reduction of the (modified)
UC hierarchy to $P(A_{2\ell -1}^{(1)})$
is studied in Sect.~\ref{sect:p5}.
Interestingly enough, we find a Lax pair of $P(A_{2\ell -1}^{(1)})$
different in both size and singularity type
from that given in Sect.~\ref{sect:p45};
cf. Tables 2 and 3. 
In Sect.~\ref{sect:p6},
we produce the $P_{\rm VI}$-chain,
a higher order analogue of $P_{\rm VI}$,
as a reduction of the UC hierarchy.
The corresponding linear equation is a Fuchsian system
with four regular singularities;
 thus, the $P_{\rm VI}$-chain is equivalent to 
 a particular case of the Schlesinger systems.
 In the appendix, we briefly indicate
a derivation of $P_{\rm I}$
from the KP hierarchy 
by using a Virasoro operator.

\section{KP hierarchy}  \label{sect:kp}
In this section, we 
recall some basic facts about 
the KP hierarchy
starting from the vertex operators associated with the Schur function.
Two contiguous solutions 
connected by the vertex operators satisfy a certain bilinear relation,
which we call the {\it modified} KP hierarchy;
it will be crucial for investigating the link to Painlev\'e equations.
We consider a sequence of homogeneous solutions of the hierarchies 
and present its Weyl group symmetry of type $A$. 
Some relevant formulae are also prepared for the following three sections.

\subsection{Schur function, vertex operator and KP hierarchy}
\label{subsec:kp}

We begin by recalling the definition of the Schur function
and then introduce the vertex operators which 
play roles of raising operators of it.
A {\it partition}
$\lambda=(\lambda_1, \lambda_2, \ldots )$
is a sequence of non-negative integers such that
$\lambda_1 \geq \lambda_2 \geq \cdots \geq 0$
and that $\lambda_i =0$ for $i \gg 0$.  
The number $l=l(\lambda)=\{ i \, | \, \lambda_i \neq 0\}$ is called the {\it length} of $\lambda$;
and the sum $|\lambda|=\lambda_1+\lambda_2+ \cdots +\lambda_l$
is called the {\it weight} of $\lambda$.
The {\it Schur function}  $S_\lambda=S_\lambda({\boldsymbol x})$
attached to a partition $\lambda$
 is a polynomial in
${\boldsymbol x}=(x_1,x_2,\ldots)$ 
determined by the Jacobi--Trudi formula (see, e.g., \cite{mac}):
\begin{equation}
S_\lambda({\boldsymbol x}) =\det \left( p_{\lambda_i-i+j}({\boldsymbol x}) \right)_{1 \leq i,j \leq l},
\end{equation}
where
$p_n$ $(n \in {\mathbb Z})$ is defined by the generating function
\begin{equation}
\sum_{n \in {\mathbb Z}} p_n({\boldsymbol x})k^n =e^{\xi({\boldsymbol x},k)}
\quad \text{and} \quad
\xi({\boldsymbol x},k)=\sum_{n=1}^\infty x_n k^n
\end{equation}
or, equivalently,
$p_{n}=0$ $(n < 0)$, $p_0=1$, and 
\[p_n = \sum_{k_1+2k_2+\cdots+n k_n= n} 
\frac{ {x_1}^{k_1} {x_2}^{k_2}  \cdots {x_n}^{k_n}   }{ k_1 ! k_2 ! \cdots  k_n !}.
\]
If we count the degree of variable $x_n$ 
as $\deg x_n=n$,
then $S_\lambda$ 
is a (weighted) homogeneous polynomial of degree $|\lambda|$.

Introduce the partial differential operators 
\begin{equation} \label{eq:vo}
X^\pm(k)
=\sum_{n \in {\mathbb Z}} X_n^\pm k^n
=e^{ \pm  \xi({\boldsymbol x},k)}e^{ \mp \xi(\tilde{\partial}_{\boldsymbol x},k^{-1})},
\end{equation}
called the {\it vertex operators}.
Here $\tilde{\partial}_{\boldsymbol x}$
stands for 
$\left( \frac{\partial}{\partial x_1}, \frac{1}{2}  \frac{\partial}{\partial x_2}, \frac{1}{3}  \frac{\partial}{\partial x_3}, \ldots
\right)$.
It is worth mentioning that
the operator $X_n^+$ is a raising operator of the Schur function
in the following sense:
\begin{equation} \label{eq:raising_schur}
S_{\lambda}({\boldsymbol x})=X^+_{\lambda_1} \ldots X^+_{\lambda_l} . 1
\end{equation}
for a partition $\lambda=(\lambda_1,\ldots,\lambda_l)$.
Let us now formulate the KP hierarchy
by using $X_n^\pm$.

\begin{dfn}  \label{dfn:kp}
\rm
For an unknown function $\tau=\tau({\boldsymbol x})$,
the bilinear equation
\begin{equation}  \label{eq:kp}
\sum_{i+j=-1} X^{-}_i \tau \otimes X^{+}_j \tau =0
\end{equation}
is called the {\it KP hierarchy}.
\end{dfn}

We regard $f \otimes g = f({\boldsymbol x'}) g({\boldsymbol x})$
as an element of ${\mathbb C}[{\boldsymbol x'}] \otimes {\mathbb C}[{\boldsymbol x}] $.
It is then  obvious that  
(\ref{eq:kp})
can be rewritten into 
the equation
\begin{equation}   \label{eq:kp-res}
\frac{1}{2 \pi \sqrt{-1}} 
\oint e^{\xi( {\boldsymbol x}-{\boldsymbol x'},z)} 
{{\rm d} z}   \,
 \tau ({\boldsymbol x'}+[z^{-1}]) 
\tau ({\boldsymbol x}-[z^{-1}])  =0
\end{equation}
with ${\boldsymbol x}$ and ${\boldsymbol x'}$
being arbitrary parameters.
Here the symbol 
$[t]$ denotes 
$( t, t^2/2, t^3/3, \ldots  )$ 
and
the integration
$\oint \frac{{\rm d} z}{2 \pi \sqrt{-1}}  $
means taking the coefficient of $1/z$ of
the integrand as a (formal) 
Laurent expansion in $z$.
By choosing appropriately
the specialization of the arbitrary parameters ${\boldsymbol x}$ and ${\boldsymbol x'}$
in (\ref{eq:kp-res}),
we can derive various functional equations for $\tau$.
For instance, 
the Taylor expansion at ${\boldsymbol x}={\boldsymbol x'}$
yields an infinite series of nonlinear differential equations
(cf. Sect.~\ref{subsect:pre_mkp}),
the first member of which is
\begin{equation} \label{eq:kp-orig}
\left(
{D_{x_1}}^4+3 {D_{x_2}}^2-4 D_{x_1} D_{x_3}
\right) \tau \cdot \tau
=0.
\end{equation}
Here 
recall the definition of the {\it Hirota differential}
\begin{equation} \label{eq:def_hirota}
P(D_{\boldsymbol x}) 
f({\boldsymbol x}) \cdot g({\boldsymbol x}) 
=
\left.
P(\partial_{\boldsymbol a})
f({\boldsymbol x}+{\boldsymbol a}) g({\boldsymbol x}-{\boldsymbol a})
\right|_{{\boldsymbol a}={\boldsymbol 0}}
\end{equation}
for  a polynomial $P(D_{\boldsymbol x})$ in 
$D_{\boldsymbol x}=(D_{x_1},D_{x_2}, \ldots)$.
Through the change of variables
\begin{equation} \label{eq:tau2u}
u=2 \left(\frac{\partial}{\partial x_1}\right)^2 \log \tau,
\end{equation}
in fact,
(\ref{eq:kp-orig}) is converted to 
the KP (Kadomtsev--Petviashvili) equation
\[
\frac{3}{4} \frac{\partial^2 u }{\partial {x_2}^2}
=  \frac{\partial }{\partial x_1}
\left(
\frac{\partial u }{\partial x_3} -\frac{3}{2} u \frac{\partial u }{\partial x_1}
-\frac{1}{4}  \frac{\partial^3 u }{\partial {x_1}^3}
\right).
\]
As one can read from (\ref{eq:tau2u}),
it is quite natural to require 
$\tau({\boldsymbol x})$
to be an entire function.
We call an entire function $\tau({\boldsymbol x})$
solving the KP hierarchy
a {\it $\tau$-function}.
Note that $\tau$-functions are distinguished up to constant multiplication.
The celebrated Sato theory revealed that
 each solution of the KP hierarchy is parameterized by a point of
 an infinite-dimensional Grassmann manifold, called the {\it Sato Grassmannian}, 
 and the  $\tau$-functions emerge from 
its Pl\"ucker coordinates;
all the Hirota differential equations, 
like (\ref{eq:kp-orig}),
constituting the KP hierarchy
are identified with the Pl\"ucker relations.
Moreover,
from the
viewpoint of infinite-dimensional Lie algebra,
 the $\tau$-functions can be described  
as the orbit of a trivial one
 $\tau \equiv 1$
under  
the action of vertex operators;
see, e.g., \cite{mjd}.

\begin{remark}[Scaling symmetry of the KP hierarchy]
\label{remark:kp_sim}
\rm
If $\tau=\tau(x_1,x_2,x_3,\ldots)$ 
is a solution of the KP hierarchy, (\ref{eq:kp}),
then 
so is $\tau(c x_1,c^2 x_2,c^3 x_3,\ldots)$ for any $c \in {\mathbb C}^\times$;
thus, it is meaningful to take interest in 
solutions invariant under this scaling symmetry up to constant multiplication.
For example, the Schur function 
$S_\lambda({\boldsymbol x})$ is a solution of the KP hierarchy, 
and possesses the homogeneity 
$S_\lambda(c x_1,c^2 x_2,c^3 x_3,\ldots)=c^{|\lambda|} S_\lambda({\boldsymbol x})$.
 In other words,
$E S_\lambda({\boldsymbol x}) = |\lambda| S_\lambda({\boldsymbol x})$
with $E$ denoting the Euler operator:
\begin{equation}  \label{eq:euler}
E= \sum_{n=1}^\infty n x_n \frac{\partial}{\partial x_n}.
\end{equation}
It should be noted that
the whole set of homogeneous polynomial solutions of the KP hierarchy
is equal to that of the Schur functions;
see  \cite{sat}.
 \end{remark}

Now let us put our attention to the
functional relations for a sequence of 
$\tau$-functions
connected by successive application of vertex operators.
Suppose $\tau_0:=\tau({\boldsymbol x})$ to be a solution of the KP hierarchy, (\ref{eq:kp}).
Let $\tau_1:= X^+(\alpha) \tau({\boldsymbol x})$  
with an arbitrary constant $\alpha \in {\mathbb C}^\times$. 
Then $\tau_1$ solves  (\ref{eq:kp}) again. 
Moreover  we can deduce 
the bilinear equation
$\sum_{i+j=-2} X^{-}_i \tau_0 \otimes X^+_j \tau_1 =0$
from (\ref{eq:kp}) multiplied by
 $1 \otimes X^+(\alpha)$ 
with the aid of 
the {\it fermionic} relations
\begin{equation} \label{eq:ferm}
X_i^{\pm} X_j^\pm+X_{j-1}^\pm X_{i+1}^\pm =0
\quad \text{and}
\quad
X_i^+ X_j^-+X_{j+1}^- X_{i-1}^+ =\delta_{i+j,0}.
\end{equation}

\begin{dfn} \label{dfn:mkp}
\rm
For a sequence 
$\tau_n=\tau_n({\boldsymbol x})$ $(n \in {\mathbb Z})$
of unknown functions,
the system of bilinear equations
\begin{equation}  \label{eq:mkp-bil}
\sum_{i+j=-2} X^-_i \tau_n \otimes X^+_j \tau_{n+1} =0
\end{equation}
is called the {\it modified KP hierarchy} or, shortly, {\it mKP hierarchy}.
\end{dfn}

 \subsection{A homogeneous $\tau$-sequence and its Weyl group symmetry}
 \label{subsec:hom}

Define a partial differential operator $V(c)$ 
$(c \in {\mathbb C})$ by
\[
V(c)= \int_\gamma X^+(k) k^{-c-1} {\rm d} k, 
\]
where the path $\gamma: [0,1] \to {\mathbb C}$ 
is well chosen such that
$\left[X^+(k) k^{-c} \right]_{\gamma(0)} ^{\gamma(1)}=0$;
thus, $\gamma$ may depend on $c$ in general. 
One can verify in the same way as Sect.~\ref{subsec:kp}
that:
if $\tau_0({\boldsymbol x})$ is a solution of the KP hierarchy
then so is $\tau_1= V(c) \tau_0$
and, 
moreover,
the pair $(\tau_0,\tau_1)$ satisfies (\ref{eq:mkp-bil}).  
In this sense we may call also $V(c)$ a vertex operator. 
An interesting feature of our vertex operator
$V(c)$
is its homogeneous property:
if a function $f=f({\boldsymbol x})$ is 
an eigenfunction of the Euler operator (see (\ref{eq:euler})),
i.e., 
$E f= d f$ for some $d \in {\mathbb C}$, 
then a new function $g=V(c) f$ satisfies
$E g= (d +c)  g$
again.
This fact is an immediate consequence of the following

\begin{lemma}  \label{lemma:EV}
It holds that
$[E,V(c)] = c V(c)$
for any $c \in {\mathbb C}$.
\end{lemma}

\pf
First recall the formula
$e^A B e^{-A} = e^{{\rm ad}(A)} B=B+[A,B]+ \frac{1}{2!} [A,[A,B]]+\cdots$
for any operators $A$ and $B$,
where
${\rm ad}(A)(B)=[A,B]$.
Since 
$\left[\xi({\boldsymbol x},k), \partial/\partial x_n \right]=-k^n$, we have
$ \left[ \xi({\boldsymbol x},k), E \right]=-\sum_{n \geq 1} n x_n k^n= -k \frac{\partial}{\partial k}  \xi({\boldsymbol x},k)$.
Therefore
\begin{equation}  \label{eq:eee1}
e^{ \xi({\boldsymbol x},k) } Ee^{ -\xi({\boldsymbol x},k) } = E - k \frac{\partial}{\partial k}  \xi({\boldsymbol x},k).
\end{equation} 
On the other hand,
from $ \left[   \xi(- \tilde{\partial}_{\boldsymbol x},k^{-1}),  x_n \right]= -k^{-n}/n$, we observe
$\left[  \xi( -\tilde{\partial}_{\boldsymbol x},k^{-1}), E \right]= 
-\sum_{n \geq 1} k^{-n}  \frac{\partial}{\partial x_n}= 
- k \frac{\partial}{\partial k} \xi( -\tilde{\partial}_{\boldsymbol x},k^{-1}) $.
Accordingly, we obtain
\begin{equation}  \label{eq:eee2}
e^{\xi(- \tilde{\partial}_{\boldsymbol x},k^{-1}) }  E  e^{-\xi( -\tilde{\partial}_{\boldsymbol x},k^{-1}) }
=E - k \frac{\partial}{\partial k} \xi( -\tilde{\partial}_{\boldsymbol x},k^{-1}). 
\end{equation}
Hence we see that
\begin{align}
E X^+(k) &= E e^{ \xi({\boldsymbol x},k)  }  e^{\xi(  -\tilde{\partial}_{\boldsymbol x},k^{-1}) }
\\
&=
e^{ \xi({\boldsymbol x},k)  }  E   e^{\xi(  -\tilde{\partial}_{\boldsymbol x},k^{-1}) } +
\left( k \frac{\partial}{\partial k}e^{ \xi({\boldsymbol x},k)} \right)
  e^{\xi(  -\tilde{\partial}_{\boldsymbol x},k^{-1})},
  \qquad  \text{using (\ref{eq:eee1})} 
\nonumber  \\
&=e^{ \xi({\boldsymbol x},k)  }
\left( e^{\xi(  -\tilde{\partial}_{\boldsymbol x},k^{-1})}  E 
+ k \frac{\partial}{\partial k}  e^{\xi(  -\tilde{\partial}_{\boldsymbol x},k^{-1})}
\right)
+ \left( k \frac{\partial}{\partial k} e^{ \xi({\boldsymbol x},k)} \right)
e^{\xi(  -\tilde{\partial}_{\boldsymbol x},k^{-1})},
 \qquad  \text{using (\ref{eq:eee2})} 
\nonumber  \\
&=X^+(k)E +  k \frac{\partial}{\partial k}  X^+(k).
 \label{eq:EX}
\end{align}
Finally, 
we conclude that
\begin{align*}
\left[  E, V(c) \right]
&=   \int_\gamma [E,  X^+(k)] k^{-c-1} {\rm d} k 
= \int_\gamma  \frac{\partial X^+(k)}{\partial k}  k^{-c} {\rm d} k, \qquad \text{using (\ref{eq:EX})} 
\\
&= \left[  X^+(k) k^{-c} \right]_{\gamma(0)}^{\gamma(1)}
+c \int_\gamma  X^+(k)  k^{-c-1} {\rm d} k
=c V(c)
\end{align*}
via integration by parts.
\qed
\\

Suppose
$\tau_0({\boldsymbol x})$
to be a solution of the KP hierarchy (\ref{eq:kp})
satisfying 
$E \tau_0 = d_0 \tau_0$.
Introduce a sequence 
$\{\tau_0, \tau_1, \tau_2, \ldots\}$ 
of solutions
defined recursively by
$\tau_{n+1}=V(c_{n}) \tau_n$
for arbitrary parameters 
$c_n \in {\mathbb C}$ given:
\begin{equation} \label{eq:tau-seq}
\cdots
\stackrel{V(c_{n-2})}{\longrightarrow} 
\tau_{n-1} \stackrel{V(c_{n-1})}{\longrightarrow} \tau_n \stackrel{V(c_{n})}{\longrightarrow} \tau_{n+1} \stackrel{V(c_{n+1})}{\longrightarrow} \cdots
\end{equation}
Since $\tau_0$ is homogeneous, 
so are all $\tau_n$ $(n \geq 1)$.
To be specific, 
we have
$E \tau_n =d_n \tau_n$ with 
$d_{n+1}=d_n+c_n$.
We shall call a sequence of solutions of the KP hierarchy of the form  (\ref{eq:tau-seq}) 
a {\it homogeneous $\tau$-sequence}.

\begin{example}[A sequence of Schur functions]\rm
\label{ex:schur}

For example, let $c=n$ be an integer and $\gamma$ a 
positively oriented small circle around $k=0$.
Then $V(n) = 2 \pi \sqrt{-1}X_n^+$ according to (\ref{eq:vo}).
Fix a partition $\lambda=(\lambda_1,\lambda_2,\ldots,\lambda_l)$,
and recall (\ref{eq:raising_schur}).
Starting from a trivial solution $\tau=S_\emptyset({\boldsymbol x}) \equiv1$ of the KP hierarchy,
we thus have a sequence of Schur functions connected by successive application of $X_n^+$
of the form 
(cf. \cite{jm83}):
\[
S_\emptyset=1
\stackrel{X^+_{\lambda_l}}{\longrightarrow} 
S_{(\lambda_l)}\stackrel{X^+_{\lambda_{l-1}}}{\longrightarrow}S_{(\lambda_{l-1}, \lambda_l)}
\stackrel{X^+_{\lambda_{l-2}}}{\longrightarrow} 
\cdots 
\stackrel{X^+_{\lambda_1}}{\longrightarrow}
S_{(\lambda_1,\lambda_{2},\ldots,\lambda_l)}
\]
This type of homogeneous $\tau$-sequence in fact yields a class of rational or algebraic solutions of Painlev\'e equations;
cf. \cite{ny99}. 
We shall explicitly demonstrate 
for the $P_{\rm III}$-chain 
in Sect.~\ref{subsec:ratsol_p3}.
\end{example}

Let us
now  concern Weyl group symmetry of the homogeneous $\tau$-sequence.
First notice that 
the fermionic relation 
\begin{equation}  \label{eq:VV}
V(a)V(b)+V(b-1)V(a+1)=0
\end{equation}
holds.
Interchange
the $(n-1)$th and $n$th operations in the chain (\ref{eq:tau-seq})
while taking (\ref{eq:VV}) into account.
We hence obtain a new sequence 
\[
\cdots  
\stackrel{V(c_{n-2})}{\longrightarrow} 
\tau_{n-1} \stackrel{V(c_{n}+1)}{\longrightarrow} \hat{\tau}_n 
\stackrel{V(c_{n-1}-1)}{\longrightarrow} \tau_{n+1} \stackrel{V(c_{n+1})}{\longrightarrow} \cdots
\]
which is identical with the original one, (\ref{eq:tau-seq}), except 
$\tau_n$ is replaced by 
\[
\hat{\tau}_n =V(c_n+1) \tau_{n-1}.
\]
Besides,
the degree of $\hat{\tau}_n$ reads as
$\hat{d}_n=d_{n-1}+c_{n}+1
=d_{n-1}-d_n+d_{n+1}+1$.
Let us refer to the above permutation of vertex operators 
as $r_n$.

We can
consider the operation $r_i$ at each $i$th site in the chain. 
With respect to the variables
$\alpha_i=  \hat{d}_i-d_i=d_{i-1}- 2 d_i+d_{i+1} +1$,
the operation $r_i$ induces the transformation
\[
r_i(\alpha_i)= -\alpha_i, \quad 
r_i(\alpha_{i \pm 1})= \alpha_{i \pm 1}+\alpha_i,
\quad  \text{and} \quad
r_i(\alpha_j)= \alpha_j \quad (j \neq i, i \pm 1).
\]
This is exactly the canonical realization of a generator of the Weyl group of type $A$
if we regard $\alpha_i$ as a simple root.
One can easily verify that $\langle r_i \rangle$ indeed fulfills 
the fundamental relations
\[
{r_i}^2=1, \quad 
r_i r_{i \pm 1} r_i = r_{i \pm  1} r_i  r_{i \pm 1},
\quad
\text{and} \quad
r_i r_j=r_j r_i  \quad (j \neq i, i \pm 1).
\]
In summary, 
we find a realization of the Weyl group of type $A$ generated by 
a permutation $r_i$
of two serial vertex operators at each $i$th site 
of the homogeneous $\tau$-sequence (\ref{eq:tau-seq}).
This is in fact one origin of 
Weyl group symmetry of Painlev\'e equations,
as demonstrated lucidly 
for $P_{\rm III}$-chain, $P(A_{\ell-1}^{(1)})$, and $P_{\rm II}$
respectively in Sects.~\ref{subsec:weyl}, 
\ref{subsect:p4_weyl},
and \ref{subsect:p2_weyl}
below.
Note that in \cite{wh06}
a similar treatment of the above Weyl group action
 was explained in the context of 
 binary Darboux transformations.
Also, there is an alternative approach
to the symmetry of Painlev\'e
equations
based on 
the Gau{\ss} decomposition
of (Lax) matrices; see  \cite{n}.

Before closing this subsection, 
we shall prepare some useful formulae 
that will be employed later.

\begin{lemma}
Let $\{\ldots,\tau_{n-1},\tau_n,\tau_{n+1},\ldots\}$ be a 
{\rm(}homogeneous{\rm)} 
$\tau$-sequence such that
$\tau_{n+1}=V(c_n) \tau_n$.
Let $\hat{\tau}_n=V(c_n+1)\tau_{n-1}$.
Then we have
\begin{subequations}
\begin{align}  \label{eq:weyl_tau_a}
\tau_{n-1} \otimes \tau_{n+1}&= \sum_{i+j=-1} X_i^- \hat{\tau}_n \otimes X_j^+ \tau_n, \\
\label{eq:weyl_tau_b}
\tau_n \otimes \hat{\tau}_n - \hat{\tau}_n \otimes \tau_n
&=
\sum_{i+j=1} X_i^- \tau_{n+1} \otimes X_j^+ \tau_{n-1}.
\end{align}
\end{subequations}
\end{lemma}

\pf
By applying $V(c_{n}+2) \otimes 1$ to (\ref{eq:mkp-bil})
one can verify (\ref{eq:weyl_tau_a}) straightforwardly via
(\ref{eq:ferm}).
Similarly it follows from (\ref{eq:kp}) applied by $V(c_{n-1}) \otimes 1$
that
\begin{equation} \label{eq:-1modkp}
\tau_{n-1} \otimes \tau_n = \sum_{i+j=0}
 X_i^- \tau_{n} \otimes X_j^+ \tau_{n-1}.
\end{equation}
In addition, applying $V(c_{n}+1) \otimes 1$ to this leads to (\ref{eq:weyl_tau_b}).
\qed


\subsection{Preliminaries for Sects.~\ref{sect:p3}--\ref{sect:p2}:
difference/differential equations inside mKP hierarchy}
\label{subsect:pre_mkp}

The mKP hierarchy (\ref{eq:mkp-bil})
can be equivalently rewritten  into 
\begin{equation}  \label{eq:mkp}
\frac{1}{2 \pi \sqrt{-1}}
\oint z e^{\xi( {\boldsymbol x}-{\boldsymbol x'},z)}  {{\rm d} z}
\, 
 \tau_n ({\boldsymbol x'}+[z^{-1}]) 
\tau_{n+1}({\boldsymbol x}-[z^{-1}])  =0
\end{equation}
for arbitrary ${\boldsymbol x}$ and ${\boldsymbol x'}$.
We first present some difference and/or differential equations arising from 
(\ref{eq:mkp});
cf. \cite{djm82}.
We henceforth require a solution 
$\tau_n({\boldsymbol x})$ of (\ref{eq:mkp})
to be an entire function.

\begin{lemma} \label{lemma:mkp}
The following functional equations hold{\rm:}
\begin{subequations}
\begin{align}  \label{eq:mkp_a}
&\left( t D_{x_1}+1 \right) 
\tau_n  \left({\boldsymbol x}-[t]\right) 
 \cdot 
 \tau_{n+1} \left({\boldsymbol x}\right) 
 -
 \tau_n  \left({\boldsymbol x}\right) 
 \tau_{n+1} \left({\boldsymbol x}-[t]\right) 
 =0, 
\\ 
 \label{eq:mkp_b}
 &\left( D_{\delta_t}-1 \right) \tau_n({\boldsymbol x}) 
 \cdot 
 \tau_{n+1}({\boldsymbol x})
+ \tau_n \left({\boldsymbol x} -[t] \right) 
\tau_{n+1} \left({\boldsymbol x} +[t]\right)
=0,
 \\
\label{eq:mkp_c}
 &(t-s) \tau_n ({\boldsymbol x}-[t]-[s])  
\tau_{n+1} ({\boldsymbol x})
-
 t \tau_n ({\boldsymbol x}-[t])  
\tau_{n+1} ({\boldsymbol x}-[s])
+s \tau_n ({\boldsymbol x}-[s])  
\tau_{n+1} ({\boldsymbol x}-[t])=0,
 \\
  \label{eq:mkp_d}
 &\left(
D_{\delta_t} +\frac{t}{s-t}
\right)
\tau_n ({\boldsymbol x}-[s])  
\cdot 
\tau_{n+1} ({\boldsymbol x}) 
+\frac{t }{t-s}
\tau_n ({\boldsymbol x}-[t])  
\tau_{n+1} ({\boldsymbol x}+[t]-[s]) 
=0, 
\end{align}
\end{subequations}
where 
$D_{\delta_t}$ denotes the Hirota differential with respect to the vector field 
\begin{equation}
\delta_t=\sum_{n=1}^\infty t^n \frac{\partial}{\partial x_n}.
\end{equation}
\end{lemma}

\pf
We shall verify  
(\ref{eq:mkp_a}),  (\ref{eq:mkp_b}),  (\ref{eq:mkp_c}), and (\ref{eq:mkp_d})
by taking the variables in (\ref{eq:mkp}) as
 \[
{\rm (a)} \  {\boldsymbol x}-{\boldsymbol x'}=[t], \quad
{\rm (b)} \  {\boldsymbol x}-{\boldsymbol x'}=2[t], \quad
{\rm (c)} \  {\boldsymbol x}-{\boldsymbol x'}=[t]+[s], \quad
{\rm (d)} \  {\boldsymbol x}-{\boldsymbol x'}=2[t]+[s],
\]
respectively.
Let $F=F(z)$ denote the integrand of (\ref{eq:mkp}) and
write $\Omega =z e^{\xi( {\boldsymbol x}-{\boldsymbol x'},z)}  {{\rm d} z}$,
for convenience.

Substitute
${\boldsymbol x'}={\boldsymbol x}-[t]$ for the case (a).
Under this specialization  we  observe 
\[\Omega = \frac{z {\rm d}z}{1-tz}\]
in view of the Taylor expansion,
$- \log(1-u)=\sum_{n=1}^\infty  u^n/n$
valid for $|u|<1$.
The integrand 
has the three singularities $z=1/t$ (simple pole), $z=\infty$ (double pole)
and $z=0$ (which may be an essential singularity).
Accordingly (\ref{eq:mkp}) becomes
\begin{align*}
&\frac{1}{2 \pi \sqrt{-1}} 
\int_C  F(z) {\rm d}z=0, \quad \text{where}
\\
&F(z)=\frac{z}{1-tz} 
 \tau_n ({\boldsymbol x}-[t]+[z^{-1}]  ) 
\tau_{n+1} ({\boldsymbol x}-[z^{-1}]  )
\end{align*}
and the integration contour $C$ is taken along a positively oriented small circle around $z=0$
such that 
$z=1/t, \infty$ are exterior to it.
The residues at the two poles are evaluated as
\[
\underset{z=1/t}{\rm{Res}} F(z)  {\rm d}z = -t^{-2}  
 \tau_n  \left({\boldsymbol x}\right) 
 \tau_{n+1} \left({\boldsymbol x}-[t]\right),
 \]
 and
\begin{align*}
\underset{z=\infty}{\rm{Res}} F(z)  {\rm d}z 
&=  - \underset{w=0}{\rm{Res}} F(w^{-1}) w^{-2}  {\rm d}w, 
\quad \text{where  $z w=1$}
\\
&=
\frac{\rm d}{{\rm d} w}  
\left.
\left(
\frac{1}{t-w} 
 \tau_n \left({\boldsymbol x}-[t]+[w]  \right) 
\tau_{n+1} \left({\boldsymbol x}-[w]  \right)
\right)
\right|_{w=0}
\\
&=
\frac{1}{(t-w)^2} 
 \tau_n \left({\boldsymbol x}-[t]+[w]  \right) 
\tau_{n+1} \left({\boldsymbol x}-[w]  \right) \\
& \qquad + \left. 
\frac{1}{w(t-w)} D_{\delta_w} 
\tau_n \left({\boldsymbol x}-[t]+[w]  \right) \cdot
\tau_{n+1} \left({\boldsymbol x}-[w]  \right)
\right|_{w=0}
\\
&=
t^{-2}\tau_n \left({\boldsymbol x}-[t]  \right) 
\tau_{n+1} \left({\boldsymbol x}  \right) 
+ t^{-1} D_{x_1}
\tau_n \left({\boldsymbol x}-[t]  \right) \cdot
\tau_{n+1} \left({\boldsymbol x}  \right). 
\end{align*}
We find through Cauchy's residue theorem that
$\underset{z=1/t}{\rm{Res}}  F(z) {\rm d}z + \underset{z=\infty}{\rm{Res}} 
F(z) {\rm d}z=0$, 
which implies (\ref{eq:mkp_a}).

For other cases the differential form of $\Omega$ and the singularities of $F(z)$
are listed below:
\[
\begin{array}{ c  c  l }
\hline
 \text{Case}     &\Omega& \text{Singularities of $F(z)$ other than $z=0$} \\
      \hline
\rm(b) &\displaystyle  \frac{z {\rm d}z}{(1-tz)^2}& z=\infty \ \text{(simple pole)}, z=1/t \ \text{(double pole)} \\
\rm(c) & \displaystyle \frac{z {\rm d}z}{(1-tz)(1-sz)}& z=1/t,1/s,\infty \  \text{(simple poles)} \\
\rm(d) & \displaystyle  \frac{z {\rm d}z}{(1-tz)^2(1-sz)}& z=1/s \  \text{(simple pole)}, z=1/t \  \text{(double pole)}\\
\hline
\end{array}
\]
We see that (\ref{eq:mkp}) takes
the form
$\frac{1}{2 \pi \sqrt{-1}} 
\int_C  F(z) {\rm d}z=0$
for each case
as well as (a).
Here $C$ encircles $z=0$ so that
all the other singularities are exterior to it.
Residue calculus again
leads to the desired results 
(\ref{eq:mkp_b})--(\ref{eq:mkp_d}).
\qed
\\

For later convenience, we summarize how to derive an infinite series of Hirota differential equations from the functional relation
\begin{equation} \label{eq:xfxg}
\sum_{i+j=-1-d} X_i^- f \otimes X_j^+ g =0
 \quad \text{for a given $d \in {\mathbb Z}$},
\end{equation}
where 
$f \otimes g=f({\boldsymbol x'}) g({\boldsymbol x})$ 
is regarded as an element of
${\mathbb C}[{\boldsymbol x'}]\otimes {\mathbb C}[{\boldsymbol x}]$.
Now let us consider the Taylor expansion of (\ref{eq:xfxg})
at ${\boldsymbol x'}={\boldsymbol x}$,
i.e., replace $({\boldsymbol x'},{\boldsymbol x})$ with  
$({\boldsymbol x}+{\boldsymbol u},{\boldsymbol x}-{\boldsymbol u})$
and then expand with respect to the variables 
${\boldsymbol u}=(u_1,u_2,\ldots)$.
We thus obtain
\[
\frac{1}{2 \pi \sqrt{-1}}\oint
z^d {\rm d}z 
e^{-2\xi({\boldsymbol u,z})}
e^{\xi(\tilde{\partial}_{\boldsymbol u},z^{-1}) }
f({\boldsymbol x}+{\boldsymbol u})g({\boldsymbol x}-{\boldsymbol u})=0,
\]
thereby,
\[
\sum_{i+j+d=-1} 
p_i(-2{\boldsymbol u})p_{-j}(\tilde{\partial}_{\boldsymbol u})
f({\boldsymbol x}+{\boldsymbol u})g({\boldsymbol x}-{\boldsymbol u})=0.
\]
If we remember the definition of the Hirota differential (\ref{eq:def_hirota}),
we can verify that
\[
p_{-j}(\tilde{\partial}_{\boldsymbol u})
f({\boldsymbol x}+{\boldsymbol u})g({\boldsymbol x}-{\boldsymbol u})
=p_{-j}(\tilde{D}_{\boldsymbol x})e^{\sum_{n=1}^\infty u_n D_{x_n}}
f({\boldsymbol x}) \cdot g({\boldsymbol x}),
\]
with $\tilde{D}_{\boldsymbol x}$ denoting 
$(D_{x_1}, D_{x_2}/2, D_{x_3}/3, \ldots)$.
Introduce the generating function
\begin{equation} \label{eq:formula_bil2hirota}
G_d(f({\boldsymbol x}),g({\boldsymbol x});{\boldsymbol u}):=
\sum_{i+j+d=-1} 
p_i(-2{\boldsymbol u})p_{-j}(\tilde{D}_{\boldsymbol x})e^{\sum_{n=1}^\infty u_n D_{x_n}}
f({\boldsymbol x}) \cdot g({\boldsymbol x})
\end{equation}
in ${\boldsymbol u}=(u_1,u_2,\ldots)$.
Therefore we conclude that
\begin{equation}  \label{eq:formula_bil2hirota=0}
G_d(f({\boldsymbol x}),g({\boldsymbol x});{\boldsymbol u})=0.
\end{equation}

\begin{example} \rm
Let us write down a few of Hirota differential equations arising from the mKP hierarchy (\ref{eq:mkp-bil}).
Put $d=1$, $f=\tau_n$, and  $g=\tau_{n+1}$ in (\ref{eq:formula_bil2hirota=0}).
Then, the coefficients of $1={\boldsymbol u}^0$ and $u_1$
give
\begin{equation}  \label{eq:mkp_hirota_1}
\left({D_{x_1}}^2 + D_{x_2} \right) \tau_n \cdot \tau_{n+1}=0
\end{equation}
and 
\begin{equation}   \label{eq:mkp_hirota_2}
\left({D_{x_1}}^3 -3  D_{x_1}D_{x_2}-4D_{x_3} \right) \tau_n \cdot \tau_{n+1}=0,
\end{equation}
respectively.
\end{example}

\section{From KP hierarchy to Painlev\'e III chain}
\label{sect:p3}

In this section,  
we derive a system of nonlinear ordinary differential equations from the $\ell$-periodic mKP hierarchy ($\ell \geq 2$)  
 through a homogeneity constraint.
This system, called the {\it $P_{\rm III}$-chain}, 
provides a higher order generalization of 
the third Painlev\'e equation,  
which indeed coincides with the original one when $\ell=2$.

\subsection{Similarity reduction}  \label{subsec:sim_p3}

Let $\tau_n(\boldsymbol x)$ be a solution of the mKP hierarchy,
(\ref{eq:mkp-bil}) or (\ref{eq:mkp}), 
satisfying
 the $\ell$-periodic condition:
$\tau_{n+\ell}=\tau_n$  ($\ell \geq 2$),
and the similarity condition:
\begin{equation}
E \tau_n(\boldsymbol x)=d_n \tau_n(\boldsymbol x) \quad (d_n \in {\mathbb C})
\label{eq:sim}
\end{equation}
where $E=\sum_{n=1}^\infty n x_n {\partial}/{\partial x_n}$. 
Introduce the functions $\sigma_n=\sigma_n(a,s)$ $(n \in {\mathbb Z}/\ell {\mathbb Z})$
defined by $\sigma_n (a,s)=\tau_n(\boldsymbol x)$
under the substitution
\begin{equation}  \label{eq:change}
x_1=s+a \quad 
\text{and} \quad
x_n=\frac{a}{n} \quad (n \geq 2).
\end{equation}
Note that $s$ will play a role of the independent variable while $a$ will be regarded as a constant parameter.

\begin{prop}  \label{prop:bil-p3}
The functions $\sigma_n=\sigma_n(a,s)$ satisfy the system of bilinear equations
 \begin{align} \label{eq:bil-p3-1}
  \left( D_s+1 \right)  \sigma_n(a-1,s) 
\cdot 
 \sigma_{n+1}(a,s) 
 - \sigma_n(a,s)  \sigma_{n+1}(a-1,s)
 &=0,
 \\
  \label{eq:bil-p3-2}
 \left( s D_s + a + d_{n+1}-d_n \right) \sigma_n(a,s) 
 \cdot 
 \sigma_{n+1}(a,s)- a \sigma_n(a-1,s) \sigma_{n+1}(a+1,s)
 &=0.
 \end{align}
\end{prop}

\pf 
Observe that
 \begin{equation}   \label{eq:dx1=ds}
\frac{ \rm d}{ {\rm d} s}= \sum_{n=1}^\infty \frac{ {\rm d} x_n}{ {\rm d} s} \frac{\partial}{\partial x_n}
= \frac{\partial}{\partial x_1}.
\end{equation}
Also we see that
 \begin{equation} \label{eq:E-p3}
 E=\sum_{n=1}^\infty   n   x_n \frac{\partial}{\partial x_n}=s\frac{\partial}{\partial x_1}+a\sum_{n=1}^\infty  \frac{\partial}{\partial x_n}
 =s\frac{ \rm d}{ {\rm d} s}+a  \delta_1.
 \end{equation}
 Hence, with the aid of the homogeneity (\ref{eq:sim}),
 we can readily verify (\ref{eq:bil-p3-1}) and (\ref{eq:bil-p3-2}) from (\ref{eq:mkp_a}) and (\ref{eq:mkp_b}), 
 respectively.
 \qed
 \\

For  appropriately chosen dependent variables,
we will derive a system of nonlinear ordinary differential equations
from the bilinear equations in Prop.~\ref{prop:bil-p3}.
  
\begin{thm}
Define the functions $f_n=f_n(a,s)$ and $g_n=g_n(a,s)$  
$(n \in {\mathbb Z}/\ell {\mathbb Z})$
by
 \begin{equation}   \label{eq:def-fg}
 f_n(a,s) = \frac{ \sigma_{n-1}(a,s) \sigma_n(a-1,s) }{  \sigma_{n-1}(a-1,s) \sigma_n(a,s) },
 \quad
 g_n(a,s) = \frac{ \sigma_{n-1}(a-1,s) \sigma_n(a+1,s) }{  \sigma_{n-1}(a,s) \sigma_n(a,s) }.
 \end{equation}
Then these functions satisfy
the system of ordinary differential equations 
\begin{subequations}   \label{subeq:p3-chain}
\begin{align}
 \label{eq:p3-chain-1}
 s \frac{{\rm d} f_n}{{\rm d} s}
 &=
\left( s (f_{n+1}-f_n ) + a (g_n-g_{n+1}) +\alpha_n-1\right)f_n,
\\
\label{eq:p3-chain-2}
\frac{{\rm d} g_n}{{\rm d} s}
 &= f_n g_n-f_{n-1} g_{n-1},
 \end{align}
 \end{subequations}
 where 
 $\alpha_n=d_{n-1}-2 d_n+d_{n+1} +1$. 
 \end{thm}

\pf 
To simplify an expression
we write as
$\overline{\sigma}_{n}=\sigma_n(a+1,s)$
and 
$\underline{\sigma}{}_{n}=\sigma_n(a-1,s)$,
while $\sigma_{n}=\sigma_n(a,s)$,
for brevity.
It follows respectively from (\ref{eq:bil-p3-1}) and (\ref{eq:bil-p3-2}) that  
\begin{align}  
  \label{eq:f-tau}
\frac{D_s \underline{\sigma}{}_{n-1}  \cdot \sigma_{n} }{ \underline{\sigma}{}_{n-1}  \sigma_n }
&=f_n-1,
\\
\label{eq:g-tau}
\frac{ s  D_s \sigma_{n-1} \cdot \sigma_{n}}{\sigma_{n-1}  \sigma_{n} }
&= a g_n-a-d_n+d_{n-1}.
\end{align}
Therefore the logarithmic derivative of $f_n$ reads
\begin{align*}
\frac{1}{f_n} \frac{ {\rm d} f_n}{ {\rm d} s}
&= \frac{ {\sigma_{n-1}}' }{ \sigma_{n-1}}+  \frac{ {\underline{\sigma}{}_n}' }{ \underline{\sigma}{}_n }
- \frac{ {\underline{\sigma}{}_{n-1} }' }{ \underline{\sigma}{}_{n-1} }-  \frac{ {\sigma_n}' }{ \sigma_n},
\qquad \text{where} \quad  
 ' =  \frac{ {\rm d}}{ {\rm d} s}
\\
&= 
- \frac{D_s \underline{\sigma}{}_{n-1}  \cdot \sigma_{n} }{ \underline{\sigma}{}_{n-1}   \sigma_{n}}
+ \frac{D_s \underline{ \sigma}{}_{n}  \cdot \sigma_{n+1}}{ \underline{\sigma}{}_{n}   \sigma_{n+1}}
+\frac{D_s \sigma_{n-1}  \cdot \sigma_{n} }{\sigma_{n-1}   \sigma_{n}}
-\frac{D_s \sigma_{n} \cdot \sigma_{n+1} }{\sigma_{n}  \sigma_{n+1}}
\\
&= f_{n+1}-f_n  + \frac{a (g_n-g_{n+1}) +d_{n-1}-2 d_n+d_{n+1}}{s},
\end{align*}
which is exactly (\ref{eq:p3-chain-1}).
Likewise we have
\[
\frac{1}{g_n} \frac{ {\rm d} g_n}{ {\rm d} s}
=
\frac{D_s \underline{\sigma}{}_{n-1}  \cdot \sigma_{n} }{ \underline{\sigma}{}_{n-1}   \sigma_{n}}
-\frac{D_s \sigma_{n-1}  \cdot \overline{\sigma}_{n} }{\sigma_{n-1}   \overline{ \sigma}_{n} }=f_n - \overline{f}_n .
\]
If we take into account that (see (\ref{eq:def-fg}))
\begin{equation}  \label{eq:overlinef}
 \overline{f}_n=\frac{f_{n-1}g_{n-1}}{g_n}, 
 \end{equation}
 then we arrive at  (\ref{eq:p3-chain-2}). 
\qed
 \\

One can find that (\ref{subeq:p3-chain}) 
possesses two conserved quantities
\begin{equation}  \label{eq:cons-p3}
\prod_{n=1}^\ell f_n=1 
\quad \text{and} \quad
\sum_{n=1}^\ell g_n =\ell,
\end{equation}
and thus it is essentially of $2 (\ell-1)$th order.
If $\ell=2$ it is indeed 
equivalent to 
the third Painlev\'e equation $P_{\rm III}$,
 which is obviously of second order,
as demonstrated in the next subsection.
For this reason
we call (\ref{subeq:p3-chain})  the {\it $P_{\rm III}$-chain}.

\begin{remark}\rm
The $P_{\rm III}$-chain 
has been investigated  in a quite different manner
by Adler et al.
in the study of Darboux transformations of a sequence of 
Schr\"odinger operators 
with quadratic eigenvalue dependence.
It is closely related to the relativistic Toda lattice equation (\cite{rui90}). 
See an excellent review article  \cite{asy00},
which involves the case of other Painlev\'e equations also.
\end{remark}

\subsection{Example: two-periodic case and $P_{\rm III}$}

Consider the case where $\ell=2$.
Introduce the canonical variables $q(s)$ and $p(s)$ 
as
\[
q =f_1= 
 \frac{ \sigma_{0} \underline{\sigma}{}_1 }{  \underline{\sigma}{}_{0} \sigma_1 },
\quad
p = -a f_0 g_0  =-a \frac{ \underline{\sigma}{}_{0} \overline{\sigma}_0 }{{\sigma_0}^2}.
\]
Thus (\ref{subeq:p3-chain}) is rewritten into 
the Hamiltonian system
\[
\frac{{\rm d} q }{ {\rm d} s } =\frac{\partial H}{ \partial p}, 
\quad
\frac{{\rm d} p }{ {\rm d} s } =-\frac{\partial H}{ \partial q},
\]
with the Hamiltonian function $H=H(q,p;s)$ 
defined by
\[
s H = q^2 p^2 - \left(  s q^2- 2 (a+d_0-d_1) q -s \right) p -2 a s q.
\]
This is identical to the Hamiltonian form of $P_{\rm III}$;
see \cite{oka}.

Alternatively, 
we can derive 
$P_{\rm III}$ more directly from the bilinear equations for $\tau$-functions. 
First we differentiate with respect to $t$ the equation (\ref{eq:mkp_a}) 
after shifting 
the variables ${\boldsymbol x}$ to ${\boldsymbol x}+[t]/2$;
thus, we have
\begin{equation}  \label{eq:mkp_a_diff}
\left( t D_{x_1} D_{\delta_t} -2 t  D_{x_1} + D_{\delta_t}  \right) 
\tau_n  \left({\boldsymbol x}-\frac{[t]}{2}\right) 
\cdot 
 \tau_{n+1} \left({\boldsymbol x}+\frac{[t]}{2}\right) 
 +
  D_{\delta_t}
 \tau_n  \left({\boldsymbol x}+\frac{[t]}{2}\right) 
 \cdot 
 \tau_{n+1} \left({\boldsymbol x}-\frac{[t]}{2}\right) 
 =0.
\end{equation}
Applying change of variables (\ref{eq:change}) and $t=1$ together with 
the homogeneity (\ref{eq:sim}),
we have from (\ref{eq:mkp_a_diff}) that
\begin{align}
&  \nonumber
\left(
\frac{1}{s}{\cal D}^2 +(s+2a+d_{n+1}-d_n-1)D_s+d_{n+1}-d_n 
\right)
\sigma_n (a-1,s)
\cdot 
\sigma_{n+1}(a,s)
\\
& \quad
+ \left({\cal D}+d_{n+1}-d_n \right)
\sigma_n (a,s) 
\cdot 
\sigma_{n+1}(a-1,s)
=0,
\label{eq:bil-p3-3}
\end{align}
where the calligraphic symbol $\cal D$ 
stands for the Hirota differential with respect to 
$s {\rm d}/{\rm d}s$.
Note that (\ref{eq:bil-p3-3}) 
is still valid without requiring the two-periodicity.
We are now interested in a second order differential equation satisfied by the variable
$q=q(s)$.
One can verify, only by using definition of the Hirota differential, the formula
\begin{equation}  \label{eq:logder}
\left(s\frac{{\rm d} }{{\rm d}s}\right)^2 \log q
=\frac{{\cal D}^2 \sigma_0\cdot \underline{\sigma}{}_1}{ \sigma_0 \underline{\sigma}{}_1}
-\frac{{\cal D}^2 \sigma_1\cdot \underline{\sigma}{}_0}{ \sigma_1 \underline{\sigma}{}_0}
- \left(\frac{{\cal D} \sigma_0\cdot \underline{\sigma}{}_1}{ \sigma_0 \underline{\sigma}{}_1}\right)^2
+\left(\frac{{\cal D} \sigma_1\cdot \underline{\sigma}{}_0}{ \sigma_1 \underline{\sigma}{}_0}\right)^2.
\end{equation}
On the other hand, we see from (\ref{eq:bil-p3-2}) and (\ref{eq:bil-p3-3}) 
in view of 
$\sigma_{n+2}=\sigma_n$
that
\begin{align*}
\frac{{\cal D} \sigma_0 \cdot \underline{\sigma}{}_1}{ \sigma_0 \underline{\sigma}{}_1}
&= s \left(1- \frac{1}{q}\right),
\quad
\frac{{\cal D} \sigma_1 \cdot \underline{\sigma}{}_0}{ \sigma_1 \underline{\sigma}{}_0}
= s (1- q),
\\
\frac{{\cal D}^2 \sigma_0\cdot \underline{\sigma}{}_1}{ \sigma_0 \underline{\sigma}{}_1}
&= s \left( 2 s +2a-1 - ( 2s + 2a+2 d_0-2d_1-1) \frac{1}{q} \right),
\\
\frac{{\cal D}^2 \sigma_1\cdot \underline{\sigma}{}_0}{ \sigma_1 \underline{\sigma}{}_0}
&= s \left(2 s+2 a-1 - ( 2s + 2a+2 d_1-2d_0-1) q \right).
\end{align*}
Substituting these into (\ref{eq:logder}), 
we finally arrive at
\[
\left(s\frac{{\rm d} }{{\rm d}s}\right)^2 \log q
=s^2 \left(q^2-\frac{1}{q^2}  \right)
+s\left( (2a + 2 d_1-2d_0-1)q-(2a+2d_0-2 d_1-1)\frac{1}{q} \right),
\]
which is equivalent to the standard form of $P_{\rm III}$:
\begin{equation} \label{eq:p3_D6}
\frac{{\rm d}^2 q}{{\rm d}s^2}=
\frac{1}{q} \left(  \frac{{\rm d} q}{{\rm d}s} \right)^2
-\frac{1}{s}  \frac{{\rm d} q}{{\rm d}s} 
+\frac{1}{s} \left( \alpha q^2+\beta\right)
+ \gamma q^3 + \frac{ \delta }{q},
\end{equation}
where
$\alpha=2a+2d_1-2d_0-1$, 
$\beta=-2a+2d_1-2d_0+1$, 
and $\gamma=-\delta=1$.

\begin{remark} \rm
To be more precise, the third Painlev\'e equation 
should be divided into three types, i.e., 
$D_6^{(1)}$, $D_7^{(1)}$, and $D_8^{(1)}$,
from the viewpoint of
space of initial conditions; see \cite{oka79, sak01}.
In this context our case (\ref{eq:p3_D6}) 
is classified as the $D_6^{(1)}$ type.
Recall also that  the $D_8^{(1)}$ type is algebraically equivalent to 
the $D_6^{(1)}$ one with special value of parameters; see \cite{tos05}. 
It still remains open how to derive the $D_7^{(1)}$ type 
from an integrable system such as the KP hierarchy.
 \end{remark}

\subsection{Affine Weyl group symmetry}
\label{subsec:weyl}

Now we shall concern birational symmetries of the $P_{\rm III}$-chain,
(\ref{subeq:p3-chain}).
Here, to be precise,  a birational transformation of the variables $(f_n, g_n,s)$
is said to be a symmetry of the $P_{\rm III}$-chain
if it keeps the system invariant except changing the constant parameters $(a, \alpha_n)$ involved.

First, we consider a symmetry shifting the parameter $a$ to $a\pm 1$.
Write as $\overline{f}_n=f_n(a+1,s)$ while $f_n=f_n(a,s)$ and so on. 
Then we have (\ref{eq:overlinef}): 
$\overline{f}_n=f_{n-1}g_{n-1}/g_n$,
 and 
\begin{align*}
\overline{g}_n
&= \frac{\sigma_{n-1}(a,s)\sigma_{n}(a+2,s)}{\sigma_{n-1}(a+1,s)\sigma_{n}(a+1,s) } \\
&= \frac{1}{a+1} 
\left(  
s \left(
\log \frac{ \overline{\sigma}_{n-1}}{\overline{\sigma}_{n}}
\right)'
+a+d_{n}-d_{n-1}+1
\right), 
\quad 
\text{where $'= \frac{\rm d}{{\rm d}s}$ and using (\ref{eq:bil-p3-2})}
\\
&= \frac{1}{a+1} 
\left(  
s \left(
\log \frac{ \overline{\sigma}_{n-1} }{ \sigma_{n-2} }
+ \log \frac{  \sigma_{n-2} }{ \sigma_{n-1} }
+ \log \frac{ \sigma_{n-1} }{ \overline{\sigma}_n }
\right)'
+a+d_{n}-d_{n-1}+1
\right)
\\
&=
 \frac{1}{a+1} 
\left(  
s \frac{\overline{\sigma}_{n-1} \sigma_n}{\sigma_{n-1} {\overline{\sigma}}_{n} }
-s \frac{\overline{\sigma}_{n-2} \sigma_{n-1}}{\sigma_{n-2} {\overline{\sigma}}_{n-1} }
+a  \frac{  \underline{\sigma}{}_{n-2} \overline{\sigma}_{n-1}}{\sigma_{n-2} \sigma_{n-1} }
+ \alpha_{n-1}
\right),
 \quad \text{using (\ref{eq:bil-p3-1}) and (\ref{eq:bil-p3-2}) }
 \\
 &=  \frac{1}{a+1} 
\left( 
 s \left(\overline{f}_n- \overline{f}_{n-1} \right) + a g_{n-1} +\alpha_{n-1}
\right).
\end{align*}
This transformation $(f_n,g_n)\mapsto \left(\overline{f}_n, \overline{g}_n\right)$
is birational and indeed  keeps (\ref{subeq:p3-chain}) invariant except shifting 
$a$ to $a+1$.

Next, we observe that the system (\ref{eq:bil-p3-1}) and (\ref{eq:bil-p3-2}) of
bilinear equations
is invariant under the transformation
$\iota:(\sigma_n(a,s),s,a,d_n) \mapsto (\sigma_{-n}(-a,s),-s,-a,d_{-n})$.
This trivial symmetry can be lifted to a birational symmetry of the $P_{\rm III}$-chain.

We shall now translate the Weyl group action $\langle r_i \rangle$
of type $A$, 
discussed in Sect~\ref{subsec:hom}, into birational transformations of $f_n$ and $g_n$.
For each $i \in {\mathbb Z}/\ell{\mathbb Z}$, 
we have that
\begin{align}
\tau_{i-1}({\boldsymbol x}-[t])\tau_{i+1}({\boldsymbol x}+[t])
&=
\frac{1}{t} D_{\delta_t} \hat{\tau}_i({\boldsymbol x}) \cdot \tau_i ({\boldsymbol x}),
\\
\tau_{i-1}({\boldsymbol x})\tau_{i+1}({\boldsymbol x})
&=
D_{x_1} \hat{\tau}_i({\boldsymbol x}) \cdot \tau_i ({\boldsymbol x}),
\end{align}
which can be deduced from (\ref{eq:weyl_tau_a})
by taking the variables as ${\boldsymbol x}-{\boldsymbol x'}=2[t]$
and as ${\boldsymbol x} = {\boldsymbol x'}$, respectively. 
If we take account of the reduction condition (\ref{eq:sim})
and (\ref{eq:change}),
we obtain
\begin{align*}
a \underline{\sigma}{}_{i-1} \overline{\sigma}_{i+1}
&= 
\left(\alpha_i-s D_s \right) \hat{\sigma}_i \cdot \sigma_i,
\\
\sigma_{i-1}\sigma_{i+1}
&= D_s \hat{\sigma}_i \cdot \sigma_i,
\end{align*}
and thereby 
\[
\alpha_i \hat{\sigma}_i \sigma_i= s \sigma_{i-1} \sigma_{i+1} +a \underline{\sigma}{}_{i-1} \overline{\sigma}_{i+1}.
\]
Namely we have 
\[
r_i(\sigma_n)= \left\{
\begin{array}{ll} \displaystyle 
\frac{s \sigma_{i-1} \sigma_{i+1} +a \underline{\sigma}{}_{i-1} \overline{\sigma}_{i+1}}{\alpha_i \sigma_i} & (n=i),
\\
\sigma_n & (n \neq i).
\end{array}
\right.
\]
It is easy to derive the associated transformation on $(f_n,g_n)$.

Finally, it is obvious that a cyclic permutation of the suffixes,
$\pi:(f_n,g_n,\alpha_n) \mapsto (f_{n+1},g_{n+1},\alpha_{n+1}) $,
preserves (\ref{subeq:p3-chain}) invariant.

Summarizing above we arrive at the
\begin{thm}
The $P_{\rm III}$-chain {\rm (\ref{subeq:p3-chain})}
 is invariant under birational transformations
$T_a$, $\iota$, $r_i$ $(i \in {\mathbb Z}/\ell {\mathbb Z})$, and $\pi$
defined by
\[
\begin{array}{|c||l|l|l|}
\hline
& \text{\rm Action on $a$, $\alpha_n$, and $s$}& \text{\rm Action on $f_n$} &  \text{\rm Action on $g_n$}
\\  \hline
T_a &a \mapsto a+1&\displaystyle f_n \mapsto \overline{f}_n& g_n \mapsto \overline{g}_n
\\  \hline
\iota & a \mapsto -a, \alpha_n \mapsto \alpha_{-n}, s \mapsto -s & f_n \mapsto \overline{f}_{-n+1}
& g_n \mapsto g_{-n+1}  
\\ \hline
r_i & \alpha_i \mapsto -\alpha_i, & \displaystyle f_i\mapsto f_i-\frac{\alpha_i f_i}{s f_i+a g_{i+1}},
& \displaystyle g_i \mapsto g_i +\frac{\alpha_i g_{i+1}}{s f_i +a g_{i+1}} ,
\\ 
& \alpha_{i \pm 1} \mapsto \alpha_{i \pm 1}+\alpha_{i}& \displaystyle
 f_{i+1}\mapsto f_{i+1}+\frac{\alpha_i f_{i+1}}{s f_i+a g_{i+1}-\alpha_i}
 & \displaystyle
  g_{i+1} \mapsto g_{i+1} -\frac{\alpha_i g_{i+1}}{s f_i +a g_{i+1}} 
  \\  \hline
  \pi & \alpha_n \mapsto \alpha_{n+1} & f_n \mapsto f_{n+1} & g_n \mapsto g_{n+1}
  \\
  \hline
\end{array}
\]
Here $\overline{f}_n=f_{n-1}g_{n-1}/g_n$ and 
$\overline{g}_n=\left(s\left(\overline{f}_n-\overline{f}_{n-1} \right) +a g_{n-1}+\alpha_{n-1}\right)/(a+1)$.
\end{thm}

As indicated in Sect.~\ref{subsec:hom}, 
we see that
$\langle r_i \ (i \in {\mathbb Z}/\ell{\mathbb Z})\rangle$
provides a realization of
an affine Weyl group of type $A_{\ell-1}^{(1)}$,
denoted by $W(A_{\ell-1}^{(1)})$.
In addition $\pi$ realizes its rotational Dynkin automorphism.
Note that the transformations $r_i$ and $\pi$ have already appeared 
in \cite{ay94}.
These symmetries are clearly understood from the view point of the KP hierarchy.
However,
the nature of $T_a$ and $\iota$ seems mysterious from this point
and, in the first place, it is still an open problem to determine the group of birational symmetries of the $P_{\rm III}$-chain.

\begin{remark}[Toda equation] \rm
We shall derive a Toda equation satisfied by a sequence of $\tau$-functions
associated with the translation symmetry $T_a$.
First, it follows from
(\ref{eq:bil-p3-3}) that 
\begin{align}
\frac{1}{s} \frac{ {\cal D}^2 \underline{\sigma}{}_{n-1} \cdot \sigma_n }{\underline{\sigma}{}_{n-1}  \sigma_n}
&=d_{n-1}-d_{n}-
\left(s+2a+d_n-d_{n-1}-1\right) 
\frac{D_s \underline{\sigma}{}_{n-1} \cdot \sigma_n }{\underline{\sigma}{}_{n-1}  \sigma_n}
\nonumber \\
& \qquad 
+(d_{n-1}-d_n) \frac{ \sigma_{n-1} \underline{\sigma}{}_n }{\underline{\sigma}{}_{n-1}  \sigma_n}
- \frac{ {\cal D} \sigma_{n-1} \cdot \underline{\sigma}{}_n }{ \underline{\sigma}{}_{n-1} \sigma_n }
\nonumber \\
&= s+2a-1-(s+2a+2d_{n}-2d_{n-1}-1)f_n
-  \frac{ {\cal D} \sigma_{n-1} \cdot \underline{\sigma}{}_n }{ \sigma_{n-1} \underline{\sigma}{}_n } f_n
\label{eq:pre-toda-p3-1}
\end{align}
by the use of (\ref{eq:bil-p3-1}) and (\ref{eq:def-fg}).
Next, we express (\ref{eq:bil-p3-1}) in the form
 \[
 \frac{{\cal D} \underline{\sigma}{}_{n-1}  \cdot  \sigma_n}{ \underline{\sigma}{}_{n-1}  \sigma_n}
 =
-s + s \frac{  \sigma_{n-1}\underline{\sigma}{}_n }{ \underline{\sigma}{}_{n-1}  \sigma_n  }.
 \]
 By differentiating this with respect to $s$,
we have
\begin{equation} \label{eq:pre-toda-p3-2}
\frac{\rm d}{{\rm d}s}\left(
 \frac{{\cal D} \underline{\sigma}{}_{n-1}  \cdot  \sigma_n}{ \underline{\sigma}{}_{n-1}  \sigma_n}\right)
 =-1
 +
 \left(
 1+\frac{{\cal D} \sigma_{n-1} \cdot \sigma_n}{\sigma_{n-1} \sigma_n}  
 -\frac{{\cal D} \underline{\sigma}{}_{n-1} \cdot \underline{\sigma}{}_n}{\underline{\sigma}{}_{n-1} \underline{\sigma}{}_n} 
 \right)f_n.
\end{equation}
Combining (\ref{eq:pre-toda-p3-1}) with (\ref{eq:pre-toda-p3-2}), we obtain
\begin{align*}
\frac{1}{s} \frac{ {\cal D}^2 \underline{\sigma}{}_{n-1} \cdot \sigma_n }{\underline{\sigma}{}_{n-1}  \sigma_n}
-
\frac{\rm d}{{\rm d}s}\left(
 \frac{{\cal D} \underline{\sigma}{}_{n-1}  \cdot  \sigma_n}{ \underline{\sigma}{}_{n-1}  \sigma_n}\right)
 &= 
  s+2a-(s+2a+2d_{n}-2d_{n-1})f_n
  \\
  & \qquad 
+\left(  \frac{ {\cal D} \underline{\sigma}{}_{n-1} \cdot \sigma_n }{ \underline{\sigma}{}_{n-1} \sigma_n } 
-2 \frac{ {\cal D} \sigma_{n-1} \cdot \sigma_n }{ \sigma_{n-1} \sigma_n }
\right)f_n
\\
&= 2a-2a f_n g_n +s (f_n-1)^2
\end{align*}
via (\ref{eq:bil-p3-1}) and  (\ref{eq:bil-p3-2}).
Hence, if we  remember (\ref{eq:f-tau}), then we find that
\begin{align*}
\frac{1}{s} \frac{{\cal D}^2 \sigma_n \cdot \sigma_n}{{\sigma_n}^2}
&=
\frac{1}{s} 
 \frac{ {\cal D}^2 \underline{\sigma}{}_{n-1} \cdot \sigma_n }{\underline{\sigma}{}_{n-1}  \sigma_n}
 -\frac{\rm d}{{\rm d}s}\left(
 \frac{{\cal D} \underline{\sigma}{}_{n-1}  \cdot  \sigma_n}{ \underline{\sigma}{}_{n-1}  \sigma_n}\right)
 - \frac{1}{s}
 \left(\frac{{\cal D} \underline{\sigma}{}_{n-1} \cdot \sigma_n }{ \underline{\sigma}{}_{n-1}  \sigma_n} \right)^2
 \\
 &=2a -2a f_n g_n \\
 &=2a-2a \frac{\underline{\sigma}{}_n \overline{\sigma}_n}{{\sigma_n}^2}.
\end{align*}
Thus, we finally arrive at the {\it Toda equation}:
\begin{equation} \label{eq:toda-p3}
\frac{1}{2a s}
{\cal D}^2 \sigma_n \cdot \sigma_n = {\sigma_n}^2
- \underline{\sigma}{}_n \overline{\sigma}_n.
\end{equation}
Recall that for the case $\ell=2$ ($P_{\rm III}$)
such a differential-difference equation of Toda-type has been 
studied in \cite{oka}.
\end{remark}

\subsection{Rational solutions in terms of Schur functions}
\label{subsec:ratsol_p3}

As previously seen in Sect.~\ref{subsec:sim_p3},
the $P_{\rm III}$-chain 
is by nature equivalent to a similarity reduction of the mKP hierarchy.
Also,  
the mKP hierarchy admits the Schur functions as its homogeneous polynomial solutions;
see
 Example~\ref{ex:schur}.
Consequently we can construct a particular solution of 
the $P_{\rm III}$-chain
in terms of the Schur function.

To state the result precisely, 
we first recall some terminology. 
A subset ${\bf m} \subset {\mathbb Z}$ is said to be a 
{\it Maya diagram} 
if
$i \in {\bf m}$ (for $i \ll 0$) 
and
$i \notin {\bf m}$ (for $i \gg 0$).
Each Maya diagram 
${\bf m} = \{ \ldots, m_3, m_2, m_1 \}$
$(m_{i+1} < m_i)$
corresponds to
a unique partition
$\lambda=(\lambda_1,\lambda_2,\ldots)$ 
such that $m_i-m_{i+1}=\lambda_i-\lambda_{i+1}+1$.
For a sequence of integers
${\boldsymbol \nu}=(\nu_1,\nu_2,\ldots,\nu_{\ell})\in{\mathbb Z}^{\ell}$,
we associate
a Maya diagram
\[
{\bf m}({\boldsymbol \nu})=
(\ell{\mathbb Z}_{<\nu_1}+1)
\cup
(\ell{\mathbb Z}_{<\nu_2}+2)
\cup
\cdots
\cup
(\ell{\mathbb Z}_{<\nu_\ell}+\ell),
\]
and denote by 
$\lambda({\boldsymbol \nu})$
its corresponding partition.
Note that 
$\lambda ({\boldsymbol \nu}+ {\boldsymbol 1}) = \lambda({\boldsymbol \nu})$ 
where
$ {\boldsymbol 1} =(1, 1,\ldots,1) $.
We call a partition of the form $\lambda({\boldsymbol \nu})$
an {\it $\ell$-core} partition.
A partition $\lambda$ is 
$\ell$-core 
if and only if
$\lambda$ 
has no hook with length of a multiple of $\ell$.
For example,
if $\ell=3$ and ${\boldsymbol \nu}=(2,0 ,3)$ then the resulting partition reads $\lambda({\boldsymbol \nu})=(4,2,1,1)$.
Next, we prepare a cyclic chain of the Schur functions attached to $\ell$-core partitions
that is
connected by successive action of vertex operators.

\begin{lemma}
It holds that
\[
X^+_{\ell \nu_i-| {\boldsymbol \nu} |} S_{\lambda({\boldsymbol \nu}(i-1))}({\boldsymbol x})
= \pm S_{\lambda({\boldsymbol \nu}(i))}({\boldsymbol x})
\]
for arbitrary ${\boldsymbol \nu}=(\nu_1,\nu_2,\ldots,\nu_\ell) \in {\mathbb Z}^\ell$,
where
${\boldsymbol \nu}(i)={\boldsymbol \nu}+(\overbrace{1,\ldots,1}^i ,\overbrace{0,\ldots,0}^{\ell-i})$
and $|{\boldsymbol \nu}|=\nu_1+\nu_2+ \cdots+\nu_\ell$.
\end{lemma}

Finally, we are led to the following expression of rational solutions 
of the $P_{\rm III}$-chain
in terms of the Schur functions attached to the $\ell$-core partitions.

\begin{thm}
For any ${\boldsymbol \nu} \in {\mathbb Z}^\ell$,
let
\[
\sigma_n(a,s)=S_{\lambda\left({\boldsymbol \nu}(n) \right)}
\left(s+a,\frac{a}{2},\frac{a}{3},\frac{a}{4}, \ldots \right).
\]
Then $\sigma_n(a,s)$ solve the system of bilinear equations 
{\rm(\ref{eq:bil-p3-1})}
and
{\rm(\ref{eq:bil-p3-2})} when
$d_{n+1}-d_n= \ell \nu_{n+1}- |{\boldsymbol \nu}|$.
Consequently, 
the $2\ell$-tuple of functions
\[
 f_n = \frac{ \sigma_{n-1}(a,s) \sigma_n(a-1,s) }{  \sigma_{n-1}(a-1,s) \sigma_n(a,s) },
 \quad
 g_n = \frac{ \sigma_{n-1}(a-1,s) \sigma_n(a+1,s) }{  \sigma_{n-1}(a,s) \sigma_n(a,s) },
 \]
gives a rational solution of the $P_{\rm III}$-chain, {\rm (\ref{subeq:p3-chain})}, with the parameters
$\alpha_n= \ell(\nu_{n+1}-\nu_n)+1$. 
\end{thm}

\begin{remark} \rm
For the case $\ell=2$ ($P_{\rm III}$)
the above rational solution has been 
studied in a different way;
cf. \cite{km99}.
\end{remark}

\begin{example}\rm
Consider the polynomial
\[
R_\lambda(a,s)=
\left( \prod_{(i,j) \in \lambda} h(i,j) \right)
S_\lambda \left(s+a,\frac{a}{2},\frac{a}{3},\frac{a}{4}, \ldots \right)
\]
for a partition $\lambda=(\lambda_1,\lambda_2,\ldots)$,
where $h(i,j)$ denotes the {\it hook-length}, 
$h(i,j)=\lambda_i+{}^{\rm T}\lambda_j-i-j+1$,
with ${}^{\rm T} \lambda$ being the transpose (or conjugate) of $\lambda$.
Interestingly enough, $R_\lambda(a,s)$
comes out to be a polynomial with integer coefficients and monic with respect to  both $s$ and $a$.
We give below some examples of $R_\lambda(a,s)$ of small degree: 
\begin{align*}
R_\emptyset =& 1, \quad
R_{\young{1}} = s+a, \quad
R_{\young{2}} = s^2+2as+a(a+1), \\
R_{\young{3}} =& s^3+3as^2+3a(a+1)s+a(a+1)(a+2),  \\
R_\young{12}  =& s^3+3a s^2+3a^2 s+(a-1)a(a+1),  \\
R_\young{4}  =& s^4+4 a s^3+6a(a+1)s^2+4 a(a+1)(a+2)s+ a(a+1)(a+2)(a+3), \\
R_\young{13}  =& s^4+4a s^3+2a(3a+1)s^2+4a^2(a+1)s+(a-1)a(a+1)(a+2),  \\
R_\young{22}  =& s^4+4 a s^3 +6 a^2 s^2 +4(a-1)a(a+1)s+(a-1)a^2 (a+1).
\end{align*}
This polynomial can be regarded as a generalization of 
the Umemura polynomial associated with  $P_{\rm III}$;
cf. \cite{noou}.
\end{example}


\subsection{Lax formalism}

We introduce the {\it wave function}: 
 \begin{equation}
 \label{eq:wavefn} 
\psi_n({\boldsymbol x},k)=
\frac{\tau_n ({\boldsymbol x}- [k^{-1}]) }{\tau_n({\boldsymbol x})}   e^{\xi({\boldsymbol x},k)},
 \end{equation}
which is a function
in ${\boldsymbol x}=(x_1,x_2,\ldots)$
equipped with an additional parameter $k$ (the {\it spectral variable}). 
In terms of the wave functions, 
the linear equations 
associated with the (modified) KP hierarchy
can be generated from
the bilinear identities 
(recall (\ref{eq:mkp-bil}) and also Sect.~\ref{subsect:pre_mkp})
in a standard manner;
cf. \cite{djm82, mjd}.
In what follows,
we show that these linear equations naturally
induce an auxiliary linear problem,
through the similarity reduction,
whose integrability condition 
amounts to the $P_{\rm III}$-chain
(Lax formalism).

We first prepare a key lemma below.

\begin{lemma}  \label{lemma:Epsi}
If $\tau_n=\tau_n({\boldsymbol x})$ obeys the homogeneity   
$E \tau_n=d_n \tau_n$.
Then it holds that
\begin{equation}  \label{eq:Epsi}
\left(E-k \frac{\partial}{\partial k} \right)\psi_n=0.
\end{equation}
\end{lemma}

\pf
The homogeneity tells us that
$\left(E - k {\partial}/{\partial k}\right) \tau_n({\boldsymbol x}-[k^{-1}])=d_n \tau_n({\boldsymbol x}-[k^{-1}])$.
Therefore, 
\[\left(E - k \frac{\partial}{\partial k}\right) 
\frac{\tau_n({\boldsymbol x}-[k^{-1}])}{ \tau_n({\boldsymbol x}) } =0.
\]
 Combining this with the formula
$\left(E-k {\partial}/{\partial k} \right) e^{\xi({\boldsymbol x},k)}=0$,
we conclude (\ref{eq:Epsi}).
\qed
\\

Now we set
\[
\phi_n(a,s,k) =\psi_n({\boldsymbol x},k)
\]
under
the reduction conditions (\ref{eq:sim}) and (\ref{eq:change}).

 \begin{lemma} The wave functions $\phi_n=\phi_n(a)=\phi_n(a,s,k)$ satisfy
  the following linear equations{\rm:}
 \begin{align} 
 \label{eq:lin1}
  \frac{\partial }{\partial s} \phi_n(a)
 &=
 \frac{ a+ d_{n+1} -d_n -a g_{n+1} }{s} 
 \phi_n(a)
 +k  \phi_{n+1}(a),
  \\
\label{eq:lin2}
\phi_n(a-1)
&=f_{n+1}  \phi_{n}(a)- k  \phi_{n+1}(a),
\\
\label{eq:lin3}
\left(k \frac{\partial }{\partial k} -  s \frac{\partial }{\partial s} \right)
\phi_n(a)
&=
a \left(
g_{n+1} - 1
\right) \phi_n(a)
+ a k g_{n+1}  \phi_{n+1}(a+1).
\end{align}
 \end{lemma}

 \pf
We shall deduce (\ref{eq:lin1}), (\ref{eq:lin2}), and (\ref{eq:lin3})
respectively
from (\ref{eq:mkp_a}), (\ref{eq:mkp_c}), and (\ref{eq:mkp_d}) in Lemma~\ref{lemma:mkp}.
First, it follows from (\ref{eq:mkp_a}) with $t$ replaced by $1/k$ that
\[
 \frac{\partial}{\partial x_1} \left( \frac{\tau_n({\boldsymbol x}-[k^{-1}])}{ \tau_n({\boldsymbol x}) } \right)
 + \left( 
 \frac{D_{x_1} \tau_n({\boldsymbol x})  \cdot  \tau_{n+1}({\boldsymbol x})}{ \tau_n({\boldsymbol x}) \tau_{n+1}({\boldsymbol x})}
 +k
 \right) \frac{\tau_n({\boldsymbol x}-[k^{-1}])}{ \tau_n({\boldsymbol x}) } 
 -k \frac{\tau_{n+1}({\boldsymbol x}-[k^{-1}])}{ \tau_{n+1}({\boldsymbol x}) } =0,
\]
i.e., 
\begin{equation}
 \frac{\partial \psi_n}{\partial x_1} + 
 \frac{D_{x_1} \tau_n  \cdot  \tau_{n+1}}{ \tau_n \tau_{n+1}} 
 \psi_n
 -k  \psi_{n+1} =0.
\end{equation}
Remembering  (\ref{eq:dx1=ds}):
${\partial}/{ \partial x_1} ={\partial}/{ \partial s}$,
we see that
\[
\frac{\partial \phi_n}{\partial s}+
\frac{D_{s}\sigma_{n} \cdot \sigma_{n+1}}{\sigma_{n} \sigma_{n+1}} \phi_n
-k \phi_{n+1}=0,
\]
which implies  (\ref{eq:lin1}) through (\ref{eq:g-tau}).

Next, 
(\ref{eq:mkp_c}) with $s$ replaced by $1/k$
shows that
\[
\psi_{n}({\boldsymbol x}-[t],k)
= \frac{\tau_{n}({\boldsymbol x})\tau_{n+1}({\boldsymbol x}-[t])}{\tau_{n}({\boldsymbol x}-[t])\tau_{n+1}({\boldsymbol x}) } 
\psi_{n}({\boldsymbol x},k)-
t k \psi_{n+1}({\boldsymbol x},k),
\]
which thus yields
\[
\phi_{n}(a-1)
= \frac{\sigma_{n}(a,s)\sigma_{n+1}(a-1,s)}{\sigma_{n}(a-1,s)\sigma_{n+1}(a,s) } 
\phi_{n}(a)-
k \phi_{n+1}(a)
\]
under (\ref{eq:change}) and $t=1$.
By (\ref{eq:def-fg}) it is immediate to obtain (\ref{eq:lin2}).

Finally, 
it follows from (\ref{eq:mkp_d}) with $s$ replaced by $1/k$
that
\begin{equation}
\delta_t \psi_n({\boldsymbol x},k)
+
 \frac{D_{\delta_t} \tau_n({\boldsymbol x})  \cdot  \tau_{n+1}({\boldsymbol x})}{ \tau_n({\boldsymbol x}) \tau_{n+1}({\boldsymbol x})}
 \psi_n({\boldsymbol x},k)
 - 
 tk
\frac{\tau_{n}({\boldsymbol x}-[t])\tau_{n+1}({\boldsymbol x}+[t])}{\tau_{n}({\boldsymbol x})\tau_{n+1}({\boldsymbol x}) } 
 \psi_{n+1}({\boldsymbol x}+[t],k)
 =0.
\label{eq:lin3-pre}
\end{equation}
Consider the substitution
(\ref{eq:change}) and $t=1$.
Applying Lemma~\ref{lemma:Epsi} together with (\ref{eq:E-p3}),
we find that
\[
\delta_t \psi_n({\boldsymbol x},k)
=
\frac{1}{a} \left( E - s \frac{\partial}{\partial s} \right)
 \phi_n
=
\frac{1}{a} \left( k \frac{\partial}{\partial k} -s \frac{\partial}{\partial s} \right)
 \phi_n.
\]
Similarly we know that
\begin{align*}
 \frac{D_{\delta_1} \sigma_n \cdot  \sigma_{n+1}}{ \sigma_n \sigma_{n+1}}
&= \delta_1 \log \frac{\sigma_n}{\sigma_{n+1}}  
=\frac{1}{a} \left( E - s \frac{\partial}{\partial s} \right)\log \frac{\sigma_n}{\sigma_{n+1}} 
\\
&=\frac{1}{a} \left( d_n-d_{n+1} - s \frac{\partial}{\partial s} \right)\log \frac{\sigma_n}{\sigma_{n+1}}, \quad \text{using  (\ref{eq:sim})} 
\\
&=1-g_{n+1}, \quad \text{using  (\ref{eq:g-tau})}. 
\end{align*}
If we put it all together, we then obtain (\ref{eq:lin3}) from (\ref{eq:lin3-pre}).
\qed
\\

The linear equations  (\ref{eq:lin2}) 
can be solved for $\phi_n(a)$
by virtue of the $\ell$-periodicity;
i.e.,
\[
 \phi_n(a) = 
\frac{1}{1-k^\ell}
\sum_{j=1}^{\ell} k^{j-1} \left(\prod_{i=1}^j \frac{1}{f_{n+i}} \right) \phi_{n+j-1}(a-1).
\]
Note here that 
the suffix $n$ should be regarded suitably as an element of 
${\mathbb Z}/\ell {\mathbb Z}$.
If we shift $a$ to $a+1$,
this expression takes the form
\begin{align}  
\phi_n(a+1) 
&= 
\frac{1}{1-k^\ell}
\sum_{j=1}^{\ell} k^{j-1} \left(\prod_{i=1}^j \frac{1}{\overline{f}_{n+i}} \right) \phi_{n+j-1}(a)
\nonumber
\\
&=
\frac{1}{1-k^\ell} \frac{1}{g_{n}}
\sum_{j=1}^{\ell} k^{j-1} g_{n+j}
\left(\prod_{i=1}^j \frac{1}{f_{n+i-1}} \right) \phi_{n+j-1}(a).
\label{eq:sol-lin2}
\end{align}
Here recall the abbreviated notation $\overline{f}_n:=f_n(a+1,s)$ and 
the contiguity relation
(\ref{eq:overlinef}).

We are now ready to present a system of 
linear differential equations for the wave functions
$\phi_n=\phi_n(a,s,k)$ $(n \in {\mathbb Z}/\ell {\mathbb Z})$,
from which the $P_{\rm III}$-chain (\ref{subeq:p3-chain}) emerges as its 
compatibility condition. 
First, 
eliminating 
$\partial \phi_n/\partial s$ and
$\phi_n(a+1)$ 
from 
(\ref{eq:lin3}) by 
(\ref{eq:lin1})
and
 (\ref{eq:sol-lin2}), 
we obtain
the linear differential equation with respect to the spectral variable $k$:
\begin{equation}  \label{eq:lin-k-p3}
k \frac{\partial }{\partial k} \phi_n
=
(d_{n+1}-d_n) \phi_n+ s k \phi_{n+1} 
+ \frac{a}{1-k^\ell} 
\sum_{j=1}^{\ell} k^j g_{n+j+1} \left(\prod_{i=1}^j \frac{1}{f_{n+i}} \right) 
\phi_{n+j}.
\end{equation}
This has 
$\ell +1$ regular singularities 
at $k=0, \exp \left(2 \pi \sqrt{-1} n/\ell\right)$ $(n \in {\mathbb Z}/\ell{\mathbb Z})$ 
and an irregular singularity at
$k=\infty$ whose Poincar\'e rank equals one.
This expression is, however, redundant in some sense;
i.e.,
we can reduce the number of singularities
appropriately  by changing the variables.
Let
 \[
 \Phi={}^{\rm T} \left(\phi_0, k \phi_1, \ldots, k^{\ell-1} \phi_{\ell-1} \right)
 \quad \text{and} \quad
 z=k^\ell.
 \]
Then (\ref{eq:lin-k-p3}) is converted to the equation 
\begin{equation} \label{eq:lax-p3-z}
\frac{\partial \Phi}{\partial z} = A \Phi
= \left( C+\frac{A_0}{z}+\frac{A_1}{z-1}\right) \Phi,
\end{equation}
where the $\ell \times \ell$ matrices $C$, $A_0$, and $A_1$ 
are given as follows:
\begin{align*}
C &=
\left(
  \begin{array}{cc}
  & \text{\LARGE $O$}  \\
 s/\ell &
 \end{array}
 \right),
\\
A_0
&= 
\left(
  \begin{array}{ccccc}
e_0 & {s}/{\ell}    &    &  & 
\\
      & e_1 &  s/\ell     &  &
\\
      &        & \ddots & \ddots &     \\
      &        &             &e_{\ell-2} & s/\ell   
\\
      &        &             &           & e_{\ell-1}
 \end{array}
 \right)
 +
   \left(
  \begin{array}{ccccc}
0& v_{0,1}    &  v_{0,2}  & \cdots &v_{0,\ell-1}
\\
      & 0 &  v_{1,2}        &  \ldots&v_{1,\ell-1}
\\
      &        & \ddots & \ddots &     \vdots
\\
      &      &             & 0 & v_{\ell-2,\ell-1}   
\\
      &        &             &           & 0
 \end{array}
 \right), 
 \\
 A_1 &= -(v_{i,j})_{0 \leq i,j \leq \ell-1}
 \end{align*}
with
 \[
e_n= \frac{d_{n+1}-d_n+n}{\ell} \quad
\text{and}
\quad 
v_{n,n+j}=
\frac{a g_{n+j+1}}{\ell}
 \prod_{i=1}^j \frac{1}{f_{n+i}}
 =\frac{a g_{n+1}}{\ell}
 \prod_{i=1}^j \frac{1}{\overline{f}_{n+i+1}} 
\]
for $0 \leq n \leq \ell-1$ and $1 \leq j \leq \ell$.
Note that we read appropriately the suffixes of dependent variables
$f_i$, $g_i$, and $v_{i,j}$ modulo $\ell$. 
Hence (\ref{eq:lax-p3-z}) has regular singularities at $z=0,1$
and an irregular singularity at $z=\infty$.
To be more precise,
the exponents at  $z=0$ and $z=1$
turn out to be $(e_0,e_1,\ldots,e_{\ell-1})$ and $(-a,0,\ldots,0)$,
which equal by definition the eigenvalues of $A_0$ and $A_1$, respectively.
The latter can be computed
in the following manner.
Let  $\boldsymbol f$ and $\boldsymbol g$ be the row vectors 
defined by
\begin{align*}
{\boldsymbol f}
&=\left(\prod_{i=1}^j \frac{1}{\overline{f}_{i+1}}\right)_{0 \leq j \leq \ell-1}
=\left( 1, \frac{1}{\overline{f}_2}, \frac{1}{\overline{f}_{2}  \overline{f}_{3}}, \ldots,
\frac{1}{\overline{f}_{2} \overline{f}_{3} \cdots \overline{f}_{\ell}}
\right), \\
{\boldsymbol g}
&=\left( g_{j+1}\prod_{i=1}^j \overline{f}_{i+1}  
\right)_{0 \leq j \leq \ell-1}
=\left( g_1, g_2 \overline{f}_{2}, g_3 \overline{f}_{2} \overline{f}_{3},
\ldots,
g_\ell \overline{f}_{2} \overline{f}_{3} \cdots \overline{f}_{\ell}
\right).
\end{align*}
We observe that
 ${}^{\rm T}{\boldsymbol g} \cdot {\boldsymbol f} = -(\ell/a) A_1$
 and
 $ {\boldsymbol f} \cdot {}^{\rm T}{\boldsymbol g} = g_1+g_2+\cdots+g_\ell =\ell$ 
 via (\ref{eq:cons-p3}),
 thereby,
the exponents at $z=1$ are evaluated as $(-a,0,\ldots,0)$.
Further,
we say $z=\infty$ to be an irregular singularity of Poincar\'e rank $1/\ell$
in the sense that 
(\ref{eq:lax-p3-z}) has a formal matrix solution of the form
\[ \Xi= {\rm diag}\left(1,z^{1/\ell},z^{2/\ell},\ldots, z^{(\ell-1)/\ell} \right)
\cdot
\left( \zeta^{i j} \right)_{0 \leq i,j \leq \ell-1}
\cdot
\left( I + \sum_{i=1}^\infty \Xi^{(i)} z^{-i/\ell}
\right) \cdot e^{T(z)}
\]
near $z=\infty$, where
\[
T(z)=
- {\rm diag}\left(1,\zeta,\zeta^2,\ldots,\zeta^{\ell-1} \right)  s  z^{1/\ell}
-\frac{a}{\ell} \log z
\]
and $\zeta$ denotes a primitive $\ell$th root of unity.
Although the series $\Xi$ is in general divergent,
it expresses the asymptotic behavior of an actual solution 
in some sectorial domain around $z=\infty$.
We refer to
\cite{huk} for the general theory of solutions around 
an irregular singularity.

On the other hand, 
we obtain from (\ref{eq:lin1})
the linear differential equation with respect to $s$:
\begin{equation}
\frac{\partial}{\partial s} \Phi
=  B \Phi=
\left(
\left(
 \begin{array}{ccccc}
u_0 & 1    &            & &
\\
      & u_1 & 1         & &
\\
      &        & \ddots & \ddots &     
\\
      &        &             &u_{\ell-2} & 1
\\
    &        &             &           & u_{\ell-1}
 \end{array}
 \right)
 + z
 \left(
 \begin{array}{ccccc}
 &     &            & &
\\
      & &        & &
\\
      &        & &  &     
\\
      &        &             && 
\\
 1    &        &             &           & 
 \end{array}
 \right)
 \right) \Phi,
  \label{eq:lax-p3-s}
\end{equation}
where 
$u_n= (a+d_{n+1}-d_n-a g_{n+1})/s$.

In the context of monodromy preserving deformations,
(\ref{eq:lax-p3-s}) governs a deformation of (\ref{eq:lax-p3-z})
along a parameter $s$ with keeping its 
monodromy matrices and Stokes multipliers invariant.
Finally, we give an alternative characterization of the $P_{\rm III}$-chain as 
a compatibility condition of the linear problem.

\begin{thm}[Lax representation]
 The integrability condition
 \[
 \left[
  \frac{\partial }{\partial s} - B, \frac{\partial }{\partial z} -A
\right]=0
 \]
 of the system
 {\rm(\ref{eq:lax-p3-z})}
 and  {\rm(\ref{eq:lax-p3-s})}
 is equivalent to the $P_{\rm III}$-chain, 
 {\rm(\ref{subeq:p3-chain})}.
 \end{thm}

\begin{remark} \rm
If $\ell=2$, the system
{\rm(\ref{eq:lax-p3-z})} and  {\rm(\ref{eq:lax-p3-s})}
is equivalent to the Lax pair for $P_{\rm III}$
that was recently found 
in a moduli-theoretical approach
\cite{ps09}
(see also \cite{oo06}). 
\end{remark}


\section{From KP hierarchy to Painlev\'e IV/V chain}
\label{sect:p45}

In this section we demonstrate a similarity reduction from the $\ell$-periodic mKP hierarchy ($\ell \geq 3$)
to the higher order Painlev\'e
equation
$P(A_{\ell-1}^{(1)})$ or, equivalently,
the Darboux chain with period $\ell$,
which coincides with the fourth and fifth Painlev\'e equations 
when 
$\ell=3$ and $\ell=4$, respectively;
cf. \cite{ny99}.

\subsection{Similarity reduction}
Fix an integer $\ell \geq 3$.
Let 
$\tau_n({\boldsymbol x})$ be a solution of 
the mKP hierarchy,
(\ref{eq:mkp-bil}) or (\ref{eq:mkp}).
Suppose it fulfills the following three conditions:
(i) homogeneity $E \tau_n=d_n \tau_n$ $(d_n \in {\mathbb C})$;
(ii) $\ell$-periodicity $\tau_{n+\ell}=\tau_n$;
(iii) specialization $x_n=0$ $(n \neq 1,2)$.
Note here that this specialization is intended to eliminate all time evolutions
except the first two.
In view of (i),
we can take
$x_2$ to be a constant without loss of generality.
For instance, we fix
\begin{equation}\label{eq:change_p45}
x_1=x, \quad x_2=-\frac{1}{2}, \quad \text{and} \quad
x_n=0 \quad (n \geq 3).
\end{equation}
Under the above constraints, we set 
$\sigma_n(x)=\tau_n({\boldsymbol x})$.

\begin{prop}    \label{prop:p45_bil}
It holds that
\begin{equation} \label{eq:bil_p45}
\left( {D_x}^2+x D_x+d_{n+1}-d_{n} \right) \sigma_n \cdot \sigma_{n+1}=0.
\end{equation}
\end{prop}

\pf
We know from
(\ref{eq:change_p45})
 that 
\begin{equation} \label{eq:p45_flow}
\frac{\partial}{\partial x_1}=\frac{\rm d}{{\rm d}x} 
\quad  \text{and} \quad
\frac{\partial}{\partial x_2}= x \frac{\rm d}{{\rm d}x} -E.
\end{equation}
Thus we deduce (\ref{eq:bil_p45}) immediately
from (\ref{eq:mkp_hirota_1}).
\qed
\\

Let us consider an
$\ell$-tuple of functions 
$w_n=w_n(x)$ $(n \in {\mathbb Z}/\ell {\mathbb Z})$
defined by
\begin{equation}  \label{eq:p45_wn}
w_n = \frac{x}{2}+\frac{D_x \sigma_n \cdot \sigma_{n+1}}{ \sigma_n \sigma_{n+1}}.
\end{equation}
By virtue of Prop.~\ref{prop:p45_bil}, we have then  the 

\begin{thm}
It holds that
\begin{equation} \label{eq:darboux}
\frac{{\rm d}}{{\rm d} x} \left(  w_n+w_{n-1}\right)
={w_n}^2-{w_{n-1}}^2+ \alpha_n \quad
(n \in {\mathbb Z}/\ell {\mathbb Z})
\end{equation}
where 
$\alpha_n=d_{n-1}-2d_n+d_{n+1}+1$.
\end{thm}

The sequence (\ref{eq:darboux}) of ordinary differential equations with quadratic nonlinearity is
known as the 
{\it Darboux chain} (with period $\ell$).
It emerged originally from the spectral theory of Schr\"odinger operators
in connection with Darboux transformations.
Moreover it provides a higher order generalization of 
$P_{\rm IV}$ and $P_{\rm V}$
which correspond to the cases 
$\ell=3$ and $4$, respectively.
For details to
\cite{adl94, vs93}.

\begin{remark} \rm
If we consider the change of variables as (cf. \cite{tak03})
\[f_n=w_n+w_{n-1}
= x+\frac{D_x \sigma_{n-1} \cdot \sigma_{n+1}}{ \sigma_{n-1} \sigma_{n+1}},
\]
then 
the Darboux chain
(\ref{eq:darboux}) takes
the following expression:
\\
(i) if $\ell=2g+1$ $(g=1,2,\ldots)$,
\[
P(A_{2g}^{(1)}): \qquad
\frac{{\rm d}f_n}{{\rm d}x}=
f_n 
\left(  \sum_{i=1}^g  f_{n+2i-1} - 
\sum_{i=1}^g  f_{n+2i}
\right)+\alpha_n;
\]
(ii)  if $\ell=2g+2$ $(g=1,2,\ldots)$,
\begin{align*}
P(A_{2g+1}^{(1)}): \qquad
\frac{\ell x}{2}
\frac{{\rm d}f_n}{{\rm d}x}
&= 
f_n 
\left(  
\sum_{1 \leq i \leq j \leq g} f_{n+2i-1} f_{n+2j} 
- \sum_{1 \leq i \leq j \leq g}  f_{n+2i} f_{n+2j+1} 
 \right)
 \\
& \quad  + \left( \frac{\ell}{2} -\sum_{i=1}^g \alpha_{n+2 i}  \right) f_n + \alpha_n \sum_{i=1}^g f_{n+2i}.
\end{align*}
This is called the {\it higher order Painlev\'e equation 
of type $A_{\ell-1}^{(1)}$};
see \cite{ny98}.
\end{remark}

\subsection{Affine Weyl group symmetry}
\label{subsect:p4_weyl}

Our goal here is to lift the Weyl group symmetry of the homogeneous 
$\tau$-sequence (see Sect.~\ref{subsec:hom})
to the level of birational transformations of the Darboux chain (\ref{eq:darboux}).

First we prepare some relevant formulae, which will be used 
in Sect.~\ref{sect:p2} ($P_{\rm II}$ case) too.
In a similar manner as Sect.~\ref{subsect:pre_mkp},
we can rewrite (\ref{eq:weyl_tau_a}) 
equivalently
into
\begin{equation}   \label{eq:weyl_tau_a'}
\tau_{i-1}({\boldsymbol x}+{\boldsymbol u}) \tau_{i+1}({\boldsymbol x}-{\boldsymbol u})
= G_0(\hat{\tau}_i({\boldsymbol x}),\tau_{i}({\boldsymbol x}); {\boldsymbol u}),
\end{equation}
where $\hat{\tau}_i=r_i(\tau_i)$.
Regarding the right hand side,
refer to  (\ref{eq:formula_bil2hirota}).
Let us consider the Taylor expansion of (\ref{eq:weyl_tau_a'})
in ${\boldsymbol u}=(u_1,u_2,\ldots)$.
Taking the coefficients of $1(={\boldsymbol u}^0)$, $u_1$, $u_2$, and ${u_1}^2$
thus yield the formulae
\begin{subequations}
\begin{align}   
&D_{x_1} \hat{\tau}_i \cdot \tau_i -\tau_{i-1} \tau_{i+1}=0,
\label{eq:weyl_tau_a_0}
\\
&D_{x_2} \hat{\tau}_i \cdot \tau_i + D_{x_1} \tau_{i-1} \cdot \tau_{i+1}=0,
\label{eq:weyl_tau_a_1}
\\
&
\left(
{D_{x_1}}^3 + 2 D_{x_3}
\right)
\hat{\tau}_i \cdot \tau_i + 3 D_{x_2} \tau_{i-1} \cdot \tau_{i+1}=0,
\label{eq:weyl_tau_a_2}
\\
&
\left(
{D_{x_1}}^3 - 4 D_{x_3}
\right)
\hat{\tau}_i \cdot \tau_i + 
3{D_{x_1}}^2 \tau_{i-1} \cdot \tau_{i+1}=0,
\label{eq:weyl_tau_a_11}
\end{align}
\end{subequations}
respectively.
Also we recall that $r_i(\tau_n)=\tau_n$ $(n \neq i)$.

Now,
if we apply the reduction conditions under consideration,
then we verify from (\ref{eq:weyl_tau_a_0}) and (\ref{eq:weyl_tau_a_1}) 
that
\begin{align} \label{eq:p45_sym_1}
D_x \hat{\sigma}_i \cdot \sigma_i &= \sigma_{i-1} \sigma_{i+1},
\\
\label{eq:p45_sym_2}
\alpha_i \hat{\sigma}_i \sigma_i &= \left(  D_x +x\right) \sigma_{i-1}  \cdot \sigma_{i+1}
\end{align}
with $\alpha_i=\hat{d}_i - d_i=d_{i-1}-2d_i+d_{i+1}+1$.
It is easy to see from (\ref{eq:p45_wn}) 
that $r_i$ should keep $w_n$ $(n \neq i, i-1)$ invariant.
The computation of $r_i(w_i)$ reads as 
\begin{align*}
r_i(w_i)&= \frac{x}{2} + \frac{D_x \hat{\sigma}_{i} \cdot \sigma_{i+1}}{\hat{\sigma_{i}} \sigma_{i+1}}
= w_i + \frac{D_x \hat{\sigma}_{i} \cdot \sigma_{i}}{\hat{\sigma_{i}} \sigma_{i}}
\\
&= w_i + \frac{\alpha_i \sigma_{i-1} \sigma_{i+1}}{  \left(  D_x +x\right) \sigma_{i-1}  \cdot \sigma_{i+1} },
\quad \text{using (\ref{eq:p45_sym_1}) and  (\ref{eq:p45_sym_2})}
\\
 &= w_i+ \frac{\alpha_i}{w_{i-1}+w_i}.
\end{align*}
Similarly we find that $r_i(w_{i-1})=w_{i-1}-\alpha_i/(w_{i-1}+w_i)$.
Moreover,
a cyclic permutation of the suffices $\pi: (\sigma_n, d_n) \mapsto (\sigma_{n+1}, d_{n+1}) $ keeps the bilinear form 
(\ref{eq:bil_p45}) invariant and so does an inversion 
$\iota:(\sigma_n, d_n,x) \mapsto \left(\sigma_{-n}, d_{-n},\sqrt{-1}x\right)$.
It is again easy to lift $\pi$ and $\iota$ to birational symmetries of (\ref{eq:darboux}).
Summarizing above, we have the

\begin{thm}
The Darboux chain  {\rm(\ref{eq:darboux})} 
is invariant under birational transformations $r_i$ $(i \in {\mathbb Z}/\ell {\mathbb Z})$, $\pi$, and $\iota$
defined by
\[
\begin{array}{|c||l|l|}
\hline
& \text{\rm Action on $\alpha_n$ and $x$}& \text{\rm Action on $w_n$} 
\\  \hline
r_i & \alpha_i \mapsto -\alpha_i, & \displaystyle w_{i-1} \mapsto w_{i-1}-\frac{\alpha_i}{w_{i-1}+w_i},
\\ 
& \alpha_{i \pm 1} \mapsto \alpha_{i \pm 1}+\alpha_{i}& \displaystyle w_{i} \mapsto w_{i}+\frac{\alpha_i}{w_{i-1}+w_i}
  \\  \hline
  \pi & \alpha_n \mapsto \alpha_{n+1} & w_n \mapsto w_{n+1} 
  \\
  \hline
  \iota& \alpha_n \mapsto \alpha_{-n}, x \mapsto \sqrt{-1} x&w_n \mapsto \sqrt{-1}w_{-n-1}
    \\
  \hline
\end{array}
\]
\end{thm}

Note that birational mappings $\langle r_i (i \in {\mathbb Z}/\ell {\mathbb Z})\rangle$ define a realization of affine Weyl group $W(A_{\ell-1}^{(1)})$.
Additionally, $\pi$ and $\iota$ realize
 respectively a rotation and an inversion of
its Dynkin diagram.

\subsection{Lax formalism}

Introduce a function 
$\rho_n({\boldsymbol x},k)=X^+(k) \tau_n({\boldsymbol x})
=
\tau_n({\boldsymbol x}-[k^{-1}])e^{\xi({\boldsymbol x},k)}$.
In order to construct linear differential equations satisfied by the 
wave function
$\psi_n({\boldsymbol x},k)=
\rho_n({\boldsymbol x},k)/\tau_n({\boldsymbol x})$,
let us first prepare some useful formulae.
It follows from 
(\ref{eq:-1modkp}) multiplied by
$1 \otimes X^+(k)$
that
\begin{equation}
k \tau_{n} \otimes \rho_{n+1}+ \sum_{i+j=-1} X_i^- \tau_{n+1} \otimes X_j^+ \rho_n
=0.
\end{equation}
Therefore, we find that (recall (\ref{eq:formula_bil2hirota}))
\[
k \tau_{n}({\boldsymbol x}+{\boldsymbol u})  \rho_{n+1}({\boldsymbol x}-{\boldsymbol u},k)+ G_0(\tau_{n+1}({\boldsymbol x}), \rho_n({\boldsymbol x},k);{\boldsymbol u})
=0.
\]
The coefficients of $1(={\boldsymbol u}^0)$, $u_1$, $u_2$, and ${u_1}^2$
in the above equation
show respectively that
\begin{align*} 
&k \tau_{n} \rho_{n+1}+ D_{x_1} \tau_{n+1} \cdot \rho_n=0,
\\
&k D_{x_1} \tau_{n} \cdot \rho_{n+1}- D_{x_2} \tau_{n+1} \cdot \rho_n=0,
\\
& 3 k D_{x_2} \tau_{n} \cdot \rho_{n+1}-\left( {D_{x_1}}^3+2 D_{x_3}  \right)\tau_{n+1} \cdot \rho_n=0,
\\
& 3 k {D_{x_1}}^2 \tau_{n} \cdot \rho_{n+1}-\left( {D_{x_1}}^3-4 D_{x_3}  \right)\tau_{n+1} \cdot \rho_n=0.
\end{align*}
Hence we have the following linear differential equations for $\psi_n$:
\begin{subequations}
\begin{align}  \label{eq:psi_a}
\frac{\partial \psi_n}{\partial x_1}
&= \frac{D_{x_1} \tau_{n+1} \cdot \tau_{n}}{\tau_{n+1} \tau_{n}} \psi_n
+k \psi_{n+1},
\\  \label{eq:psi_b}
\frac{\partial \psi_n}{\partial x_2}
&= \frac{D_{x_2} \tau_{n+1} \cdot \tau_{n}}{\tau_{n+1} \tau_{n}} \psi_n
+k \frac{D_{x_1} \tau_{n+2} \cdot \tau_{n}}{\tau_{n+2} \tau_{n}} \psi_{n+1}
+k^2 \psi_{n+2},
\\  \nonumber
\frac{\partial \psi_n}{\partial x_3}
&= \frac{1}{2}\frac{D_{x_3} \tau_{n+1} \cdot \tau_{n}}{\tau_{n+1} \tau_{n}} \psi_n
 + \frac{k}{2}\frac{ \left({D_{x_1}}^2+D_{x_2} \right)  \tau_{n+2} \cdot \tau_{n}}{\tau_{n+2} \tau_{n}} \psi_{n+1}
 \\
&\quad  +k^2 \frac{D_{x_1} \tau_{n+3} \cdot \tau_{n}}{\tau_{n+3} \tau_{n}} \psi_{n+2}
 +k^3 \psi_{n+3}.
 \label{eq:psi_c}
\end{align}
\end{subequations}
Note that it is possible to express 
${\partial \psi_n}/{\partial x_a}$ 
as a linear sum of $\psi_{n},\psi_{n+1},\ldots, \psi_{n+a}$ for general $a \in {\mathbb Z}_{>0}$.

Now we require the homogeneity condition
 $E \tau_n = d_n \tau_n$
 and  set 
\[ \phi_n(x,k)=\psi_n((x,-1/2,0,0, \ldots), k).
\]
In view of Lemma~\ref{lemma:Epsi} and
(\ref{eq:p45_flow}),
we deduce from
(\ref{eq:psi_a}) and (\ref{eq:psi_b}) the following

\begin{lemma} 
The wave functions $\phi_n=\phi_n(x,k)$
satisfy the following linear differential equations{\rm:}
\begin{align} 
k \frac{\partial}{\partial k} \phi_n 
&= (d_{n+1}-d_n) \phi_n + k (w_n+w_{n+1}) \phi_{n+1} -k^2 \phi_{n+2},
\label{p45_lax_1}
\\
\frac{\partial}{\partial x} \phi_n 
&= \left( \frac{x}{2} - w_n \right) \phi_n + k  \phi_{n+1}.
\label{p45_lax_2}
\end{align}
\end{lemma}

The linear differential equation (\ref{p45_lax_1})
with respect to the spectral variable $k$
has a regular singularity at $k=0$
and an irregular singularity (of Poincar\'e rank two) at $k=\infty$.
While the latter (\ref{p45_lax_2}) describes the monodromy preserving deformation
of the former (\ref{p45_lax_1}).
We shall slightly modify this system.
Let
$ \Phi={}^{\rm T} \left(\phi_0, k \phi_1, \ldots, k^{\ell-1} \phi_{\ell-1} \right)$
and
$z=k^\ell$.
Then (\ref{p45_lax_1}) is converted to the $\ell \times \ell$ matrix equation 
\begin{equation} \label{eq:p45_lax_A}
\frac{\partial \Phi}{\partial z} = A \Phi = \left( C+ \frac{A_0}{z}\right)\Phi,
\end{equation}
where
\begin{align*}
C
&= \left(  \begin{array}{ccc} 
&& \text{\LARGE $O$}
\\ 
-1 & & \\
h_{0} & -1 & 
 \end{array}  \right),
 \\
 A_0 &=  \left(  \begin{array}{ccccc} 
e_0 & h_1 & -1 && 
\\ 
 & e_1& h_2 & \ddots &
 \\
 &&  e_2& \ddots& -1
\\
&&&\ddots&h_{\ell-1}
\\
&&&& e_{\ell-1}
 \end{array}  \right)
 \end{align*}
 with $e_n=(d_{n+1}-d_n+n)/\ell$
 and
 $h_n =(w_{n}+w_{n-1})/\ell$.
 Note that the two singularities of (\ref{eq:p45_lax_A})
 are $z=0$ (regular singularity)
 and $z=\infty$ (irregular singularity of Poincar\'e rank $2/\ell$).
On the other hand, the deformation equation
(\ref{p45_lax_2}) becomes
\begin{equation}
\frac{\partial \Phi}{\partial x}= B \Phi = 
\left( \left(  \begin{array}{ccccc} 
\frac{x}{2}-w_0 & 1 &  &&
\\ 
 &   \frac{x}{2}-w_1 & 1    &&
 \\
 & &\ddots& \ddots&
\\
&&&& 1
\\
&&&&    \frac{x}{2}-w_{\ell-1} 
 \end{array}  \right)
 +z  
 \left(
 \begin{array}{ccccc}
 &     &            & &
\\
      & &        & &
\\
      &        & &  &     
\\
      &        &             && 
\\
 1    &        &             &           & 
 \end{array}
 \right)
 \right) \Phi.
\end{equation}
We can recover the Darboux chain (\ref{eq:darboux}),
as well as the case of $P_{\rm III}$-chain (Sect.~\ref{sect:p3}), 
from the integrability condition 
$\left[\frac{\partial}{\partial x}-B, \frac{\partial}{\partial z}-A\right]=0$.

\section{From KP hierarchy to Painlev\'e II equation}
\label{sect:p2}
In this section we briefly review the derivation of the second Painlev\'e equation from the two-periodic mKP hierarchy, i.e., mKdV hierarchy, through a similarity reduction; cf. \cite{as77}.

\subsection{Similarity reduction}

Let $\tau_n({\boldsymbol x})$ be a solution of the mKP hierarchy,
(\ref{eq:mkp-bil}) or (\ref{eq:mkp}),
satisfying the following conditions:
(i) homogeneity $E \tau_n=d_n \tau_n$ $(d_n \in {\mathbb C})$;
(ii) two-periodicity $\tau_{n+2}=\tau_n$;
(iii) specialization $x_n=0$ $(n \neq 1,3)$.
We see that this specialization is meant to 
eliminate all time evolutions except the first nontrivial two. 
No generality is lost by taking 
\begin{equation}  \label{eq:p2_special}
x_1=x, \quad
x_3= - \frac{4}{3}, \quad \text{and}  \quad  x_n=0  \quad
(n \neq 1,3).
\end{equation}
Set $\sigma_{n}(x)=\tau_n({\boldsymbol x})$.

\begin{prop} \label{prop:p2_bil}
A pair of functions $\sigma_n=\sigma_n(x)$
$(n \in {\mathbb Z}/2 {\mathbb Z})$
satisfies
\begin{align} \label{eq:bil_p2_1}
&{D_x}^2  \sigma_n \cdot \sigma_{n+1}=0,
\\
\label{eq:bil_p2_2}
&\left( {D_x}^3-x D_x+d_{n}-d_{n+1} \right) \sigma_n \cdot \sigma_{n+1}=0.
\end{align}
\end{prop}

\pf 
Notice by (\ref{eq:p2_special}) that 
\begin{equation} \label{eq:p2_flow}
\frac{\partial}{\partial x_1}= \frac{\rm d}{{\rm d}x} \quad
\text{and} \quad
4 \frac{\partial}{\partial x_3}= x \frac{\rm d}{{\rm d}x} -E.
\end{equation}
It is immediate to
verify (\ref{eq:bil_p2_1}) and (\ref{eq:bil_p2_2}) 
from (\ref{eq:mkp_hirota_1}) and  (\ref{eq:mkp_hirota_2}), respectively.
\qed
\\

Define functions $q=q(x)$ and  
$f_n=f_n(x)$ $(n\in{\mathbb Z}/2 {\mathbb Z})$ as
\begin{equation}
q=\frac{D_x \sigma_0 \cdot \sigma_1}{\sigma_0 \sigma_1},
\quad
f_n= \frac{x}{2} -\frac{{D_x}^2 \sigma_{n+1} \cdot \sigma_{n+1}}{{\sigma_{n+1}}^2}.
\end{equation}
Let
$\alpha_n=d_{n-1}-2d_n+d_{n+1}+1$.
Due to Prop.~\ref{prop:p2_bil} we then arrive at the 
\begin{thm} It holds that
\begin{equation}  \label{eq:p2}
\frac{{\rm d} q}{{\rm d} x} = \frac{f_0-f_1}{2}, \quad
\frac{{\rm d} f_0}{{\rm d} x}= 2 f_0 q + \frac{\alpha_0}{2}, \quad
\frac{{\rm d} f_1}{{\rm d} x} = -2 f_1 q + \frac{\alpha_1}{2}, 
\end{equation}
and $f_0+f_1-2q^2=x$.
\end{thm}

Let $p=f_0$. Then (\ref{eq:p2}) is converted to
the Hamiltonian system
\[
\frac{{\rm d} q}{{\rm d} x}=\frac{\partial H}{\partial p}
\quad \text{and}  \quad
\frac{{\rm d} p}{{\rm d} x}=-\frac{\partial H}{\partial q}
\quad \text{with} \quad 
H=\frac{p^2}{2}-\left(q^2+\frac{x}{2}\right)p - \frac{\alpha_0 q}{2},
\]
which is equivalent to 
 the second Painlev\'e equation $P_{\rm II}$:
 \[
 \frac{{\rm d}^2 q}{{\rm d}x^2}=2 q^3 + x q +\frac{\alpha_0-1}{2}.
 \]

\subsection{Affine Weyl group symmetry}
\label{subsect:p2_weyl}

First, it is clear that the permutation $\sigma_0 \leftrightarrow \sigma_1$ 
simultaneous with $d_0  \leftrightarrow d_1$ 
leaves the bilinear form (\ref{eq:bil_p2_1}) and (\ref{eq:bil_p2_2}) of $P_{\rm II}$ 
invariant.
As a consequence we obtain a birational symmetry
$\pi:(q,f_0,f_1,\alpha_0,\alpha_1)\mapsto (-q,f_1,f_0,\alpha_1,\alpha_0)$.

The next task is to realize the action of $r_1$, 
given in Sect~\ref{subsec:hom},
as a birational symmetry of $P_{\rm II}$. 
Applying the reduction conditions to the formulae (\ref{eq:weyl_tau_a_0}),
(\ref{eq:weyl_tau_a_2}), and (\ref{eq:weyl_tau_a_11}),
we thus observe that
\begin{align} \label{p2_sym_1}
D_x \hat{\sigma}_1 \cdot \sigma_1
&= {\sigma_0}^2,
\\  \label{p2_sym_2}
\alpha_1 \hat{\sigma}_1  \sigma_1
&= \left(-2 {D_x}^2  +x \right) \sigma_0 \cdot \sigma_0.
\end{align}
Since $f_1$ and $f_0+f_1-2 q^2(=x)$ are unchanged under $r_1$,
it is enough to compute only $r_1(q)$.
We have
\begin{align*}
r_1(q) &= \frac{D_x \sigma_0 \cdot   \hat{\sigma}_1}{\sigma_0  \hat{\sigma}_1}
= q + \frac{D_x \sigma_1 \cdot   \hat{\sigma}_1}{\sigma_1  \hat{\sigma}_1}
=q - \frac{ {\sigma_0}^2}{ \sigma_1  \hat{\sigma}_1}, \quad \text{using (\ref{p2_sym_1})}
\\
&=q- \frac{\alpha_1 {\sigma_0}^2}{ \left(-2 {D_x}^2  +x \right) 
\sigma_0 \cdot \sigma_0},
 \quad \text{using (\ref{p2_sym_2})}
\\
&=q -\frac{\alpha_1}{2f_1}.
\end{align*}
The action of $r_0$ is obtained as a composition 
$\pi r_1 \pi^{-1}$.
The result is summarized in the

\begin{thm}
$P_{\rm II}$ is invariant under birational transformations $r_i$ $(i=0,1)$ and $\pi$
defined by
\[
\begin{array}{|c||l|l|l|}
\hline
& \text{\rm Action on $\alpha_n$}& \text{\rm Action on $q$}& \text{\rm Action on $f_n$} 
\\  \hline
r_0 & (\alpha_0, \alpha_1)  \mapsto (-\alpha_0,   \alpha_1+2\alpha_0) & \displaystyle q \mapsto q+\frac{\alpha_0}{2f_0}
& 
 \displaystyle  f_1 \mapsto f_1 + \frac{2\alpha_0 q}{f_0}+ \frac{{\alpha_0}^2}{2{f_0}^2}
\\  \hline
r_1 & (\alpha_0, \alpha_1)  \mapsto (\alpha_0+2 \alpha_1,   -\alpha_1) & \displaystyle q \mapsto q-\frac{\alpha_1}{2f_1}
& 
 \displaystyle  f_0 \mapsto f_0 - \frac{2\alpha_1 q}{f_1}+ \frac{{\alpha_1}^2}{2{f_1}^2}
  \\  \hline
  \pi & \alpha_0 \leftrightarrow \alpha_{1} & q \mapsto -q
  & f_0 \leftrightarrow f_1
  \\
  \hline
\end{array}
\]
\end{thm}

It can be verified straightforwardly
that birational mappings 
$\langle r_0,r_1,\pi\rangle$ certainly fulfill 
the fundamental relations, 
${r_i}^2=\pi^2=1$ and $r_0 \pi=\pi r_1$,
of affine Weyl group $W(A_1^{(1)})$,
where $\pi$ corresponds to the diagram automorphism.

\subsection{Lax formalism}

The homogeneity 
$E \tau_n=d_n \tau_n$
implies that $\left( E- k {\partial }/{\partial k} \right)\psi_n({\boldsymbol x},k)=0$;
see Lemma~\ref{lemma:Epsi}.
Let 
$\phi_n(x,k)=\psi_n((x,0,-3/4,0,0,\ldots),k)$.
By virtue of the two-periodicity $\tau_{n+2}=\tau_n$
and (\ref{eq:p2_flow}),
the lemma below is immediate 
from 
(\ref{eq:psi_a}) and (\ref{eq:psi_c}).

\begin{lemma}
The wave functions $\phi_n=\phi_n(x,k)$ $(n \in {\mathbb Z}/2{\mathbb Z})$
satisfy the following linear equations{\rm:}
\begin{align} 
\label{eq:p2_lax_1}
k \frac{\partial}{\partial k} \phi_n 
&= (d_{n+1}-d_n) \phi_n + 2 k f_{n+1} \phi_{n+1} 
+ 4 (-1)^n k^2 q\phi_{n}-4 k^3 \phi_{n+1},
\\
 \label{eq:p2_lax_2}
\frac{\partial}{\partial x} \phi_n 
&= (-1)^{n+1} q \phi_n + k  \phi_{n+1}.
\end{align}
\end{lemma}

In fact,
the above linear system  
(\ref{eq:p2_lax_1}) and (\ref{eq:p2_lax_2}) 
coincides with the Lax pair found 
by Flaschka and Newell  \cite{fn80}.
We shall slightly modify it
by
taking $\Phi= {}^{\rm T}(\phi_0,k \phi_1)$ and $z=k^2$.
Thus we have
\begin{align} \label{eq:p2_lax_A}
\frac{\partial \Phi}{\partial z} 
&= A \Phi =
\left(
\frac{1}{z}
\left( 
\begin{array}{cc}
e_0 & f_1 \\
   0  & e_1 
\end{array}
\right)
+
\left( 
\begin{array}{cc}
2 q & -2 \\
f_0 & -2 q 
\end{array}
\right)
+
z
\left( 
\begin{array}{cc}
0  & 0  \\
-2 & 0
\end{array}
\right)
\right) \Phi,
\\
\label{eq:p2_lax_B}
\frac{\partial \Phi}{\partial x}
&= B \Phi = 
\left(
\left( 
\begin{array}{cc}
-q & 1 \\
 0 & q 
\end{array}
\right)
+
z
\left( 
\begin{array}{cc}
0  & 0  \\
1 & 0
\end{array}
\right)
\right) \Phi,
\end{align}
where $e_0=(d_1-d_0)/2$ and $e_1=(d_0-d_1+1)/2$.
The two singularities of (\ref{eq:p2_lax_A}) 
are $z=0$ (regular singularity)
and $z=\infty$ (irregular singularity of Poincar\'e rank $3/2$).
The latter (\ref{eq:p2_lax_B}) represents the monodromy preserving deformation of the former  (\ref{eq:p2_lax_A}).
Again one can derive $P_{\rm II}$, (\ref{eq:p2}), from the integrability condition
$\left[\frac{\partial}{\partial x}-B, \frac{\partial}{\partial z}-A\right]=0$
 of the system.

\section{UC hierarchy} \label{sect:uc}

In this section
we review the UC hierarchy, 
which is an extension of the KP hierarchy
proposed in \cite{tsu04}.
We also present some
functional equations arising from the UC hierarchy 
as preliminaries to the following two sections.

\subsection{Universal character and UC hierarchy}  \label{subsec:uc_uc}

For a pair of partitions $\lambda$ and $\mu$,
the {\it universal character} $S_{[\lambda,\mu]}=S_{[\lambda,\mu ]}({\boldsymbol x},{\boldsymbol y})$ is 
a polynomial in  
${\boldsymbol x}=(x_1,x_2,\ldots)$ and ${\boldsymbol y}= (y_1,y_2, \ldots)$
defined by the {\it twisted} Jacobi--Trudi formula
(see \cite{koi89}):
\begin{equation}  \label{eq:def-of-uc}
S_{[\lambda,\mu ]}({\boldsymbol x},{\boldsymbol y})
= \det 
\left(
  \begin{array}{ll}
 p_{\mu_{l'-i+1}  +i - j }({\boldsymbol y}),  &  1 \leq i \leq l'  \\
 p_{\lambda_{i-l'}-i+j}({\boldsymbol x}),     &   l'+ 1 \leq i \leq l+ l'   \\
  \end{array}
\right)_{1 \leq i,j \leq l+l'}
\end{equation}
with $l=l(\lambda)$ and $l'=l(\mu)$.
If we count the degree of variables as $\deg x_n =n$ and
$\deg y_n =-n$,
then  
$S_{[\lambda,\mu ]}$
is homogeneous of degree $|\lambda|-|\mu|$;
i.e.,
$E S_{[\lambda,\mu ]}=(|\lambda|-|\mu|) S_{[\lambda,\mu ]}$.
Here, we henceforth let $E$ denote the Euler operator
given as
\[
E= \sum_{n=1}^\infty 
\left(n x_n \frac{\partial}{\partial x_n} - n y_n \frac{\partial}{\partial y_n}
\right).
\]
Note that the Schur function 
$S_\lambda$ 
is a special case of the universal character: 
$S_\lambda({\boldsymbol x}) 
= \det \bigl( p_{\lambda_i-i+j}({\boldsymbol x}) \bigr)
= S_{[\lambda, \emptyset ]}({\boldsymbol x},{\boldsymbol y})$.

Let us introduce
 the {\it vertex operators}
\begin{align*}
X^\pm(k)&= \sum_{n \in {\mathbb Z}} X_n^\pm k^n =e^{\pm \xi({\boldsymbol x}-\widetilde{\partial}_{\boldsymbol y},k)} e^{\mp \xi(\widetilde{\partial}_{\boldsymbol x},k^{-1})}, \\
Y^\pm(k^{-1})&= \sum_{n \in {\mathbb Z}} Y_n^\pm k^{-n} =e^{\pm \xi({\boldsymbol y}-\widetilde{\partial}_{\boldsymbol x},k^{-1})} e^{\mp \xi(\widetilde{\partial}_{\boldsymbol y},k)}. 
\end{align*}
The operators $X_i^\pm$ ($i \in {\mathbb Z}$)
then satisfy the fermionic relations:
$X_i^\pm X_j^\pm+X_{j-1}^\pm X_{i+1}^\pm =0$ and 
$X_i^+X_j^- +X_{j+1}^-X_{i-1}^+ =\delta_{i+j,0}$.
The same relations hold also for  $Y_i^\pm$, of course.
Interestingly enough,
$X_i^\pm$  and $Y_j^\pm$ mutually commute.
By means of these operators,
the universal character admits 
the following expression
(cf. (\ref{eq:raising_schur})):
\begin{equation}   \label{eq:raising_uc}
S_{[\lambda,\mu]}({\boldsymbol x},{\boldsymbol y})=
X^+_{\lambda_1} \ldots X^+_{\lambda_l} Y^+_{\mu_1} \ldots Y^+_{\mu_{l'}} . 1.
\end{equation}
Now we are ready to formulate the UC hierarchy.

\begin{dfn} \rm
For an unknown function $\tau=\tau({\boldsymbol x},{\boldsymbol y})$,
the system of bilinear relations
\begin{equation} \label{eq:UCH}
\sum_{i+j=-1}X_i^{-} \tau \otimes X_j^{+} \tau =
\sum_{i+j=-1}Y_i^{-} \tau \otimes Y_j^{+} \tau =0
\end{equation}
is called the {\it UC hierarchy}.
\end{dfn}

If $\tau=\tau({\boldsymbol x},{\boldsymbol y})$
does not depend on ${\boldsymbol y}=(y_1,y_2,\ldots)$, then the latter equality of 
(\ref{eq:UCH}) trivially holds and the former is 
reduced to the bilinear expression  (\ref{eq:kp})
of the KP hierarchy.
From this aspect
the UC hierarchy is literally an extension of the KP hierarchy.
Moreover, as shown in \cite{tsu04}
the totality of solutions of  
(\ref{eq:UCH})
forms a direct-product of two Sato Grassmannians
and, in particular, the set of homogeneous polynomial solutions 
is equal to that of the universal characters.

It is obvious that  (\ref{eq:UCH}) can be rewritten into the form
\begin{subequations}  \label{subeq:uc-res}
\begin{align}
 &\frac{1}{2 \pi \sqrt{-1}} \oint 
 e^{\xi({\boldsymbol x}-{\boldsymbol x'},z)} {\rm d}z \,
\tau ({\boldsymbol x'}+[z^{-1}],{\boldsymbol y'}+[z]) 
\tau({\boldsymbol x}-[z^{-1}],{\boldsymbol y}-[z])
=0,  \label{eq:uc-res1}
\\
&\frac{1}{2 \pi \sqrt{-1}} \oint 
e^{\xi({\boldsymbol y}-{\boldsymbol y'},w)} {\rm d}w \,
\tau({\boldsymbol x'}+[w],{\boldsymbol y'}+[w^{-1}]) 
\tau({\boldsymbol x}-[w],{\boldsymbol y}-[w^{-1}])
=0   \label{eq:uc-res2}
\end{align}
\end{subequations}
for arbitrary 
${\boldsymbol x}$, ${\boldsymbol y}$, ${\boldsymbol x'}$, and ${\boldsymbol y'}$.
Let us try to write down a Hirota differential equation {\it naively} after the case of the KP hierarchy; cf. Sect.~\ref{subsect:pre_mkp}.
Namely, 
consider the Taylor expansion of (\ref{eq:uc-res1}) at 
$\{{\boldsymbol x'}={\boldsymbol x}, {\boldsymbol y'}={\boldsymbol y} \}$,
i.e., replace 
$({\boldsymbol x'},{\boldsymbol x}, {\boldsymbol y'},{\boldsymbol y} )$
with 
$({\boldsymbol x}+ {\boldsymbol u},{\boldsymbol x}- {\boldsymbol u}, {\boldsymbol y}+{\boldsymbol v},{\boldsymbol y}-{\boldsymbol v})$
and then expand with respect to $({\boldsymbol u},{\boldsymbol v})=(u_1,u_2,\ldots,v_1,v_2,\ldots)$.
Hence we obtain
\[
\sum_{i+j+k=-1} 
p_i(-2{\boldsymbol u})p_{-j}(\tilde{\partial}_{\boldsymbol u})
p_{k}(\tilde{\partial}_{\boldsymbol v})
\tau({\boldsymbol x}+{\boldsymbol u}, {\boldsymbol y}+{\boldsymbol v})\tau({\boldsymbol x}-{\boldsymbol u}, {\boldsymbol y}-{\boldsymbol v})=0.
\]
Taking the coefficient of $1={\boldsymbol u}^0{\boldsymbol v}^0$, for example,
leads to 
\[ \sum_{i=0}^\infty p_{i+1}(\tilde{D}_{\boldsymbol x}) p_i(\tilde{D}_{\boldsymbol y})
\tau \cdot \tau=0.
\]
Unfortunately, every differential equation with respect to ${\boldsymbol x}$ and $\boldsymbol y$ contained in 
the UC hierarchy is of infinite order as well as the above one.
This fact reflects that 
the integrand of (\ref{eq:uc-res1}) with 
${\boldsymbol x'}={\boldsymbol x}$ and ${\boldsymbol y'}={\boldsymbol y}$
may be singular not only at $z=0$ but also at $z=\infty$,
unlike the case of the KP hierarchy; cf. (\ref{eq:kp-res}).
But, however, it is possible to derive a closed functional equation from the 
UC hierarchy,
(\ref{eq:UCH}) or
(\ref{subeq:uc-res}),
 by a certain appropriate choice of parameters  
${\boldsymbol x}$, ${\boldsymbol y}$,
${\boldsymbol x'}$, and ${\boldsymbol y'}$;
see Sect.~\ref{subsec:pre-uc} below.

It is also known that
 if $\tau=\tau({\boldsymbol x},{\boldsymbol y})$ 
is a solution of (\ref{eq:UCH}), then so are $X^+(\alpha)\tau$ and $Y^+(\beta)\tau$
for arbitrary constants
$\alpha, \beta \in {\mathbb C}^\times$.
Now we are interested
 in the bilinear relations
 among the contiguous solutions connected by the vertex operators.
For the UC hierarchy, 
a counterpart of 
the modified KP hierarchy (cf. Definition.~\ref{dfn:mkp})
is introduced as follows:

\begin{dfn} \rm
Suppose 
$\tau_{m,n}=\tau_{m,n}({\boldsymbol x},{\boldsymbol y})$ 
to be a solution of the UC hierarchy.
Let 
\[\tau_{m+1,n}=X^+(\alpha_{m}) \tau_{m,n},  \quad
\tau_{m,n+1}=Y^+(\beta_n) \tau_{m,n}, \quad
\tau_{m+1,n+1}=X^+(\alpha_m) Y^+(\beta_n) \tau_{m,n}= Y^+(\beta_n)X^+(\alpha_m) \tau_{m,n},
\] 
for arbitrary constants $\alpha_m, \beta_n \in {\mathbb C}^\times$.
The whole set of functional equations satisfied by $\tau_{m,n}$'s
are called the {\it modified UC hierarchy} or, shortly, {\it mUC hierarchy}.
\end{dfn}

\begin{example}  \label{example:muc}
\rm
The mUC hierarchy includes 
the bilinear equations
\begin{align}  
& \sum_{i+j=-2} X_i^{-} \tau_{m,n} \otimes X_j^{+} \tau_{m+1,n}
=
\sum_{i+j=-1} Y_i^{-} \tau_{m,n} \otimes Y_j^{+} \tau_{m+1,n}
=0,
\label{eq:MUCH_1}
\\
&\tau_{m,n} \otimes \tau_{m+1,n+1}
- \sum_{i+j=0} X_i^{-} \tau_{m+1,n} \otimes X_j^{+} \tau_{m,n+1}
=
 \sum_{i+j=-2} Y_i^{-} \tau_{m+1,n} \otimes Y_j^{+} \tau_{m,n+1}
 =0.
 \label{eq:MUCH_2}
\end{align}
Here the former and the latter 
can be deduced from (\ref{eq:UCH})
by applying
$1 \otimes X^+(\alpha_m)$
and 
$X^+(\alpha_m) \otimes Y^+(\beta_n)$,
respectively.
\end{example}

Next we construct a sequence of 
homogeneous solutions of the UC hierarchy,
which is crucial to reach for Painlev\'e equations.
The argument will be proceeded 
along a parallel way with the case of the KP hierarchy; cf. Sect.~\ref{subsec:hom}.
Let us first introduce
partial differential operators 
$V_X(c)$ and $V_Y(c)$
equipped with a constant parameter $c\in {\mathbb C}$,
defined by
\[
V_X(c)= \int_\gamma X^+(k)k^{-c-1} {\rm d}k
\quad \text{and} \quad
V_Y(c)= \int_\gamma Y^+(k^{-1})k^{c-1} {\rm d}k,
\]
where the path $\gamma$ is again taken as well as  
Sect.~\ref{subsec:hom}.
Let $\tau$ be a solution of  the UC hierarchy,
(\ref{eq:UCH}).
Then one can verify that both
$V_X(c)\tau$ and $V_Y(c) \tau$ solve
 (\ref{eq:UCH}); 
 moreover, they satisfy the mUC hierarchy.

\begin{lemma}
It holds that
$[E, V_X(c)]=c V_X(c)$
and 
$[E, V_Y(c)]= -c V_Y(c)$.
\end{lemma}

This  lemma guarantees that $V_X$ and $V_Y$
preserve the homogeneity.
Suppose $\tau_{0,0}$ to be a solution of the UC hierarchy (\ref{eq:UCH})
satisfying $E \tau_{0,0}=d_{0,0}\tau_{0,0}$.
Define a sequence 
$\{ \tau_{0,0},\tau_{1,0},\tau_{0,1}, \tau_{1,1},\ldots \}$
recursively by
$\tau_{m+1,n}=V_X(c_m)\tau_{m,n}$ and 
$\tau_{m,n+1}=V_Y(c'_n)\tau_{m,n}$ 
for arbitrary 
$c_m,c'_n \in {\mathbb C}$ given.
Then $\tau_{m,n}$ becomes a solution of the mUC hierarchy 
still
satisfying the homogeneity
$E \tau_{m,n}=d_{m,n}\tau_{m,n}$,
where $d_{m+1,n}=d_{m,n}+c_m$ and 
$d_{m,n+1}=d_{m,n}-c'_n$;
thereby,
the {\it balancing} condition
\begin{equation} \label{eq:balancing}
d_{m,n}+d_{m+1,n+1}=d_{m,n+1}+d_{m+1,n}
\end{equation} 
is fulfilled. 
Though we do not enter into details,
we can find a mutually commuting pair of  Weyl group actions of type $A$ 
on 
 this  homogeneous $\tau$-sequence;
 see \cite{tsu09b}.

\begin{remark} \rm
\label{remark:balance}
Let us demonstrate one alternative to derive 
(\ref{eq:balancing}).
Suppose $\tau_{m,n} $ to be a homogeneous solution of
the mUC hierarchy.
Therefore we have
$\tau_{m,n}=c^{-d_{m,n}}\tau_{m,n}(c x_1,c^2x_2, \ldots, c^{-1}y_1,c^{-2}y_2,\ldots)$ for some $d_{m,n} \in {\mathbb C}$ and
arbitrary $c \in {\mathbb C}^\times$.
Substituting this into (\ref{eq:MUCH_2}),
we thus obtain (\ref{eq:balancing}) as a necessary condition.
\end{remark}

\begin{example}[A sequence of universal characters]
\rm
If we take $c=n$ to be an integer and $\gamma$ a positively oriented small circle around $k=0$, it follows that $V_X(n)=2 \pi \sqrt{-1}X_n^+$ and $V_Y(n)=2 \pi \sqrt{-1}Y_n^+$.
Hence, starting from a trivial solution $\tau \equiv 1$ of the UC hierarchy,
we can construct via (\ref{eq:raising_uc})
a homogeneous $\tau$-sequence 
in terms of the universal characters.
This type of polynomial 
solutions of the (modified) UC hierarchy
gives rise to rational or algebraic solutions of Painlev\'e equations
through the similarity reduction;
cf. \cite{tsu05, tsu06}.
\end{example}

\subsection{Preliminaries for Sects.~\ref{sect:p5}{}  and
\ref{sect:p6}: difference/differential equations inside mUC hierarchy}
\label{subsec:pre-uc}

Let $\delta_t$ and $\tilde{\delta}_t$ denote
the vector fields  
\begin{equation}
\delta_t=\sum_{n=1}^\infty \left(
 t^n \frac{\partial}{\partial x_n} - t^{-n} \frac{\partial}{\partial y_n}
 \right)
 \quad \text{and} \quad 
 \tilde{\delta}_t=\sum_{n=1}^\infty \left(
 n t^n \frac{\partial}{\partial x_n} + n  t^{-n} \frac{\partial}{\partial y_n}
 \right)
\end{equation}
hereafter.
For a vector field $\boldsymbol v$, let
$D_{\boldsymbol v}$ stand for its corresponding Hirota differential.
We first summarize some difference and/or differential equations 
arising from the mUC hierarchy.

\begin{lemma} \label{lemma:muc}
The following functional equations  hold{\rm:}
\begin{align}  
\nonumber
 &(t-s) \tau_{m,n} ({\boldsymbol x}-[t]-[s],{\boldsymbol y}-[t^{-1}]-[s^{-1}])  
\tau_{m+1,n+1} ({\boldsymbol x},{\boldsymbol y}) \\
\nonumber
& \qquad 
- t \tau_{m,n+1} ({\boldsymbol x}-[t], {\boldsymbol y}-[t^{-1}])  
\tau_{m+1,n} ({\boldsymbol x}-[s],{\boldsymbol y}-[s^{-1}])  \\
& \qquad
+s \tau_{m,n+1} ({\boldsymbol x}-[s],{\boldsymbol y}-[s^{-1}]) 
\tau_{m+1,n} ({\boldsymbol x}-[t],{\boldsymbol y}-[t^{-1}])=0,
 \label{eq:muc_a}
 \\
 \nonumber
 &\left( D_{\delta_t}-1 \right) \tau_{m,n+1}({\boldsymbol x},{\boldsymbol y}) 
 \cdot 
 \tau_{m+1,n}({\boldsymbol x},{\boldsymbol y}) 
 \\
  \label{eq:muc_b}
& \qquad + \tau_{m,n} ({\boldsymbol x} -[t] , {\boldsymbol y} -[t^{-1}] ) 
\tau_{m+1,n+1} ({\boldsymbol x} +[t],{\boldsymbol y} +[t^{-1}] )
=0,
 \\
 &  \nonumber
 \left(
D_{\delta_t} +\frac{t}{s-t}
\right)
\tau_{m,n} ({\boldsymbol x}-[s],  {\boldsymbol y}-[s^{-1}])  
\cdot 
\tau_{m+1,n} ({\boldsymbol x},{\boldsymbol y} ) 
\\
& \qquad
+\frac{t }{t-s}
\tau_{m,n} ({\boldsymbol x}-[t],{\boldsymbol y}-[t^{-1}])  
\tau_{m+1,n} ({\boldsymbol x}+[t]-[s], {\boldsymbol y}+[t^{-1}]-[s^{-1}]) 
=0.
 \label{eq:muc_c}
\end{align}
\end{lemma}

We refer the reader to \cite{tsu09b} where a more general class of 
functional relations of the UC hierarchy
including the above is established.
See also \cite[Appendix]{tsu09a} 
regarding (\ref{eq:muc_a}).
We mention that the first two equations and the last are originated from 
(\ref{eq:MUCH_2}) and  (\ref{eq:MUCH_1}), respectively.
Furthermore, if we differentiate 
(\ref{eq:muc_b}) with respect to $t$,
then we get 
\begin{align}  
 \nonumber
 & D_{\tilde{\delta}_t} \tau_{m,n+1}({\boldsymbol x},{\boldsymbol y}) 
 \cdot 
 \tau_{m+1,n}({\boldsymbol x},{\boldsymbol y}) 
 \\
& \qquad -D_{\delta_t} \tau_{m,n} ({\boldsymbol x} -[t] , {\boldsymbol y} -[t^{-1}] ) 
\cdot
\tau_{m+1,n+1} ({\boldsymbol x} +[t],{\boldsymbol y} +[t^{-1}] )
=0.
 \label{eq:muc_d}
\end{align}

Such bilinear relations as (\ref{eq:muc_a})--(\ref{eq:muc_d})
generate the auxiliary system of linear equations
associated with the UC hierarchy.
Let us introduce the {\it wave function}
$\psi_{m,n}=
\psi_{m,n}({\boldsymbol x},{\boldsymbol y},k)$
defined by
\begin{equation}  \label{eq:uc_wave}
\psi_{m,n}({\boldsymbol x},{\boldsymbol y},k)= \frac{ \tau_{m,n-1}({\boldsymbol x}-[k^{-1}],{\boldsymbol y}-[k])  }{ \tau_{m,n}({\boldsymbol x},{\boldsymbol y}) }e^{\xi({\boldsymbol x},k)}.
\end{equation}
Note that if $\tau_{m,n}$ does not depend on 
variables ${\boldsymbol y}=(y_1,y_2,\ldots)$ and $n$
then $\psi_{m,n}$ reduces to the wave function of the KP hierarchy;
cf. (\ref{eq:wavefn}).
We list below a few of the linear equations
that are relevant to us.

\begin{lemma}  \label{lemma:uc_lax}
The following linear functional equations hold{\rm:}
\begin{align}
&
\psi_{m,n}
=\frac{\overline{\tau}_{m,n+1}  \tau_{m+1,n-1} }{ \overline{\tau}_{m+1,n}   \tau_{m,n} }
\overline{\psi}_{m,n+1} -t k \overline{\psi}_{m+1,n},
 \label{eq:uc_lax_1}
\\
&
\delta_t \psi_{m,n}+ \frac{D_{\delta_t} \tau_{m,n} \cdot  \tau_{m+1,n-1}}{ \tau_{m,n} \tau_{m+1,n-1}} \psi_{m,n}
-t k 
\frac{ \underline{\tau}{}_{m,n-1}   \overline{\tau}_{m+1,n} }{\tau_{m,n} \tau_{m+1,n-1}}
\overline{\psi}_{m+1,n}
=0,  \label{eq:uc_lax_2}
\\
&
\tilde{\delta}_t \psi_{m,n}+ \frac{D_{\tilde{\delta}_t} \tau_{m,n} \cdot  \tau_{m+1,n-1}}{ \tau_{m,n} \tau_{m+1,n-1}} \psi_{m,n}
+t k 
\frac{ \left( D_{\delta_t}  -1 \right)
\underline{\tau}{}_{m,n-1} \cdot  \overline{\tau}_{m+1,n} }{\tau_{m,n} \tau_{m+1,n-1}}
\overline{\psi}_{m+1,n}
\nonumber
\\
& \quad
-t k 
\frac{ \underline{\tau}{}_{m,n-1}   \overline{\tau}_{m+1,n} }{\tau_{m,n} \tau_{m+1,n-1}}
\delta_t 
\overline{\psi}_{m+1,n}
=0.
\label{eq:uc_lax_3}
\end{align}
Here, for a function  $f=f({\boldsymbol x},{\boldsymbol y})$
we abbreviate 
$f({\boldsymbol x}+[t], {\boldsymbol y}+[t^{-1}])$ and
$f({\boldsymbol x}-[t], {\boldsymbol y}-[t^{-1}])$
 to 
 $\overline{f}$ and $\underline{f}$, respectively.
\end{lemma}

\pf If we put $s=1/k$ in (\ref{eq:muc_a}) and (\ref{eq:muc_c}), then we verify
(\ref{eq:uc_lax_1}) and (\ref{eq:uc_lax_2}) immediately.
Moreover, differentiating (\ref{eq:uc_lax_2}) with respect to $t$
yields (\ref{eq:uc_lax_3}).
\qed

\section{From UC hierarchy to Painlev\'e V chain}
\label{sect:p5}

This section concerns a similarity reduction of the periodic mUC hierarchy.
As a result we obtain the higher order Painlev\'e equation $P(A_{2\ell-1}^{(1)})$ or, equivalently,
the Darboux chain 
with period $2 \ell$
$(\ell \geq 2)$;
cf. Sect.~\ref{sect:p45}.

\subsection{Similarity reduction}
\label{subsec:p5_sim}

Let $\tau_{m,n}=\tau_{m,n}(\boldsymbol x, \boldsymbol y)$ 
be a solution of
the mUC hierarchy.
Assume the $(\ell_1, \ell_2)$-periodicity
$\tau_{m+\ell_1,n}=\tau_{m,n+\ell_2}=\tau_{m,n}$
and the homogeneity
\begin{equation} \label{eq:sim_p5}
E \tau_{m,n}(\boldsymbol x, \boldsymbol y)=d_{m,n} \tau_{m,n}(\boldsymbol x, \boldsymbol y) 
\quad (d_{m,n} \in {\mathbb C})
\end{equation}
where $E=\sum_{n=1}^\infty (n x_n \partial/\partial x_n - n y_n \partial/\partial y_n)$.
Note that the constants  
$d_{m,n}$ necessarily satisfy the balancing condition
$d_{m,n}+d_{m+1,n+1}=d_{m,n+1}+d_{m+1,n}$;
see Remark~\ref{remark:balance}. 
As seen later we can in fact restrict ourselves to the case
where
$\ell_1=\ell_2$,
without loss of generality;
nonetheless we shall consider a general case for a while.
Now, we introduce the functions $\sigma_{m,n}=\sigma_{m,n}(a,s)$ $(m,n \in {\mathbb Z}/\ell {\mathbb Z})$
defined by $\sigma_{m,n} (a,s)=\tau_{m,n}(\boldsymbol x, \boldsymbol y)$
under the substitution
\begin{equation}  \label{eq:change_p5}
x_n=s+\frac{a}{n} \quad 
\text{and} \quad
y_n=-s+\frac{a}{n}.
\end{equation}
Note that $s$ will play a role of the independent variable.
From now on,
we use abbreviated notations $\overline{\sigma}_{m,n}$ and $\underline{\sigma}{}_{m,n}$ which stand respectively for 
$\sigma_{m,n}(a+1,s)$ and $\sigma_{m,n}(a-1,s)$,
while $\sigma_{m,n}=\sigma_{m,n}(a,s)$.

\begin{prop}  \label{prop:bil_p5}
The functions $\sigma_{m,n}=\sigma_{m,n}(a,s)$ satisfy the system of bilinear equations 
 \begin{align} \label{eq:bil_p5_1}
  \left( D_s+1 \right)  \sigma_{m,n-1} 
\cdot 
 \sigma_{m-1,n} 
 &= \underline{\sigma}{}_{m-1,n-1}  \overline{\sigma}_{m,n},
 \\
\label{eq:bil_p5_2}
 s D_s  \underline{\sigma}{}_{m-1,n-1}  \cdot \overline{\sigma}_{m,n}
 &=
 \left( a D_s +  d_{m-1,n}-d_{m, n-1} \right) \sigma_{m,n-1} 
 \cdot 
 \sigma_{m-1,n}. 
 \end{align}
\end{prop}

\pf
We observe that
 \begin{equation}    \label{eq:flow_p5}
\frac{ \rm d}{ {\rm d} s}
= \sum_{n=1}^\infty 
\left(
\frac{ {\rm d} x_n}{ {\rm d} s} \frac{\partial}{\partial x_n}+\frac{ {\rm d} y_n}{ {\rm d} s} \frac{\partial}{\partial y_n}
\right)
=  \sum_{n=1}^\infty 
\left( \frac{\partial}{\partial x_n}- \frac{\partial}{\partial y_n}
\right)
= \delta_1,
\end{equation}
and also that
 \begin{equation} \label{eq:E_p5}
 E=\sum_{n=1}^\infty  
 \left( n   x_n \frac{\partial}{\partial x_n} - n   y_n \frac{\partial}{\partial y_n}
 \right)
 =\sum_{n=1}^\infty  
 \left( (n s+a) \frac{\partial}{\partial x_n} +(n s-a) \frac{\partial}{\partial y_n}
 \right) 
 =s \tilde{\delta}_1+a  \delta_1.
 \end{equation}
 Via the homogeneity (\ref{eq:sim_p5}), 
 we rapidly verify 
  (\ref{eq:bil_p5_1}) and (\ref{eq:bil_p5_2}) from (\ref{eq:muc_b}) and (\ref{eq:muc_d}), 
 respectively.
 \qed
 \\

The next task is to derive a system of nonlinear ordinary differential equations from
the bilinear equations given in Prop.~\ref{prop:bil_p5}. 
 We introduce a set of dependent variables
 $f_{m,n}=f_{m,n}(a,s)$ and
 $g_{m,n}=g_{m,n}(a,s)$
defined by
\begin{equation}   \label{eq:def_fg_p5}
 f_{m,n}= \frac{ \overline{\sigma}_{m,n-1} \sigma_{m-1,n-1} }{  \overline{\sigma}_{m-1,n} \sigma_{m,n-2} },
 \quad
 g_{m,n} = \frac{ \underline{\sigma}{}_{m-1,n-1} \overline{\sigma}_{m,n} }{  \sigma_{m-1,n} \sigma_{m,n-1} }.
\end{equation}
Let $\ell$ denote the least common multiple of $\ell_1$ and $\ell_2$.
One can then find 
the conservation laws:
\begin{equation}
\label{eq:p5_cons}
\prod_{i=1}^\ell f_{m+i,n-i}=1
\quad \text{and} \quad
\sum_{i=1}^\ell g_{m+i,n-i}=\ell, 
 \end{equation}
 where the former is trivial
and the latter a consequence of (\ref{eq:bil_p5_1}).
Additionally, we consider another set of variables
 $u_{m,n}=u_{m,n}(a,s)$ and
 $v_{m,n}=v_{m,n}(a,s)$
defined by
 \begin{equation} \label{eq:def_uv_p5}
 u_{m,n}=\frac{D_s \sigma_{m,n} \cdot \overline{\sigma}_{m+1,n} }{ \sigma_{m,n} \overline{\sigma}_{m+1,n} },
 \quad 
 v_{m,n}=- \frac{D_s \sigma_{m,n} \cdot \overline{\sigma}_{m,n+1} }{ \sigma_{m,n} \overline{\sigma}_{m,n+1} }.
 \end{equation}
Note that
 $u_{m,n}$ and  $v_{m,n}$ 
 are identical each other 
 if we interchange the roles of  suffixes $m$ and $n$
 and replace $s$ with $-s$.

We know
from (\ref{eq:bil_p5_1}) that
\begin{equation}  \label{eq:g&gup}
g_{m,n}=1-u_{m-1,n}-v_{m,n-1} \quad \text{and} \quad
\overline{g}_{m,n}=1-u_{m-1,n-1}-v_{m-1,n-1}.
\end{equation}
Although the above linear equations themselves can not be solved for 
$u_{m,n}$ and $v_{m,n}$,
one can express $u_{m,n}$ and $v_{m,n}$ 
in terms of $g_{m,n}$ and $\overline{g}_{m,n}$ conversely.
To be specific, we state the

\begin{lemma}
Variables $u_{m,n}$ and $v_{m,n}$ are quadratic  polynomials 
in $g_{m,n}$ and $\overline{g}_{m,n}$.
\end{lemma}

\pf
First,
the logarithmic derivative of $g_{m,n}$ reads
\begin{align}  \nonumber
 \frac{1}{g_{m,n}}\frac{{\rm d} g_{m,n}}{{\rm d}s}
&=
-\frac{D_s \sigma_{m,n-1} \cdot  \overline{\sigma}_{m,n}}{ \sigma_{m,n-1} \overline{\sigma}_{m,n}}
+
\frac{D_s  \underline{\sigma}{}_{m-1,n-1} \cdot  \sigma_{m-1,n}}{ \underline{\sigma}{}_{m-1,n-1} \sigma_{m-1,n}}
\\
&=v_{m,n-1}-\underline{v}{}_{m-1,n-1}.
\label{eq:p5_g_ld}
\end{align}
Observe that
 \begin{align}  \nonumber
 u_{m-1,n}-\underline{v}{}_{m-1,n-1}
 &= \frac{ D_s\underline{\sigma}{}_{m-1,n-1}  \cdot \overline{\sigma}_{m,n} }{\underline{\sigma}{}_{m-1,n-1}  \overline{\sigma}_{m,n}}
 \\  \nonumber 
 &=\frac{\left( a D_s +  d_{m-1,n}-d_{m, n-1} \right) \sigma_{m,n-1} 
 \cdot 
 \sigma_{m-1,n}}{ s \underline{\sigma}{}_{m-1,n-1}  \overline{\sigma}_{m,n}},
 \quad \text{using (\ref{eq:bil_p5_2})}
 \\
 &= \frac{a (g_{m,n}-1)+ d_{m-1,n}-d_{m, n-1}  }{s g_{m,n}},
 \quad \text{using (\ref{eq:bil_p5_1}) and  (\ref{eq:def_fg_p5})}.
 \label{eq:p5_v_under}
 \end{align}
Thereby, with (\ref{eq:g&gup}) in mind, each $u_{m,n}(a \pm 1)$ and $v_{m,n}(a \pm 1)$ can be written as a rational function in $u_{m,n}(a)$ and 
$v_{m,n}(a)$.
Now, (\ref{eq:p5_g_ld}) combined with  (\ref{eq:p5_v_under}) yields
\begin{align}
 \frac{{\rm d} g_{m,n}}{{\rm d}s}
&=
(v_{m,n-1}-u_{m-1,n})g_{m,n}
+ \frac{a (g_{m,n}-1)+ d_{m-1,n}-d_{m, n-1}}{s}
\nonumber
\\
&=(2v_{m,n-1} +g_{m,n}-1)g_{m,n}
+ \frac{a (g_{m,n}-1)+ d_{m-1,n}-d_{m, n-1}}{s},
\quad \text{using (\ref{eq:g&gup}).}
\label{eq:p5_g'}
 \end{align}
Due to the conservation (\ref{eq:p5_cons}) and the periodicity,
we hence obtain the equality 
\begin{equation} \label{eq:p5_sumofg'}
0 = \sum_{j=1}^{\ell} \frac{{\rm d} g_{m+j,n-j}}{{\rm d}s}
= \sum_{j=1}^{\ell}(2 v_{m+j,n-j-1}+g_{m+j,n-j}-1)g_{m+j,n-j}.
\end{equation}
On the other hand, it follows again from (\ref{eq:g&gup})
that
\[
v_{m+j,n-j-1}=v_{m,n-1}
+\sum_{i=1}^j 
\left( \overline{g}_{m+i,n-i+1} -g_{m+i,n-i}
\right) 
\]
for $1 \leq j \leq \ell$.
Substituting this into (\ref{eq:p5_sumofg'}) we obtain a linear equation for $v_{m, n-1}$;
thus, we have
\[
v_{m,n-1}=\frac{\ell+1}{2} 
- \frac{1}{\ell} \sum_{1 \leq i \leq j \leq \ell}
\overline{g}_{m+i,n-i+1}g_{m+j,n-j}.
\]
By (\ref{eq:g&gup}), we can modify this into the form
\begin{equation}  \label{eq:p5_u_g}
u_{m,n}=\frac{1-\ell}{2}+
\frac{1}{\ell} \sum_{1 \leq i \leq j \leq \ell-1}
\overline{g}_{m+i+1,n-i+1}g_{m+j+1,n-j}.
\end{equation}  
Interchanging the roles of suffixes $m$ and $n$ in the above formula,
we see also that
\begin{align}
v_{m,n}&=\frac{1-\ell}{2}+
\frac{1}{\ell} \sum_{1 \leq i \leq j \leq \ell-1}
\overline{g}_{m-i+1,n+i+1}g_{m-j,n+j+1}
\nonumber
\\
&=\frac{1-\ell}{2}+
\frac{1}{\ell} \sum_{1 \leq i \leq j \leq \ell-1}
g_{m+i,n-i+1} \overline{g}_{m+j+1,n-j+1},
\quad \text{using the periodicity.}
 \label{eq:p5_v_g}
\end{align}
The lemma is proven. \qed
\\

In fact, one can adopt 
$(u_{m,n},v_{m,n})$ or, alternatively, $(g_{m,n}, \overline{g}_{m,n})$
as essential dependent variables
because: 
if once $(u_{m,n},v_{m,n})$ or $(g_{m,n},\overline{g}_{m,n})$
are given then 
$f_{m,n}$ is determined from the first order linear equation
\begin{align*} 
\frac{1}{f_{m,n}}\frac{{\rm d} f_{m,n}}{{\rm d}s}
&= 
\frac{D_s  \sigma_{m-1,n-1} \cdot  \overline{\sigma}_{m-1,n}}{ \sigma_{m-1,n-1} \overline{\sigma}_{m-1,n}}
-\frac{D_s  \sigma_{m,n-2} \cdot  \overline{\sigma}_{m,n-1}}{ \sigma_{m,n-2}  \overline{\sigma}_{m,n-1}} \\
&=v_{m,n-2}-v_{m-1,n-1} \\
&= \overline{g}_{m,n}-g_{m,n-1}
\end{align*}
by quadrature.

Starting from the logarithmic derivative of $\overline{g}_{m,n}$:
\[
\frac{1}{ \overline{g}_{m,n}} \frac{{\rm d}  \overline{g}_{m,n} }{{\rm d} s}
=
u_{m-1,n-1}- \overline{u}_{m-1,n},
\] 
we arrive at
\begin{equation}  \label{eq:p5_gup'}
\frac{{\rm d}  \overline{g}_{m,n} }{{\rm d} s}
=
(u_{m-1,n-1}-v_{m-1,n-1})\overline{g}_{m,n}
- \frac{(a+1)(\overline{g}_{m,n}-1)+ d_{m-1,n}-d_{m, n-1}}{s}
\end{equation}
via (\ref{eq:p5_v_under})
in the same manner as before; cf. (\ref{eq:p5_g'}).
Now the necessary differential equations
 have been all present.
Since (\ref{eq:p5_u_g}) and (\ref{eq:p5_v_g}) hold,
one can eliminate $u_{m,n}$ and $v_{m,n}$ 
from (\ref{eq:p5_g'}) and (\ref{eq:p5_gup'})
and
thus obtain the differential equations for unknowns 
$(g_{m,n}, \overline{g}_{m,n})$.
Alternatively, 
eliminating
$g_{m,n}$ and $\overline{g}_{m,n}$ 
from (\ref{eq:p5_g'}) and (\ref{eq:p5_gup'})
by  (\ref{eq:g&gup})
yields the system of differential equations for $(u_{m,n},v_{m,n})$:
\begin{subequations}  \label{subeq:dar_uc}
\begin{align} 
\frac{\rm d}{{\rm d}s} \left(  u_{m-1,n+1} + v_{m,n}  \right)
 &=
{v_{m,n}}^2- {u_{m-1,n+1}}^2 -
\left(  1 -\frac{a}{s} \right)v_{m,n} +\left(  1 +\frac{a}{s} \right)u_{m-1,n+1}
\nonumber
\\
& \quad + \frac{d_{m,n}-d_{m-1,n+1}}{s},
\label{eq:dar_uc_1}
\\
  \frac{\rm d}{{\rm d}s} \left(  v_{m,n} +u_{m,n} \right)
 &= {u_{m,n}}^2- {v_{m,n}}^2
 -\left(  1 +\frac{a+1}{s} \right)u_{m,n}+\left(  1 -\frac{a+1}{s} \right)v_{m,n}
 \nonumber \\
& \quad  + \frac{d_{m,n+1}-d_{m+1,n}}{s}.
\label{eq:dar_uc_2}
\end{align}
\end{subequations}
Interestingly enough,
for each $(m,n)$ given
 (\ref{subeq:dar_uc})
 is closed with respect to the $2 \ell$-tuple of variables 
 $(u_{m+i,n-i}, v_{m+i,n-i})$ where 
 $i \in {\mathbb Z}/\ell {\mathbb Z}$.
 For instance, we fix
 $m=n=0$ hereafter.
 To improve the representation  
 we consider further change of variables 
\begin{equation} 
 x=  - \sqrt{-2 s}, \quad
 w_{2i} = \sqrt{-2 s}  u_{i,-i}+ \frac{s+a+\frac{1}{2}}{\sqrt{-2 s}},
 \quad
 w_{2i-1} = \sqrt{-2 s}v_{i,-i} + \frac{s-a-\frac{1}{2}}{  \sqrt{-2 s}}.
\end{equation}
Then we are led to the

 \begin{thm}
 The $2 \ell$-tuple of functions $w_n=w_n(x)$ 
 $(n \in {\mathbb Z}/2\ell {\mathbb Z})$
 satisfies
  \begin{equation}
\frac{{\rm d}}{{\rm d} x}(w_n +w_{n-1})={w_n}^2 -{w_{n-1}}^2 +\alpha_n,
\end{equation}
where $\alpha_{2i}=2( d_{i+1,-i}-d_{i,-i+1}+a+1)$ and
$\alpha_{2i-1}=2(d_{i-1,-i+1} -  d_{i,-i} - a)$.
\end{thm}
 
This is exactly the Darboux chain with period $2\ell$ (see Sect.~\ref{sect:p45})
and thus it is equivalent to the fifth Painlev\'e equation $P_{\rm V}$ when $\ell=2$.

\begin{remark} \rm   \label{remark:period}
No generality is lost by assuming  
$\ell_1=\ell_2$
 because the system with general $(\ell_1, \ell_2)$-periodicity 
 is obviously a special case of that with 
 $(\ell, \ell)$-periodicity,
 provided $\ell$ is the least common multiple of $\ell_1$ and $\ell_2$.
\end{remark}

\subsection{Lax formalism}  
\label{subsec:lax_p5}
In view of Remark~\ref{remark:period},
we let $\ell_1=\ell_2=\ell$ from now on.
Our goal here is to derive an auxiliary linear problem
for $P(A_{2 \ell-1}^{(1)})$ from the UC hierarchy via the similarity reduction.
Recall the definition (\ref{eq:uc_wave}) of the wave function:
\[
\psi_{m,n}({\boldsymbol x},{\boldsymbol y},k)=
\frac{\tau_{m,n-1}({\boldsymbol x}-[k^{-1}],{\boldsymbol y}-[k])}{\tau_{m,n}({\boldsymbol x},{\boldsymbol y})} e^{\xi({\boldsymbol x},k)}.
 \]
As analogous to Lemma~\ref{lemma:Epsi}, 
we have the

\begin{lemma}  \label{lemma:p5_eular}
If $\tau_{m,n}=\tau_{m,n}({\boldsymbol x},{\boldsymbol y})$ obeys the homogeneity
$E \tau_{m,n}=d_{m,n} \tau_{m,n}$.
Then it holds that
\begin{equation}  \label{eq:p5_eular}
\left(E- k\frac{\partial}{\partial k}\right)
 \psi_{m,n}=(d_{m,n-1}-d_{m,n})\psi_{m,n}.
\end{equation}
\end{lemma}

Set
$\phi_{m,n}(a,s,k)=\psi_{m,n}({\boldsymbol x},{\boldsymbol y},k)$
under the substitution (\ref{eq:change_p5}).

\begin{lemma} 
The wave functions 
$\phi_{m,n}=\phi_{m,n}(a)=\phi_{m,n}(a,s,k)$
satisfy the following linear equations{\rm:}
\begin{align} 
\phi_{m,n}(a)&= \frac{1}{f_{m+1,n+1}} \phi_{m,n+1}(a+1)-k \phi_{m+1,n}(a+1),
 \label{eq:p5_lax_1}
\\
\frac{\partial }{\partial s}\phi_{m,n}(a)
&= (g_{m+1,n}-1) \phi_{m,n}(a)+k g_{m+1,n} \phi_{m+1,n}(a+1),
 \label{eq:p5_lax_2}
\\
k \frac{\partial }{\partial k} 
\phi_{m,n}(a)
&=
(d_{m+1,n-1}-d_{m,n-1}) \phi_{m,n}(a) 
\nonumber \\
& \quad  
+ k \left(s g_{m+1,n} +a+ d_{m+1,n-1}-d_{m,n} \right) \phi_{m+1,n}(a+1)
\nonumber \\ 
& \quad
+k s g_{m+1,n} 
\frac{\partial }{\partial s} \phi_{m+1,n}(a+1).
 \label{eq:p5_lax_3}
\end{align}
\end{lemma}

\pf
It is immediate to deduce (\ref{eq:p5_lax_1}) and (\ref{eq:p5_lax_2}) 
respectively from (\ref{eq:uc_lax_1}) and (\ref{eq:uc_lax_2}) with $t=1$
through (\ref{eq:flow_p5}).
We can verify  (\ref{eq:p5_lax_3}) from (\ref{eq:uc_lax_3})
by taking into account (\ref{eq:E_p5}) and Lemma~\ref{lemma:p5_eular}.
\qed
\\

Notice the equality
\begin{equation}  \label{eq:p5_f_shift}
f_{m,n}= \frac{  \overline{f}_{m+1,n} \overline{g}_{m,n} }{ \overline{g}_{m+1,n-1} },
\end{equation}
which follows clearly from (\ref{eq:def_fg_p5}).
The periodicity $\phi_{m+\ell,n}=\phi_{m,n+\ell}=\phi_{m,n}$
allows us to solve the linear equations (\ref{eq:p5_lax_1}) for $\overline{\phi}_{m,n}=\phi_{m,n}(a+1)$;
we thus find that
\begin{equation} \label{eq:p5_phi_a+1}
\overline{\phi}_{m,n}= 
\frac{1}{1-k^\ell}
\sum_{j=1}^{\ell} k^{j-1}  
\left(
 \prod_{i=1}^j  f_{m+i,n-i+1}
\right)
\phi_{m+j-1,n-j}.
\end{equation}
Moreover, applying the above formula twice shows that
\begin{align}
\overline{\overline{\phi}}_{m,n}
&=\phi_{m,n}(a+2)
\nonumber
\\
&=\frac{1}{(1-k^\ell)^2}
\sum_{j=1}^{\ell} 
\sum_{j'=1}^{\ell} k^{j+j'-2} 
\left(
 \prod_{i=1}^j  \overline{f}_{m+i,n-i+1}
\right)
\left(
 \prod_{i'=1}^{j'}  f_{m+j+i'-1,n-j-i'+1}
\right)
\phi_{m+j+j'-2,n-j-j'}
\nonumber
\\
&=
\frac{f_{m,n}}{(1-k^\ell)^2 \overline{g}_{m,n} }
\sum_{j=1}^{\ell} 
\sum_{j'=1}^{\ell} k^{j+j'-2} 
\overline{g}_{m+j,n-j}
\left(
 \prod_{i=1}^{j+j'-1}  f_{m+i,n-i}
\right)
\phi_{m+j+j'-2,n-j-j'},
\quad \text{using (\ref{eq:p5_f_shift})}
\nonumber
\\
&= \frac{f_{m,n}}{(1-k^\ell)^2 \overline{g}_{m,n} }
\sum_{r=1}^{2\ell-1} k^{r-1} C_{r} \phi_{m+r-1,n-r-1},
\quad \text{taking  $r=j+j'-1$},
 \label{eq:p5_phi_a+2}
\end{align}
with
\[
C_r=
\left\{
  \begin{array}{ll}
\sum_{j=1}^{r} \overline{g}_{m+j,n-j}\prod_{i=1}^{r}  f_{m+i,n-i}
& \text{(if $1 \leq r \leq \ell$)}
\\
\sum_{j=r-\ell+1}^{\ell} \overline{g}_{m+j,n-j}
\prod_{i=1}^{r}  f_{m+i,n-i} 
&  \text{(if $\ell \leq r \leq 2 \ell-1$)}.
  \end{array}
\right.
\]
Here we have used the conservation 
(\ref{eq:p5_cons}).

Firstly, 
we shall take interest in a differential equation with respect to the spectral variable $k$.
With the aid of 
(\ref{eq:p5_lax_1}),   (\ref{eq:p5_lax_2}), 
and  (\ref{eq:p5_f_shift}),
one can rewrite (\ref{eq:p5_lax_3}) into 
\begin{align*}
k \frac{\partial }{\partial k} 
\phi_{m,n}
&= (d_{m+1,n-1}-d_{m,n-1}) \phi_{m,n} 
+ k \left(a+ d_{m+1,n-1}-d_{m,n} \right) \overline{\phi}_{m+1,n}
\\
&  \quad 
+k  s  \frac{g_{m+1,n} \overline{g}_{m+1,n+1}}{ f_{m+1,n+1}}
\overline{\overline{\phi}}_{m+1,n+1}.
\end{align*}
Hence,
eliminating 
$\overline{\phi}_{m+1,n}$
and 
$\overline{\overline{\phi}}_{m+1,n+1}$ 
by 
(\ref{eq:p5_phi_a+1}) and  (\ref{eq:p5_phi_a+2})
yields the equation 
\begin{align}
k \frac{\partial }{\partial k} 
\phi_{m,n}
&= (d_{m+1,n-1}-d_{m,n-1}) \phi_{m,n}
\nonumber
\\ 
& \quad +
\frac{a+ d_{m+1,n-1}-d_{m,n} }{1-k^\ell}
\sum_{j=1}^{\ell} k^j  
\left(
 \prod_{i=1}^j  f_{m+i+1,n-i+1}
\right)
\phi_{m+j,n-j}
\nonumber
\\
& \quad
+ \frac{ s g_{m+1,n}}{(1-k^\ell)^2}
\sum_{j=1}^{\ell} k^{j} 
\sum_{i'=1}^{j} \overline{g}_{m+i' +1,n-i' +1} 
\left(\prod_{i=1}^{j}  f_{m+i+1,n-i+1} \right)
\phi_{m+j,n-j}
\nonumber
\\
& \quad + 
 \frac{s g_{m+1,n}}{(1-k^\ell)^2}
\sum_{j=\ell+1 }^{2 \ell-1} k^{j} 
\sum_{i'=j-\ell+1}^{\ell} \overline{g}_{m+i' +1,n-i' +1} 
\left(\prod_{i=1}^{j}  f_{m+i+1,n-i+1} \right)
\phi_{m+j,n-j}.
\label{eq:p5_lax_A_pre}
\end{align}
This has 
two regular singularities 
at $k=0,\infty$ and $\ell$ irregular singularities at
$k=\exp \left(2 \pi \sqrt{-1} n/\ell\right)$ $(n \in {\mathbb Z}/\ell{\mathbb Z})$. 
Notice that for each $(m,n)$ fixed 
(\ref{eq:p5_lax_A_pre})
is closed with respect to
$\phi_{m+i,n-i}$ ($i \in {\mathbb Z}/\ell{\mathbb Z}$) as before.
Let us consider
\[\Phi={}^{\rm T}\left( \phi_{0,0}, k \phi_{1,-1},   k^2 \phi_{2,-2}, \ldots, k^{\ell-1}  \phi_{\ell-1, -\ell +1} \right)
\quad
\text{and} \quad
z=k^\ell.
\]
Then (\ref{eq:p5_lax_A_pre}) takes the form 
\begin{equation}  \label{eq:p5_lax_A}
\frac{\partial \Phi}{\partial z}=A \Phi
=
\left(\frac{A_0}{z} + \frac{A_{1,0}}{z-1}+\frac{A_{1,-1}}{(z-1)^2} \right) \Phi,
\end{equation}  
where the $\ell \times \ell$ matrices $A_0$, $A_{1,0}$, and $A_{1,-1}$
are given by
\begin{align*}
A_0 &= 
  \left(
  \begin{array}{cccc}
  e_0&\zeta_{0,1}+\eta_{0,1}& \cdots & \zeta_{0,\ell-1}+\eta_{0,\ell-1} \\
        & e_1 & \ddots&\vdots   \\
        &        & \ddots    &\zeta_{\ell-2,\ell-1}+\eta_{\ell-2,\ell-1} \\
        &&&e_{\ell-1}
  \end{array}
  \right),
  \\
  A_{1,0} &=
   -\left( \zeta_{i,j} + \eta_{i,j}  \right)_{0 \leq i,j \leq \ell-1}
  +
    \left(
  \begin{array}{cccc}
   \xi_{0,0} && &\text{\LARGE $O$}  \\
   \xi_{1,0} & \xi_{1,1} & &   \\
  \vdots  &  \ddots  &  \ddots  &  \\
    \xi_{\ell-1,0} &\xi_{\ell-1,1} & \cdots & \xi_{\ell-1,\ell-1}
  \end{array}
  \right),
  \\
  A_{1,-1}&= \left( \xi_{i,j} \right)_{0 \leq i,j \leq \ell-1}
\end{align*}
with 
\begin{align*}
e_n &= \frac{d_{n+1,-n-1}-d_{n,-n-1}+n}{\ell},
\\
\zeta_{n,n+j}&= \frac{a+d_{n+1,-n-1}-d_{n,-n}}{\ell} 
\prod_{i=1}^j f_{n+i+1,-n-i+1},
\\
\eta_{n,n+j}&= \frac{s g_{n+1,-n}}{\ell}
\sum_{i'=1}^j \overline{g}_{n+i' +1,-n-i'+1}
\prod_{i=1}^j f_{n+i+1,-n-i+1},
\\
\xi_{n,n+j} &= s g_{n+1,-n}\prod_{i=1}^j f_{n+i+1,-n-i+1}
\end{align*}
for $0 \leq n \leq \ell-1$ and  $1 \leq j \leq \ell$.
Note that
the suffix of each variable should be regarded as an element of 
${\mathbb Z}/\ell {\mathbb Z}$.
We observe that 
(\ref{eq:p5_lax_A}) has an irregular singularity (of Poincar\'e rank one)
at $z=1$, and 
has two regular singularities $z=0$ and $z=\infty$ whose exponents read respectively
$(e_0,e_1,\ldots,e_{\ell-1})$ and $(\rho_0,\rho_1,\ldots, \rho_{\ell-1})$
with $\rho_n = (a+d_{n,-n-1}-d_{n,-n}-n)/\ell$.
From a standpoint of monodromy preserving deformations,
(\ref{eq:p5_lax_A}) is the original linear differential equation 
that will be deformed with keeping
its monodromy matrices and Stokes multipliers invariant.

Secondly, we derive from (\ref{eq:p5_lax_2}) 
the deformation equation
\begin{equation}  \label{eq:p5_lax_B}
\frac{\partial}{\partial s} \Phi
=  B \Phi,
\end{equation}
where
\begin{align*}
B
&={\rm diag} \left( g_{i+1,-i} -1 \right)_{0 \leq i \leq \ell-1}
\\
& \quad + \frac{1}{s(1-z)} 
 \left(  
   \begin{array}{cccc}
   0 &   \xi_{0,1}      & \cdots  & \xi_{0,\ell-1}
   \\
        & 0 &  \ddots       & \vdots
   \\     
          &        &\ddots& \xi_{\ell-2,\ell-1}
   \\
          &        &          & 0    
   \end{array}
 \right)
 +
  \frac{z}{s(1-z)} 
 \left(  
   \begin{array}{cccc}
   \xi_{0,0} &      &  &\text{\LARGE $O$} 
   \\
  \xi_{1,0}        & \xi_{1,1} &        & 
   \\     
    \vdots      &   \ddots     &\ddots& 
   \\
 \xi_{\ell-1,0}  &  \xi_{\ell-1,1}        &   \cdots      &    \xi_{\ell-1,\ell-1}    
   \end{array}
 \right).
\end{align*}

Finally, all the differential equations appearing in Sect.~\ref{subsec:p5_sim}
emerge as the integrability condition of 
the system
(\ref{eq:p5_lax_A}) and (\ref{eq:p5_lax_B}).

\begin{remark} \rm
If $\ell=2$, this system is equivalent to the Lax pair for $P_{\rm V}$
given in \cite{jm81}.
\end{remark}

\section{From UC hierarchy to Painlev\'e VI chain}
\label{sect:p6}

In this section, we 
consider a certain similarity reduction of the $(\ell,\ell)$-periodic mUC hierarchy.
The resulting system of nonlinear ordinary differential equations 
describes
a monodromy preserving deformation of 
an $\ell \times \ell$ Fuchsian system 
with four regular singularities 
and, thus, it is regarded as a generalization of the sixth Painlev\'e equation.

\subsection{Similarity reduction}
Fix an integer $\ell \geq 2$. 
Let us require 
a solution $\tau_{m,n}=\tau_{m,n}({\boldsymbol x},{\boldsymbol y})$ 
of the mUC hierarchy
to
fulfill the following conditions:
(i) homogeneity  $E \tau_{m,n}=d_{m,n} \tau_{m,n}$ $(d_{m,n} \in {\mathbb C})$;
(ii) $(\ell,\ell)$-periodicity $\tau_{m+\ell,n}=\tau_{m,n+\ell}=\tau_{m,n}$;
(iii) specialization of variables as
\begin{equation}  \label{eq:special_p6}
x_n=\frac{a+b t^n}{n} \quad \text{and} \quad y_n=\frac{a+b t^{-n}}{n}. 
\end{equation}
Concerning (i), we note that
$d_{m,n}+d_{m+1,n+1}=d_{m,n+1}+d_{m+1,n}$
automatically holds.
Under these conditions, set 
$\sigma_{m,n}(a,b,t)=\tau_{m,n}({\boldsymbol x},{\boldsymbol y})$.
First we prepare a system of bilinear relations satisfied by $\sigma_{m,n}$.
To simplify an expression, we shall write for brevity
$\overline{\sigma}_{m,n}=\sigma_{m,n} (a+1,b,t)$
and 
$\underline{\sigma}{}_{m,n}=\sigma_{m,n} (a-1,b,t)$,
while 
$\sigma_{m,n}=\sigma_{m,n}(a,b,t)$.
Likewise let $\check{\sigma}_{m,n}$ and $\underaccent{\check}{\sigma}_{m,n}$
stand for 
$\sigma_{m,n} (a,b+1,t)$ and $\sigma_{m,n} (a,b-1,t)$,
respectively.

\begin{prop} The functions $\sigma_{m,n}=\sigma_{m,n}(a,b,t)$ satisfy the following bilinear equations{\rm:}
\begin{align}
&(t-1)\sigma_{m,n} \check{\overline{\sigma}}_{m+1,n+1}
- t \check{\sigma}_{m+1,n} \overline{\sigma}_{m,n+1}
+\overline{\sigma}_{m+1,n}\check{\sigma}_{m,n+1}
=0,  \label{eq:bil_p6_1}
\\
& \left( t D_t + b \right)
\sigma_{m+1,n} \cdot \sigma_{m,n+1}
=
b \underaccent{\check}{\sigma}_{m,n} \check{\sigma}_{m+1,n+1},
 \label{eq:bil_p6_2}
 \\
&
\left( -t D_t + d_{m+1,n}-d_{m,n+1}+a \right)
\sigma_{m+1,n} \cdot \sigma_{m,n+1}
=
a \underline{\sigma}{}_{m,n} \overline{\sigma}_{m+1,n+1},
 \label{eq:bil_p6_3}
\\
&
\left(  (t-1)D_t -b \right)
\underline{\sigma}{}_{m,n} \cdot \sigma_{m+1,n}
+b \underaccent{\check}{\sigma}_{m,n} \check{\underline{\sigma}}{}_{m+1,n}
=0,
 \label{eq:bil_p6_4}
\\
&
\left(  t(t-1)D_t +(t-1)(d_{m+1,n}-d_{m,n})-a \right) 
\underaccent{\check}{\sigma}_{m,n} \cdot \sigma_{m+1,n}
+a \underline{\sigma}{}_{m,n}
\underaccent{\check}{\overline{\sigma}}_{m+1,n}
=0.
 \label{eq:bil_p6_5}
\end{align}
\end{prop}

\pf 
First
(\ref{eq:bil_p6_1}) is immediate from (\ref{eq:muc_a}) with
$s=1$.
We see that (\ref{eq:special_p6}) implies
\begin{equation}  \label{eq:p6_flow}
t \frac{{\rm d}}{{\rm d}t} 
=
b \delta_t
\quad 
\text{and} \quad
E= a \delta_1+b \delta_t.
\end{equation}
Hence, (\ref{eq:bil_p6_2}) and (\ref{eq:bil_p6_4}) are direct consequences of (\ref{eq:muc_b}) and (\ref{eq:muc_c}) with $s=1$,
respectively.
By virtue of the homogeneity condition 
$E \tau_{m,n}=d_{m,n} \tau_{m,n}$,
also
(\ref{eq:bil_p6_3}) and (\ref{eq:bil_p6_5})
can be deduced from respectively
(\ref{eq:muc_b}) and (\ref{eq:muc_c}) with 
$(t,s)$ replaced by $(1,t)$.
\qed
\\

Next we shall write down a system of nonlinear differential equations
for appropriately chosen variables.
Consider the functions
$f^{(i)}_{m,n}=f^{(i)}_{m,n}(a,b,t)$ and $g^{(i)}_{m,n}=g^{(i)}_{m,n}(a,b,t)$
$(i=0,1)$
defined by
\begin{align}
f^{(0)}_{m,n}
&= \frac{\overline{\sigma}_{m,n-1}  \sigma_{m-1,n-1}}{\overline{\sigma}_{m-1,n}  \sigma_{m,n-2} }, \quad
f^{(1)}_{m,n}
= \frac{\check{\sigma}_{m,n-1}  \sigma_{m-1,n-1}}{\check{\sigma}_{m-1,n}  \sigma_{m,n-2} }, 
\label{eq:p6_f}
\\
g^{(0)}_{m,n}
&= a \frac{\underline{\sigma}{}_{m-1,n-1}  \overline{\sigma}_{m,n}}{   \sigma_{m-1,n} \sigma_{m,n-1} }, \quad
g^{(1)}_{m,n}
= b \frac{\underaccent{\check}{\sigma}_{m-1,n-1}  \check{\sigma}_{m,n}}{  \sigma_{m-1,n}  \sigma_{m,n-1}  }.
\label{eq:p6_g}
\end{align}
One can find the following conserved quantities
\begin{align}
\prod_{j=1}^\ell f_{m+j,n-j}^{(i)}&=1, \label{eq:preserve_p6_f} \\  
\sum_{j=1}^\ell g_{m+j,n-j}^{(0)}&= \ell a, \quad
\sum_{j=1}^\ell g_{m+j,n-j}^{(1)}= \ell b.
 \label{eq:preserve_p6_g}
\end{align}
Here the first line is immediate by definition of $f$-variables and the second 
can be verified from (\ref{eq:bil_p6_2}) and (\ref{eq:bil_p6_3}).
Furthermore we introduce auxiliary variables $U_{m,n}^{(i,j)}$ and $V_{m,n}^{(i,j)}$
($i,j \in \{0,1\}$, $i \neq j$)  
given as 
\begin{align}
U_{m,n}^{(0,1)}
&= \frac{a t}{1-t} \frac{ \check{\underline{\sigma}}{}_{m,n-1} \overline{\sigma}_{m,n} }{ \sigma_{m,n-1} \check{\sigma}_{m,n} },
\quad
U_{m,n}^{(1,0)}
= \frac{b }{t-1} \frac{ \underaccent{\check}{\overline{\sigma}}{}_{m,n-1} \check{\sigma}_{m,n} }{ \sigma_{m,n-1} \overline{\sigma}_{m,n} },
\\
V_{m,n}^{(0,1)}
&= \frac{a}{1-t} \frac{ \check{\underline{\sigma}}{}_{m-1,n} \overline{\sigma}_{m,n} }{ \sigma_{m-1,n} \check{\sigma}_{m,n} },
\quad
V_{m,n}^{(1,0)}
= \frac{b t }{t-1} \frac{ \underaccent{\check}{\overline{\sigma}}{}_{m-1,n} \check{\sigma}_{m,n} }{ \sigma_{m-1,n} \overline{\sigma}_{m,n} }.
\end{align}
Then we observe that 
\begin{equation}
V_{m,n}^{(i,j)}-U_{m,n}^{(i,j)}=g_{m,n}^{(i)} \quad
\text{and} \quad
\frac{U_{m-1,n}^{(i,j)}}{ V_{m,n-1}^{(i,j)} }
= \frac{t_j f_{m,n}^{(j)}}{t_i f_{m,n}^{(i)} }
\end{equation}
with $(t_0,t_1)=(1,t)$.
Here the former is a consequence of (\ref{eq:bil_p6_1}) and the latter follows just from (\ref{eq:p6_f}). 
Solving the above linear equations leads us to 
\begin{align*}
U_{m,n}^{(i,j)}
&= \frac{1}{\left( \frac{t_i}{t_j}\right)^\ell -1}
\sum_{\alpha=1}^\ell   g^{(i)}_{m-\alpha+1,n+\alpha-1}  
\prod_{\beta=1}^{\alpha-1} \frac{t_i f^{(i)}_{m-\beta+1,n+\beta}}{t_j f^{(j)}_{m-\beta+1,n+\beta}}, \\
V_{m,n}^{(i,j)}
&= \frac{1}{\left( \frac{t_i}{t_j}\right)^\ell -1}
\sum_{\alpha=1}^\ell   g^{(i)}_{m-\alpha,n+\alpha}  
\prod_{\beta=0}^{\alpha-1} \frac{t_i f^{(i)}_{m-\beta,n+\beta+1}}{t_j f^{(j)}_{m-\beta,n+\beta+1}}.
\end{align*}
Namely, in view of (\ref{eq:preserve_p6_f}),
we can actually express  $U_{m,n}^{(i,j)}$ and $V_{m,n}^{(i,j)}$ as polynomials
in $f^{(i)}_{m,n}$ and $g^{(i)}_{m,n}$.

\begin{thm}  \label{thm:p6-chain}
The functions
$f^{(i)}_{m,n}$ and $g^{(i)}_{m,n}$
$(i=0,1)$ satisfy the system of ordinary differential equations
\begin{subequations} \label{subeq:p6-chain}
\begin{align}
t \frac{{\rm d} f_{m,n}^{(1)}}{{\rm d} t}
&= \left(
\alpha_{m,n} -g_{m,n-1}^{(1)} +U_{m-1,n}^{(0,1)} - V_{m,n-1}^{(0,1)}
\right)  f_{m,n}^{(1)},
 \label{eq:p6-chain_a}
\\
t \frac{{\rm d} f_{m,n}^{(0)}}{{\rm d} t}
&=\left(-g_{m,n-1}^{(1)}-U_{m-1,n}^{(1,0)} +V_{m,n-1}^{(1,0)}
\right) f_{m,n}^{(0)},
 \label{eq:p6-chain_b}
\\
t \frac{{\rm d} g_{m,n}^{(1)}}{{\rm d} t}
&= -U_{m,n}^{(1,0)}g_{m,n}^{(0)}-V_{m,n}^{(0,1)} g_{m,n}^{(1)},
 \label{eq:p6-chain_c}
\\
t \frac{{\rm d} g_{m,n}^{(0)}}{{\rm d} t}
&= U_{m,n}^{(0,1)}g_{m,n}^{(1)}+V_{m,n}^{(1,0)} g_{m,n}^{(0)},
 \label{eq:p6-chain_d}
\end{align}
\end{subequations}
where
\begin{equation}
 \label{eq:p6_alpha}
\alpha_{m,n}=a+b+d_{m,n-1}-d_{m-1,n}
= g_{m,n}^{(0)}+g_{m,n}^{(1)} \in {\mathbb C} 
\end{equation}
are constant parameters.
\end{thm}

\pf
We shall demonstrate only (\ref{eq:p6-chain_a}) and  (\ref{eq:p6-chain_d}) 
because the others 
can be verified in a similar manner.
First we notice from (\ref{eq:bil_p6_1}) that the formulae
\begin{equation}
\check{g}_{m,n}^{(0)}= V_{m,n-1}^{(0,1)}-U_{m-1,n}^{(0,1)}
\quad 
\text{and}
\quad
\overline{g}_{m,n}^{(1)}= V_{m,n-1}^{(1,0)}-U_{m-1,n}^{(1,0)}
\end{equation}
hold.
Taking the logarithmic derivative of $f_{m,n}^{(1)}$ shows 
\begin{align*}
\frac{t}{f_{m,n}^{(1)}}  \frac{{\rm d} f_{m,n}^{(1)}}{{\rm d} t}
&= \frac{t D_t \check{\sigma}_{m,n-1} \cdot  \check{\sigma}_{m-1,n}}{\check{\sigma}_{m,n-1} \check{\sigma}_{m-1,n}}
-\frac{t D_t \sigma_{m,n-2} \cdot \sigma_{m-1,n-1}}{\sigma_{m,n-2}\sigma_{m-1,n-1}}\\
&= \left( d_{m,n-1}-d_{m-1,n}+a
-a 
\frac{ \check{\underline{\sigma}}{}_{m-1,n-1} \check{\overline{\sigma}}_{m,n}}{ \check{\sigma}_{m,n-1} \check{\sigma}_{m-1,n} }
\right)
\\
& \quad + 
\left(b - b \frac{ \underaccent{\check}{\sigma}_{m-1,n-2} \check{\sigma}_{m,n-1} }{ \sigma_{m,n-2}\sigma_{m-1,n-1} }
\right),
\quad \text{using (\ref{eq:bil_p6_2}) and (\ref{eq:bil_p6_3})}
\\
&
= \alpha_{m,n}-g_{m,n-1}^{(1)}-\check{g}_{m,n}^{(0)},
\end{align*}
which thus implies (\ref{eq:p6-chain_a}).
Likewise we observe that
\begin{align*}
\frac{t}{g_{m,n}^{(0)}} \frac{{\rm d} g_{m,n}^{(0)}}{{\rm d} t}
&=  
\frac{t D_t \underline{\sigma}{}_{m-1,n-1} \cdot \sigma_{m,n-1}}{\underline{\sigma}{}_{m-1,n-1}\sigma_{m,n-1}}
-
\frac{ t D_t \sigma_{m-1,n} \cdot \overline{\sigma}_{m,n}}{ \sigma_{m-1,n} \overline{\sigma}_{m,n}}
\\
&=\frac{bt}{t-1} \left(- 
\frac{ \underaccent{\check}{\sigma}_{m-1,n-1}  \check{\underline{\sigma}}{}_{m,n-1}  }{   \underline{\sigma}{}_{m-1,n-1} \sigma_{m,n-1}  }
+
\frac{ \underaccent{\check}{\overline{\sigma}}_{m-1,n}  \check{\sigma}{}_{m,n}  }{   \sigma_{m-1,n} \overline{\sigma}_{m,n}  }
\right),
\quad \text{using (\ref{eq:bil_p6_4})}
\\
&=\frac{bt}{t-1} \left(- 
 \frac{a g_{m,n}^{(1)}}{b g_{m,n}^{(0)}}
\frac{ \check{\underline{\sigma}}{}_{m,n-1} \overline{\sigma}_{m,n} }{ \sigma_{m,n-1} \check{\sigma}_{m,n} }
+
\frac{ \underaccent{\check}{\overline{\sigma}}_{m-1,n}  \check{\sigma}{}_{m,n}  }{   \sigma_{m-1,n} \overline{\sigma}_{m,n}  }
\right)
\\
&= U_{m,n}^{(0,1)} \frac{g_{m,n}^{(1)}}{g_{m,n}^{(0)}}+V_{m,n}^{(1,0)},
\end{align*}
which coincides with (\ref{eq:p6-chain_d}).
\qed
\\

Remark that for each $(m,n)$ given the system is closed with respect to 
the $2 \ell$-tuple of variables: 
\[g_{m+j,n-j}=g_{m+j,n-j}^{(1)} \quad
\text{and} \quad 
h_{m+j,n-j+1}=\frac{f_{m+j,n-j+1}^{(1)}}{f_{m+j,n-j+1}^{(0)}}
\quad
\text{for $j \in {\mathbb Z}/\ell {\mathbb Z}$}.
\]
If we take into account the conservation 
(\ref{eq:preserve_p6_f}) and (\ref{eq:preserve_p6_g}),
then the essential dimension of the phase space turns out to be 
$2 \ell -2$.
As demonstrated below the case $\ell=2$ is equivalent to the sixth Painlev\'e equation  
$P_{\rm VI}$, 
which is, needless to say, of second order.
We shall call (\ref{subeq:p6-chain}) the {\it $P_{\rm VI}$-chain}.

\subsection{Example: $(2,2)$-periodic case and $P_{\rm VI}$}
Consider the case where $\ell=2$.
We are interested in a system of differential equations
satisfied by the variables 
$g=g_{1,0}^{(1)}$ and $h=f_{1,1}^{(1)}/f_{1,1}^{(0)}$.
From Theorem~\ref{thm:p6-chain} we observe that 
\begin{align*}
(t^2-1) \frac{{\rm d} g}{{\rm d}t} 
&=
-(h-h^{-1})g(g-2b)
+(\alpha_{1,0} h+ \alpha_{0,1} h^{-1})g
-2 \alpha_{1,0}b h,
\\
(t^2-1) \frac{{\rm d} h}{{\rm d}t} &=
(h-t)(h-t^{-1})(2g-2b-\alpha_{1,0} )
-(\alpha_{0,0} t +\alpha_{1,1} t^{-1})h
+2(a+b).
\end{align*}
Let us take the change of variables $(g,h) \mapsto (q,p)$ defined by
\begin{align*}  
q&= \frac{h(g-2 b)}{t(g+2a-\alpha_{1,0}) }
=\frac{-b}{a t} \frac{ \check{\sigma}_{1,0} \underaccent{\check}{\sigma}_{1,0} }{ \overline{\sigma}_{1,0}\underline{\sigma}{}_{1,0} }, 
\\
p&= \frac{g}{2 q} 
= \frac{-a t}{2} 
\frac{ \overline{\sigma}_{1,0} \underline{\sigma}{}_{1,0} \underaccent{\check}{\sigma}_{0,1}}{ \sigma_{0,0} \sigma_{1,1} \underaccent{\check}{\sigma}_{1,0}}.
\end{align*}
Put $s=1/t^2$.
We then  arrive at the Hamiltonian system 
\[
\frac{{\rm d} q }{ {\rm d} s } =\frac{\partial H}{ \partial p}
\quad
\text{and} \quad 
\frac{{\rm d} p }{ {\rm d} s } =-\frac{\partial H}{ \partial q}
\]
with the Hamiltonian function $H=H(q,p;s)$ 
given as 
\[
s(s-1) H
= q(q-1)(q-s)p^2
-\left(
(\kappa_0-1)q(q-1) +
\kappa_1 q(q-s)
+ \theta (q-1)(q-s)
\right) p
+ \kappa q,
\]
where
$\kappa_0=d_{1,0}-d_{0,0}+{1}/{2}$,
$\kappa_1=d_{0,1}-d_{0,0}+{1}/{2}$, 
$\theta= b$,  and $\kappa= \alpha_{1,0}(\alpha_{1,0}-2a)/4$.
This is exactly the Hamiltonian form of $P_{\rm VI}$;
see \cite{mal22, oka_p6}.

Note that also the $P_{\rm VI}$-chain, (\ref{subeq:p6-chain}), 
for a general $\ell$ case can be transformed into a Hamiltonian system
whose Hamiltonian function is a polynomial in the canonical variables.
For details to \cite{tsu09b}.

\subsection{Lax formalism} \label{subsect:p6_lax}
The homogeneity $E \tau_{m,n}=d_{m,n} \tau_{m,n}$
ensures the formula
$(E-k {\partial}/{\partial k})\psi_{m,n}=(d_{m,n-1}-d_{m,n}) \psi_{m,n}$;
see Lemma~\ref{lemma:p5_eular}.
Set 
$\phi_{m,n}(a,b,t,k)=\psi_{m,n}({\boldsymbol x},{\boldsymbol y},k)$ under the substitution (\ref{eq:special_p6}).
Then, 
Lemma~\ref{lemma:uc_lax}
 together with (\ref{eq:p6_flow})
leads us to the 
\begin{lemma} 
The wave functions 
$\phi_{m,n}=\phi_{m,n}(a,b,t,k)$
satisfy the following linear equations{\rm:}
\begin{align}
\phi_{m,n}
&=\frac{1}{f_{m+1,n+1}^{(0)}} \overline{\phi}_{m,n+1}
-k \overline{\phi}_{m+1,n},
\label{eq:p6_lax_1}
\\
\phi_{m,n}
&=\frac{1}{f_{m+1,n+1}^{(1)}} \check{\phi}_{m,n+1}
-t k \check{\phi}_{m+1,n},
\label{eq:p6_lax_2}
\\
t \frac{\partial }{\partial t} \phi_{m,n}
&= \left(  g_{m+1,n}^{(1)} -b \right) \phi_{m,n} + t k  g_{m+1,n}^{(1)} \check{\phi}_{m+1,n},
\label{eq:p6_lax_3}
\\
\left(
k \frac{\partial }{\partial k}-t \frac{\partial }{\partial t}
\right) \phi_{m,n}
&= (d_{m,n}-d_{m,n-1}) \phi_{m,n}+
\left(  g_{m+1,n}^{(0)} -a\right) \phi_{m,n} + k  g_{m+1,n}^{(0)} \overline{\phi}_{m+1,n}.
\label{eq:p6_lax_4}
\end{align}
Here we have used the abbreviations 
$\overline{\phi}_{m,n}=\phi_{m,n}(a+1,b,t,k)$
and 
$\check{\phi}_{m,n}=\phi_{m,n}(a,b+1,t,k)$.
\end{lemma}

Due to the $(\ell,\ell)$-periodicity
one can solve the linear equations (\ref{eq:p6_lax_1}) for $\overline{\phi}_{m,n}$;
thus,
\[
\overline{\phi}_{m,n}= 
\frac{1}{1-k^\ell}
\sum_{j=1}^{\ell} k^{j-1}  
\left(
 \prod_{i=1}^j  f_{m+i,n-i+1}^{(0)}
\right)
\phi_{m+j-1,n-j} .
\]
Likewise (\ref{eq:p6_lax_2}) tells us that
\[
\check{\phi}_{m,n}= \frac{1}{1-(t k)^\ell}
\sum_{j=1}^{\ell} (t k)^{j-1}  
\left(
 \prod_{i=1}^j  f_{m+i,n-i+1}^{(1)}
\right)
\phi_{m+j-1,n-j}.
\]

Firstly, we put our attention to  
the linear differential equation with respect to  the spectral variable $k$.
Noticing (\ref{eq:p6_alpha}) and combining  
(\ref{eq:p6_lax_3})
with (\ref{eq:p6_lax_4}),
we therefore obtain 
\begin{align*}
k \frac{\partial}{\partial k} \phi_{m,n}
&= (d_{m+1,n-1}-d_{m,n-1}) \phi_{m,n}
+\frac{g_{m+1,n}^{(0)}}{1-k^\ell}
\sum_{j=1}^{\ell} k^{j}  
\left(
 \prod_{i=1}^j  f_{m+i+1,n-i+1}^{(0)}
\right)
\phi_{m+j,n-j} 
\\
& \quad
+\frac{g_{m+1,n}^{(1)}}{1-(t k)^\ell}
\sum_{j=1}^{\ell} (t k)^{j}  
\left(
 \prod_{i=1}^j  f_{m+i+1,n-i+1}^{(1)}
\right)
\phi_{m+j,n-j},
\end{align*}
which has the $2 \ell +2$ regular singularities at
$k=0, \infty, 
\exp \left( 2 \pi \sqrt{-1}n/\ell \right),
t^{-1} \exp \left( 2 \pi \sqrt{-1}n/\ell\right)$
$(n \in {\mathbb Z}/\ell{\mathbb Z})$.
But, however, this expression is quite redundant.
Let 
\[
\Phi={}^{\rm T}\left( \phi_{0,0}, k \phi_{1,-1},   k^2 \phi_{2,-2}, \ldots, k^{\ell-1}  \phi_{\ell-1, -\ell +1} \right),
\quad
z=k^\ell, \quad
\text{and} \quad 
s=1/t^\ell.
\]
Then
we obtain the equation
\begin{equation}  \label{eq:p6_lax_A}
\frac{\partial \Phi}{\partial z}=A \Phi
=
\left(\frac{A_0}{z} + \frac{A_1}{z-1}+ \frac{A_{s}}{z-s} \right) \Phi,
\end{equation}  
where
the $\ell \times \ell$ matrices 
$A_0$, $A_1$, and $A_s$
read
\begin{align*}
A_0 &= 
 \left(  
   \begin{array}{cccc}
   e_0 &   w_{0,1}      & \cdots  & w_{0,\ell-1}
   \\
        & e_1 &  \ddots       & \vdots
   \\     
          &        &\ddots& w_{\ell-2,\ell-1}
   \\
          &        &          & e_{\ell-1}    
   \end{array}
 \right),
 \\
 A_1&= -\left(v_{i,j}^{(0)}\right)_{0 \leq i,j \leq \ell-1}, \\
 A_s &= 
- \left(  
   \begin{array}{cccc}
   0 & v_{0,1}^{(1)}   &    \cdots & v_{0,\ell-1}^{(1)}  
   \\
       & 0&  \ddots    & \vdots
   \\      
          &        & \ddots & v_{\ell-2,\ell-1}^{(1)} 
   \\
          &        &          & 0
   \end{array}
 \right) 
 - s 
  \left(  
   \begin{array}{cccc}
   v_{0,0}^{(1)}  &    & &  \text{\LARGE $O$} 
   \\
    v_{1,0}^{(1)}    &v_{1,1}^{(1)}  &        & 
   \\      
     \vdots     &  \ddots  & \ddots& 
   \\
  v_{\ell-1,0}^{(1)}   & v_{\ell-1,1}^{(1)} &\cdots& v_{\ell-1,\ell-1}^{(1)} 
   \end{array}
 \right)
\end{align*}
with
$e_n= (d_{n+1,-n-1}-d_{n,-n-1}+n)/\ell$ and
$w_{i,j}=  v_{i,j}^{(0)}+v_{i,j}^{(1)}$; we set
\begin{align*}
v_{n,n+j}^{(0)}&=
\frac{g_{n+1,-n}^{(0)}}{\ell}
 \prod_{i=1}^j  f_{n+i+1,-n-i+1}^{(0)},
\\
 v_{n,n+j}^{(1)}&=
\frac{g_{n+1,-n}^{(1)}}{\ell}
t^j \prod_{i=1}^j  f_{n+i+1,-n-i+1}^{(1)}
\end{align*}
for 
$0 \leq n \leq \ell-1$
and
$1 \leq j \leq \ell$.
Note that we read appropriately the suffixes of dependent variables modulo $\ell$ as before.
The equation (\ref{eq:p6_lax_A})  still remains Fuchsian and has 
the four regular singularities  $z=0,1,s,\infty$.
The exponents at each singularity are listed in the following table
(Riemann scheme):
\begin{equation} \label{eq:riemann}
\begin{array}{cc}
\hline
\text{Singularity} & \text{Exponents}  \\ \hline
z=0 & (e_0,e_1,\ldots,e_{\ell-1})  \\ 
z=1 & (-a, 0, \ldots,0)  \\ 
z=s & (-b, 0, \ldots,0)  \\       
z=\infty & \left(\frac{\alpha_{1,0}}{\ell}-e_0, \frac{\alpha_{2,-1}}{\ell}-e_1, \ldots,\frac{\alpha_{\ell,-\ell+1}}{\ell}-e_{\ell-1}
\right)  \\ \hline
   \end{array}
\end{equation}
Note that both $A_1$ and $A_s$ are not full rank (but rank one),
unlike $A_0$ and $A_\infty =-(A_0+A_1+A_s)$.
For example let us look at $A_s$.
Fix the row vectors 
\begin{align*}
{\boldsymbol f}
&=\left(t^j \prod_{i=1}^j  f_{ i+1,-i+1 }^{(1)}
\right)_{0 \leq j \leq \ell-1}
=
\left(1,t f_{2,0}^{(1)}, t^2  f_{2,0}^{(1)}f_{3,-1}^{(1)}, \ldots
, t^{\ell-1}  f_{2,0}^{(1)}f_{3,-1}^{(1)} \cdots f_{0,2}^{(1)} 
\right), \\
{\boldsymbol g}
&=\left( \frac{g_{j+1,-j}^{(1)} }{t^{j}\prod_{i=1}^j f_{ i+1,-i+1 }^{(1)} }
\right)_{0 \leq j \leq \ell-1}
=\left(
g_{1,0}^{(1)}, \frac{ g_{2,-1}^{(1)} }{ t f_{2,0}^{(1)}},
 \frac{ g_{3,-2}^{(1)} }{ t^2 f_{2,0}^{(1)} f_{3,-1}^{(1)}},
 \ldots,
  \frac{ g_{\ell,-\ell+1}^{(1)} }{ t^{\ell-1} f_{2,0}^{(1)} f_{3,-1}^{(1)}  \cdots f_{0,2}^{(1)} }
\right).
\end{align*}
We see indeed that $A_s$ is expressed as
$-\ell A_s={}^{\rm T}{\boldsymbol g} \cdot {\boldsymbol f}$
and that
$ {\boldsymbol f} \cdot {}^{\rm T}{\boldsymbol g} = 
\sum_{j=0}^{\ell-1} g_{j+1,-j}^{(1)} = \ell b$.
Also, the case of $A_1$ can be checked in the same way.

Secondly, we obtain from (\ref{eq:p6_lax_3}) the deformation equation
with respect to $s=1/t^\ell$:
\begin{equation}  \label{eq:p6_lax_B}
\frac{\partial}{\partial s} \Phi
=  B \Phi,
\end{equation}
where
\begin{align*}
B&= {\rm diag} \left( \frac{b}{\ell s} -v_{i,i}^{(1)} \right)_{0 \leq i \leq \ell-1}
\\
& \quad
+ \frac{1}{z-s}
 \left(  
   \begin{array}{cccc}
   0 & v_{0,1}^{(1)}   &    \cdots & v_{0,\ell-1}^{(1)}  
   \\
       & 0&  \ddots    & \vdots
   \\      
          &        & \ddots & v_{\ell-2,\ell-1}^{(1)} 
   \\
          &        &          & 0
   \end{array}
 \right) 
 +\frac{z}{z-s} 
  \left(  
   \begin{array}{cccc}
   v_{0,0}^{(1)}  &    & &  \text{\LARGE $O$} 
   \\
    v_{1,0}^{(1)}    &v_{1,1}^{(1)}  &        & 
   \\      
     \vdots     &  \ddots  & \ddots& 
   \\
  v_{\ell-1,0}^{(1)}   & v_{\ell-1,1}^{(1)} &\cdots& v_{\ell-1,\ell-1}^{(1)} 
   \end{array}
 \right).
\end{align*}

Finally, we can again recover the $P_{\rm VI}$-chain from the integrability condition 
$\left[\frac{\partial}{\partial s}-B, \frac{\partial}{\partial z}-A\right]=0$
of the system (\ref{eq:p6_lax_A}) and (\ref{eq:p6_lax_B}).
Since (\ref{eq:p6_lax_A}) is Fuchsian, the $P_{\rm VI}$-chain
turns out to be equivalent to a Schlesinger system 
(\cite{sch12})
specified by the Riemann scheme (\ref{eq:riemann}).

\begin{remark} \rm
If $\ell=2$, the above system 
is equivalent to the Lax pair for $P_{\rm VI}$ given in \cite{jm81, sch12}.

\end{remark}

\appendix
\section{From KP hierarchy to Painlev\'e I equation}
\label{sect:p1}

In this appendix, we demonstrate 
the derivation of the first Painlev\'e
equation ($P_{\rm I}$)
from the two-reduced KP hierarchy, i.e., KdV hierarchy,
through a certain reduction procedure by the use of a Virasoro operator.
Such a relationship between $P_{\rm I}$ and the KP hierarchy was first 
recognized in the context of `two-dimensional quantum gravity';
see, e.g., \cite{am92, fkn92, kon92}.
We also show that the Lax pair for $P_{\rm I}$ 
naturally 
arises from the associated linear equations of the KP hierarchy.

\subsection{Reduction by using a Virasoro operator}
Introduce a differential operator 
(a {\it Virasoro operator}):
\begin{equation*}
L(=L_{-2})=\frac{{x_1}^2}{2}+\sum_{n=1}^\infty (n+2)x_{n+2} \frac{\partial}{\partial x_n}.
\end{equation*}
Let $\tau=\tau({\boldsymbol x})$ be a solution of 
the KP hierarchy, (\ref{eq:kp}) or (\ref{eq:kp-res}),
that fulfills  
the following conditions:
\begin{align}
&\frac{\partial \tau}{\partial x_{2 n}} \equiv 0 \quad (n=1,2,\ldots),
\label{eq:kdv_cond}
\\
&L \tau({\boldsymbol x})= c\tau({\boldsymbol x}) \quad (c \in {\mathbb C}).
\label{eq:virasoro_cond}
\end{align}
The first condition means that $\tau$ is a solution of the KdV hierarchy.
For instance, it satisfies
the KdV equation 
(cf. (\ref{eq:kp-orig})):
\begin{equation}  \label{eq:kdv}
\left(  {D_{x_1}}^4 -4 D_{x_1} D_{x_3}
\right) \tau \cdot \tau=0.
\end{equation}
Set 
$\sigma(x)=\tau({\boldsymbol x})$
under the specialization 
\begin{equation}  \label{eq:p1_special}
x_1=x, \quad x_5=\frac{4}{5}, \quad  \text{and} \quad  x_n =0 \quad (n \neq 1,5).
\end{equation}

\begin{prop} A function $\sigma=\sigma(x)$ satisfies 
the bilinear differential equation
\begin{equation}  \label{eq:p1_bil}
\left({D_x}^4+2x \right) \sigma \cdot \sigma =0.
\end{equation}
\end{prop}

\pf
Observing that 
\begin{equation} \label{eq:p1_flow}
\frac{\partial}{\partial x_1}=\frac{\rm d}{{\rm d} x}
\quad \text{and}
 \quad
4  \frac{\partial}{\partial x_3}= -\frac{{x}^2}{2}+L,
\end{equation}
we can deduce (\ref{eq:p1_bil}) 
from (\ref{eq:kdv}) immediately. 
\qed
\\

If we take the variable
$q = -({\rm d}/{\rm d} x)^2 \log \sigma$,
then
(\ref{eq:p1_bil}) is equivalently rewritten into 
the first Painlev\'e equation $P_{\rm I}$:
\[
\frac{{\rm d}^2 q}{{\rm d} x^2} =6 q^2+x.
\]

\subsection{Lax formalism}

We first prepare some formulae for the wave function.
Set
\[
\rho({\boldsymbol x},k)= X^+(k) \tau({\boldsymbol x})
=\tau({\boldsymbol x}-[k^{-1}])e^{\xi({\boldsymbol x},k)}.
\]
The KP hierarchy (\ref{eq:kp}) multiplied by 
 $1 \otimes X^+(k)$
yields
$\sum_{i+j=-2}  X_i^- \tau \otimes X_j^+ \rho =0$,
namely,
\[
G_1(\tau({\boldsymbol x}) ,  \rho({\boldsymbol x},k ); {\boldsymbol u})=0;
\]
recall (\ref{eq:formula_bil2hirota}).
The coefficients of $1={\boldsymbol u}^0$
and $u_1$ of the above equation 
show respectively that 
\begin{align*}
&\left(  {D_{x_1}}^2 +D_{x_2}   \right) \tau \cdot \rho =0,
\\
&
\left(  {D_{x_1}}^3 -3 D_{x_1} D_{x_2} -4  D_{x_3}  \right) \tau \cdot \rho =0.
\end{align*}
Hence, for the wave function 
$\psi({\boldsymbol x},k)=\rho({\boldsymbol x},k)/\tau({\boldsymbol x})$,
we find that
\begin{align}
&\left(
\frac{\partial^2}{\partial {x_1}^2} -\frac{\partial}{\partial x_2}
+ \frac{{D_{x_1}}^2 \tau \cdot \tau}{\tau^2} \right) \psi =0,
\label{eq:kp_wave_formula1}
\\
&\left(
\frac{\partial^3}{\partial {x_1}^3} 
+3 \frac{\partial^2}{\partial x_1 \partial x_2}
-4 \frac{\partial}{\partial x_3}
+3 \frac{{D_{x_1}}^2 \tau \cdot \tau}{\tau^2}  \frac{\partial}{\partial x_1}
+3  \frac{D_{x_1} D_{x_2} \tau \cdot \tau}{\tau^2} 
\right)\psi=0.
\label{eq:kp_wave_formula2}
\end{align}

Now, by applying the constraint (\ref{eq:virasoro_cond}),
we can verify the formula
\begin{equation} \label{eq:p1_lax_L}
\left(L- \frac{{x_1}^2}{2} -k^{-1} \frac{\partial}{\partial k} \right) \psi
= \left( -2 x_2 -  \frac{k^{-2}}{2} \right) \psi.
\end{equation}
Under the substitution (\ref{eq:p1_special}), 
we write
$\phi(x,k)=\psi({\boldsymbol x},k)$.
By virtue of (\ref{eq:p1_lax_L}) together with (\ref{eq:p1_flow}),
the lemma below follows readily from 
(\ref{eq:kp_wave_formula1}) and
(\ref{eq:kp_wave_formula2}).

\begin{lemma}
The wave function $\phi=\phi(x,k)$ satisfies the following
linear equations{\rm:}
\begin{align*}
k^{-1}\frac{\partial  \phi}{\partial k}  &=
\left(2 p +\frac{k^{-2}}{2}\right) \phi
+
4 \left( - q +k^2
\right) \frac{\partial \phi}{\partial x}, 
\\
\frac{\partial^2  \phi}{\partial x^2}  &= \left( 2 q+  k^2 \right)\phi,
\end{align*}
where 
$p={\rm d} q/{\rm d}x$.
\end{lemma}

Let 
$\Phi={}^{\rm T}\left(\phi,\partial\phi/\partial x \right)  k^{-1/2}$
and $z=k^2$.
We then arrive at 
the system of $2 \times 2$ matrix equations
\begin{align}  \label{eq:p1_lax_1}
\frac{\partial}{\partial z} \Phi 
&=
A \Phi
= \left(
 \left( \begin{array}{cc}
p & -2 q \\
2 q^2+x & -p
\end{array}\right)
+z 
\left( 
\begin{array}{cc}
0 & 2 \\
2 q & 0
\end{array}\right)
+z^2 
\left( 
\begin{array}{cc}
0 & 0 \\
2  & 0
\end{array}\right)
\right) \Phi,
\\
 \label{eq:p1_lax_2}
\frac{\partial}{\partial x} \Phi 
&=
B \Phi
= \left( 
\left( 
\begin{array}{cc}
0 & 1 \\
2 q & 0
\end{array}\right)
+ z \left( 
\begin{array}{cc}
0 & 0 \\
1  & 0
\end{array}\right)  \right) \Phi.
\end{align}
We see that (\ref{eq:p1_lax_1}) has only one irregular singularity at
$z=\infty$
whose Poincar\'e rank is $5/2$.
The latter, (\ref{eq:p1_lax_2}),
governs the monodromy preserving deformation
of the former.
This system is identical to the Lax pair 
for $P_{\rm I}$ given in \cite{jm81}. 
Indeed, 
$P_{\rm I}$ can be recovered 
as
its integrability condition.

\small
\paragraph{\it Acknowledgement.}
I would like to express my sincere gratitude
to Saburo Kakei, Yuji Terashima,
and Yasuhiko Yamada 
for their helpful comments/suggestions especially on the content of 
Sect.~\ref{subsec:hom}.
I wish to thank Hidetaka Sakai
for his kind information about Hukuhara's result \cite{huk}. 
I have also benefited from discussions with 
Kenji Kajiwara, Tetsu Masuda, 
Masatoshi Noumi, Yasuhiro Ohta,  Marius van der Put,
Kanehisa Takasaki,
 and Takashi Takebe.
 This manuscript was prepared during my stay in 
 the Issac Newton Institute for Mathematical Sciences 
at the program ``Discrete Integrable Systems" 
(2009).
My research is supported by a grant-in-aid from the Japan Society for the Promotion of Science (JSPS).


\end{document}